\documentclass{article}
\usepackage{amsmath, amsthm, amssymb}
\usepackage[pdftex]{graphicx}
\usepackage{multirow}
\usepackage{enumerate}
\usepackage{subfig}
\usepackage{hyperref}

\title{Harmonic functions and the spectrum of the Laplacian on the Sierpinski carpet}
\author{Matthew Begu\'{e}, Tristan Kalloniatis \\ and Robert S. Strichartz
\footnote{The research of the first author was supported by the National Science Foundation through the Research Experiences for Undergraduates program at Cornell University.  The research of the third author was supported in part by the National Science Foundation, grant DMS-0652440.}}

\numberwithin{equation}{section}
\numberwithin{figure}{section}
\numberwithin{table}{section}

\def \Sum{\displaystyle\sum}

\begin{document}
\maketitle

\begin{center}
\small
Contacts:\\
begue@math.umd.edu\\
Matthew Begu\'{e}\\
Department of Mathematics\\
University of Maryland\\
College Park, MD 20742 USA
\vskip 3mm
tk346@cam.ac.uk\\
Tristan Kalloniatis\\
Faculty of Mathematics\\
University of Cambridge\\
Cambridge  CB3 0WA UK
\vskip 3mm
str@math.cornell.edu\\
Robert S. Strichartz\\
Department of Mathematics, Malott Hall\\
Cornell University\\
Ithaca, NY 14853 USA
\end{center}

\begin{abstract}
Kusuoka and Zhou have defined the Laplacian on the Sierpinski carpet using average values of functions on small cells and the graph structure of cell adjacency. We have implemented an algorithm that uses their method to approximate solutions to boundary value problems. As a result we have a wealth of data concerning harmonic functions with prescribed boundary values, and eigenfunctions of the Laplacian with either Neumann or Dirichlet boundary conditions. We will present some of this data and discuss some ideas for defining normal derivatives on the boundary of the carpet.
\end{abstract}

\section{Introduction}

\begin{figure}[h]
\begin{center}
\includegraphics[scale=.4]{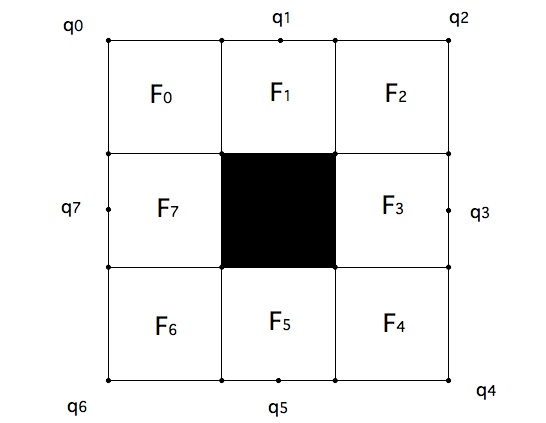}
\end{center}
\caption{The level 1 graph approximation of SC with the eight cells labeled.}\label{SC1figures}
\end{figure}

The Sierpinski carpet, SC, is the self similar fractal in the plane determined by the identity
\begin{equation}
SC=\bigcup_{i=0}^7 F_i (SC)
\end{equation}
where $\{F_i\}$ are contraction mappings defined by 
\begin{equation}
F_i(x)=1/3(x-q_i)+q_i,
\end{equation}
where $\{q_i\}$ are the eight points shown in Figure \ref{SC1figures}.  

By convention, the boundary of SC, $\partial$SC, is defined to be the boundary of the unit square containing all of SC.

We follow the usual convention for iterated function systems like $\{F_i\}$, let $\omega=(\omega_1,...,\omega_m)$ denote a word of length $m$ whose each $\omega_j$ takes on values in $\{0,1,2,3,4,5,6,7\}$ and $F_\omega=F_{\omega_1} \circ F_{\omega_2} \circ \cdots \circ F_{\omega_m}$.  We call $F_\omega(\text{SC})$ an $m$-cell, and let SC$_m$ denote the graph whose vertices are the $m$-cells and with edges connecting two $m$-cells that intersect along an edge.

Originally there were two different constructions of a self-similar energy and Laplacian on SC with the full $D_4$ group of symmetries, one due to Barlow and Bass \cite{BB} and one due to Kusuoka and Zhou \cite{KZ}.  It was recently shown that the two constructions are equivalent \cite{BBKT}, and in fact the energy and Laplacians are unique up to a constant multiple.  In particular this implies that the graph Laplacians on SC$_m$, when appropriately normalized, converge to the Laplacian on SC$_m$ (in \cite{KZ} this was only proved for a subsequence of $m$'s).  In this paper we systematically use this observation to approximate many of the basic functions on SC: eigenfunctions of the Laplacian (and their eigenvalues) under various boundary conditions, harmonic functions with prescribed boundary conditions, Poisson kernels, heat kernels, and Dirichlet kernels. (A different approximation method was used in \cite{BHS}, but it does not appear to be as versatile.)  In addition we study analogous problems on fractafolds and covering spaces of SC.  For example, one of our most striking experimental observations is that the fractafolds obtained from SC by gluing the boundaries according to the torus, Klein bottle, or projective plane model, have eigenvalue counting functions that differ from each other by uniformly bounded amounts.  This is a stronger statement than is true for the standard Laplacian on the torus, Klein bottle, and projective plane.

We note that SC is quite different from the PCF fractals (\cite{Ki,S3}) such as the Sierpinski gasket SG, where there is a simpler construction of a Laplacian.  In particular, for SG there is a theory of spectral decimation (\cite{FS}) that allows the exact computation of eigenfunctions of the Laplacian.  In absence of precise theoretical tools on SC, the experimental approach becomes even more important.  We will discuss the implications of our numerical results in the concluding section of this paper.

We present a brief discussion of the method of averages in section 2.  In section 3 we present our results on harmonic functions, including a discussion of the Poisson kernel and the effective resistance metric.  In section 4 we discuss the spectrum of the Laplacian under Dirichlet and Neumann boundary conditions, and also for fractafolds \cite{S1} obtained by identifying the boundary as in the construction of the torus, Klein bottle, and projective plane from the square.  We discuss many properties of the eigenvalues, and also present some graphs of the eigenfunctions.  The actual eigenfunctions are crucial to the subsequent computations in this paper.  In section 5 we discuss the heat kernel, both the on-diagonal and off-diagonal behaviors.  In section 6 we examine the decay rates at the boundary of some of the functions we have constructed.  Section 7 is devoted to the spectrum of two examples of covering spaces with a $\mathbb{Z}$-action, in the spirit of results for covering spaces of Julia sets in \cite{ADS}.  Section 8 contains a concluding discussion. 

One of the motivating problems for our work was the challenge of trying to define a normal derivative on the boundary of SC for functions in the domain of the Laplacian in such a way that a Gauss-Green formula is valid.  (Indeed this is one of the basic tools for studying Laplacians on PCF fractals.)  In section 6 we discuss our experimental observations that indicate why this is a very challenging problem.  So, although our work does not resolve this problem, it may provide guidance for future work.  We present a small sampling of our numerical results here.  More data, and the programs used to generate them, may be found on the website \url{http://www.math.cornell.edu/~reu/sierpinski-carpet/}.

\section {Method of Averages}
The \emph{method of averages} was introduced by Kusuoka and Zhou in \cite{KZ} for functions on SC.  Instead of function values defined at vertices in fractal approximation graphs, we consider average values on entire $m$-cells.  Let $\{A_j^{(m)}\}$ be the collection of all $m$-cells.  Then we define $f(A_j^{(m)})= 8^m \int_{A_j^{(m)}} f\, d\mu$.

We reference a specific $m$-cell by the \emph{cell address} which is the sequence of integers corresponding to the contraction maps that map the unit square to the particular cell.  For notation, we list the cell address in single quotes and in the opposite order that it would be read as function composition.  For example, if $I$ is the unit square, then the $3$-cell given by $F_0 \circ F_3 \circ F_2 (I)$ is `230'.

We write $A_j^{(m)}\sim A_k^{(m)}$ if the two cells have a boundary edge in common.  We say that $A_j^{(m)}$ and $A_k^{(m)}$ are neighbors.
It can be verified that any cell in SC$_m$ will have either 2, 3, or 4 neighbors.  There exist algorithms that give the number and specific cell addresses of the neighbors of any $m$-cell and are available on the website.

Using this method of averages we can now define discrete approximations using graph energy and graph Laplacian on level $m$:

\begin{equation}\label{graphenergy}
E_m(f) = \Sum_{A_j^{(m)}\sim A_k^{(m)}} \left( f(A_j^{(m)})-f(A_k^{(m)})\right)^2,
\end{equation}
\begin{equation}\label{graphlaplacian}
\Delta_m f(A_j^{(m)})=\Sum_{A_j^{(m)}\sim A_k^{(m)}} \left( f(A_k^{(m)})-f(A_j^{(m)})\right).
\end{equation}

To compare these approximations on different levels we need to introduce renormalization factors.  For the energy there is a factor $r$ such that 
\begin{equation}\label{2.3}
\mathcal{E}_m(f)=r^{-m} E_m(f) \end{equation}
and 
\begin{equation}\label{2.4}
\mathcal{E}(f)=\lim_{m\to\infty} E_m(f) \end{equation}
constructs a local, regular Dirichlet form (energy) on SC.  Experimentally, $r^{-1}\approx 1.25$.  Then
\begin{equation}\label{2.5}
\Delta f = \lim_{m\to\infty} (8r^{-1})^m \Delta_m f \end{equation}
is the Laplacian on SC.  We denote by $\rho$ the factor $8r^{-1}\approx 10.01$.  The energy and Laplacian are related by 
\begin{equation}\label{2.6}
\int_{\text{SC}} (\Delta f) g\,d\mu = \mathcal{E}(f,g) \end{equation}
for $g\in \text{dom} \mathcal{E}$ with $g$ vanishing on $\partial$SC (here $\mathcal{E}(f,g)$ is obtained from $\mathcal{E}(f)$ by polarization), where $\mu$ is the symmetric self-similar measure on SC.  The factor 8 in \eqref{2.5} is of course the scaling factor for $\mu$.

We will be imposing various boundary conditions on SC.  Since the boundary of SC is the boundary of the outer most unit square, we have to handle cells that intersect with the boundary of SC separately.  To encode the behavior on the boundary to our method of averages on cells we introduce the notion of \emph{virtual cells}.  Every cell that borders the boundary has exactly one virtual cell on the exterior of SC except for the four corner cell which each have 2 virtual cells.

\begin{figure}[ht]
\begin{center}
\begin{picture}(40,60)(-20,10)
\setlength{\unitlength}{8pt} \thicklines
\put(-4,8){\line(-1,0){2}}
\put(-2,6){\line(-1,0){4}}
\put(-2,4){\line(-1,0){4}}
\put(-4,2){\line(-1,0){2}}
\put(-6,2){\line(0,1){6}}
\put(-4,2){\line(0,1){6}}
\put(-2,4){\line(0,1){2}}
\put(-8.05,5.2){$\blacksquare$}
\put(-6.9,5.2){$\blacksquare$}
\put(-8.05,4){$\blacksquare$}
\put(-6.9,4){$\blacksquare$}
\put(-6.9,5.1){\circle*{4}}
\put(-7.05,4.95){\circle*{4}}
\setlength{\unitlength}{8pt} \thicklines
\put(-5.5,4.7){$a$}
\put(-5.5,6.7){$b$}
\put(-5.5,2.7){$c$}
\put(-3.5,4.7){$d$}
\thinlines \multiput(-9,2)(0,2){4}{\line(1,0){7.5}}
\thinlines \multiput(-8,1)(2,0){4}{\line(0,1){7.5}}

\thicklines
\put(4,6){\line(1,0){6}}
\put(4,4){\line(1,0){6}}
\put(4,2){\line(1,0){2}}
\multiput(4,6)(2,0){3}{\line(0,-1){5}}
\put(5.9,3.2){$\blacksquare$}
\put(7.1,3.2){$\blacksquare$}
\put(5.9,2){$\blacksquare$}
\put(7.1,2){$\blacksquare$}
\put(7.1,3.1){\circle*{4}}
\put(6.9,2.9){\circle*{4}}
\put(2,4){\dashbox{.2}(2,2)}
\put(2,2){\dashbox{.2}(2,2)}
\put(4,6){\dashbox{.2}(2,2)}
\put(6,6){\dashbox{.2}(2,2)}
\put(4.5,4.7){$x$}
\put(4.6,6.5){$\tilde{x}$}
\put(2.6,4.5){$\tilde{x'}$}
\put(6.6,4.7){$y$}
\put(4.5,2.7){$z$}
\put(8,8){\dashbox{.2}(1,0)}
\put(2,1){\dashbox{.2}(0,1)}
\end{picture}
\caption{Every cell in SC will have exactly two to four neighbor cells.  For example, on the left blowup of SC, the interior cell $a$ has neighbors $b,c,$ and $d$.  On the right hand picture, $x$ is the top-left corner cell.  It has four neighbor cells: two interior cells, $y$ and $z$; and two virtual cells $\tilde{x}$ and $\tilde{x'}$.}
\end{center}
\end{figure}
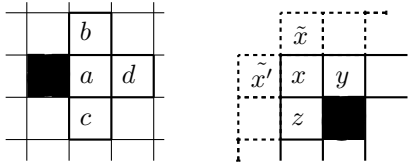

\section {Harmonic Functions}
In this section we study harmonic functions on SC by approximating them by harmonic functions on SC$_m$.  At interior cells $x$, the harmonic condition is simply that $h(x)$ is the average value of $h(y)$ over the neighbors $y$ (there may be two, three, or four neighbors).  We would like to impose boundary conditions $h|_{\partial\text{SC}}=f$ where $f$ is a function defined on $\partial$SC.  It would be clumsy to try to impose these conditions on cells in $\partial$SC$_m$, since these cells extend into the interior of SC.  Instead we introduce ``virtual cells" $x^*$ for each boundary cell $x$ in SC$_m$ that reflect outside SC across the boundary line segment (for the four corner cells there are two such reflections, but we will not introduce any new notation to handle these cases).  The function $h$ will be extended to be defined on the reflected cells.  We then require the harmonic condition
\begin{equation}\label{Bob3.1}
\Sum_{\stackrel{y\sim x}{_m}} \left( h(x)-h(x)\right)=0 \quad \text{ for each } x\in \text{SC}_m
\end{equation}
including virtual neighbors and the boundary conditions become
\begin{equation}\label{Bob3.2}
\frac{1}{2}(h(x)+h(x^*))=\text{ Average value of }f \text{ on } x\cap x^*
\end{equation}
for every virtual cell $x^*$.  Note that together \eqref{Bob3.1} and \eqref{Bob3.2} give the same number of equations as unknowns $(8^m+4\cdot3^m)$, and it is easy to see that there is a unique solution for any given $f$.  If we wish to impose Neumann boundary conditions on all or portions of the boundary, we replace \eqref{Bob3.2} by the even extension 
\begin{equation}\label{Bob3.3}
h(x^*)=h(x) \end{equation}
for the appropriate cells.  Note that this has the effect of imposing \eqref{Bob3.1} without adding the virtual cell neighbor.

\begin{figure}[p]
\begin{center}
\includegraphics[scale=.3]{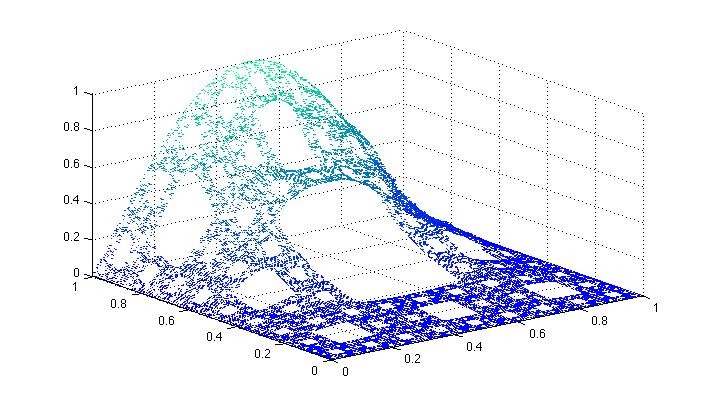}\includegraphics[scale=.25]{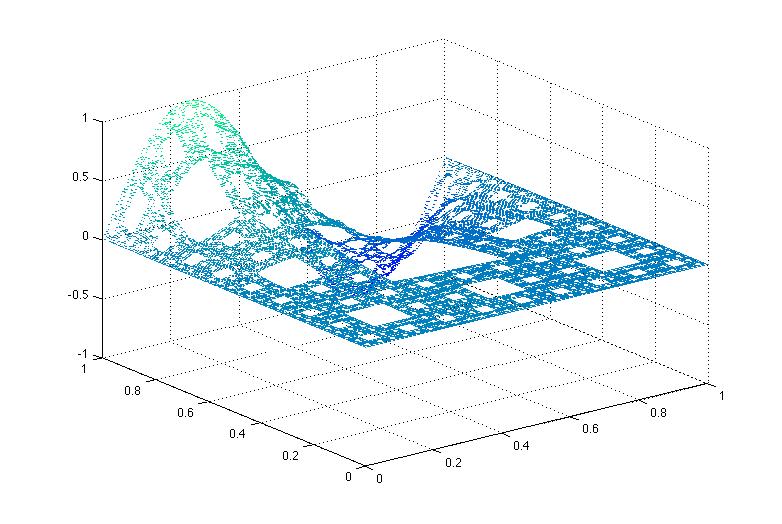}\\
\includegraphics[scale=.25]{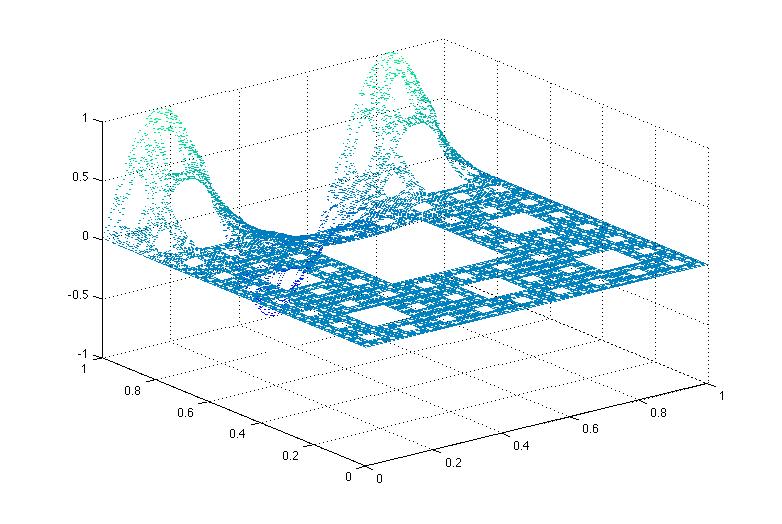}\includegraphics[scale=.35]{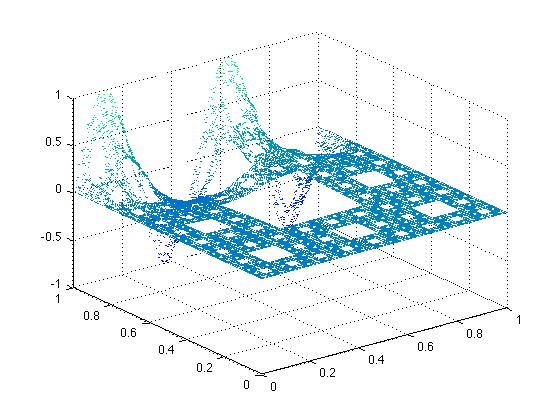}\\
\includegraphics[scale=.35]{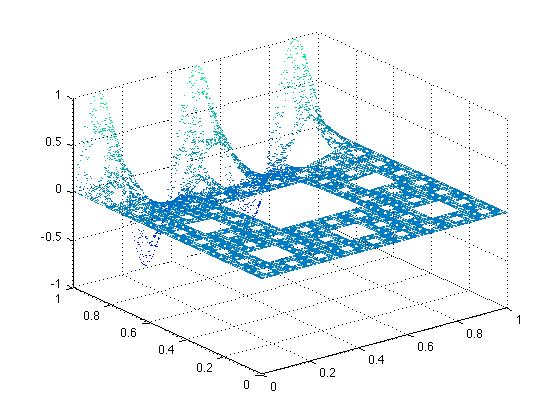}\includegraphics[scale=.35]{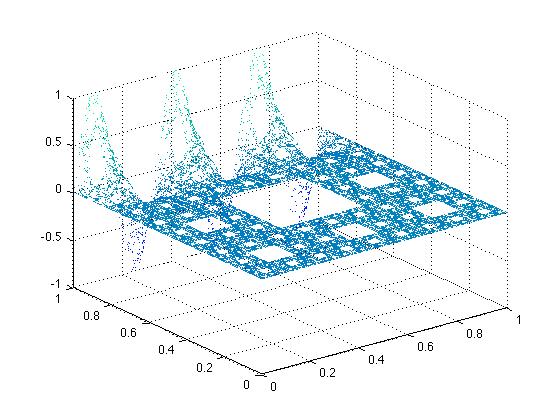}
\end{center}
\caption{Graphs of the harmonic functions on SC$_6$ with $f(t)=\sin \pi k t$ defined along one edge for $k=1,2,3,4,5,6$ and zero on the other three edges.}\label{sinharmonicfig}
\end{figure}

\begin{figure}[h]
\begin{center}
\includegraphics[scale=.3]{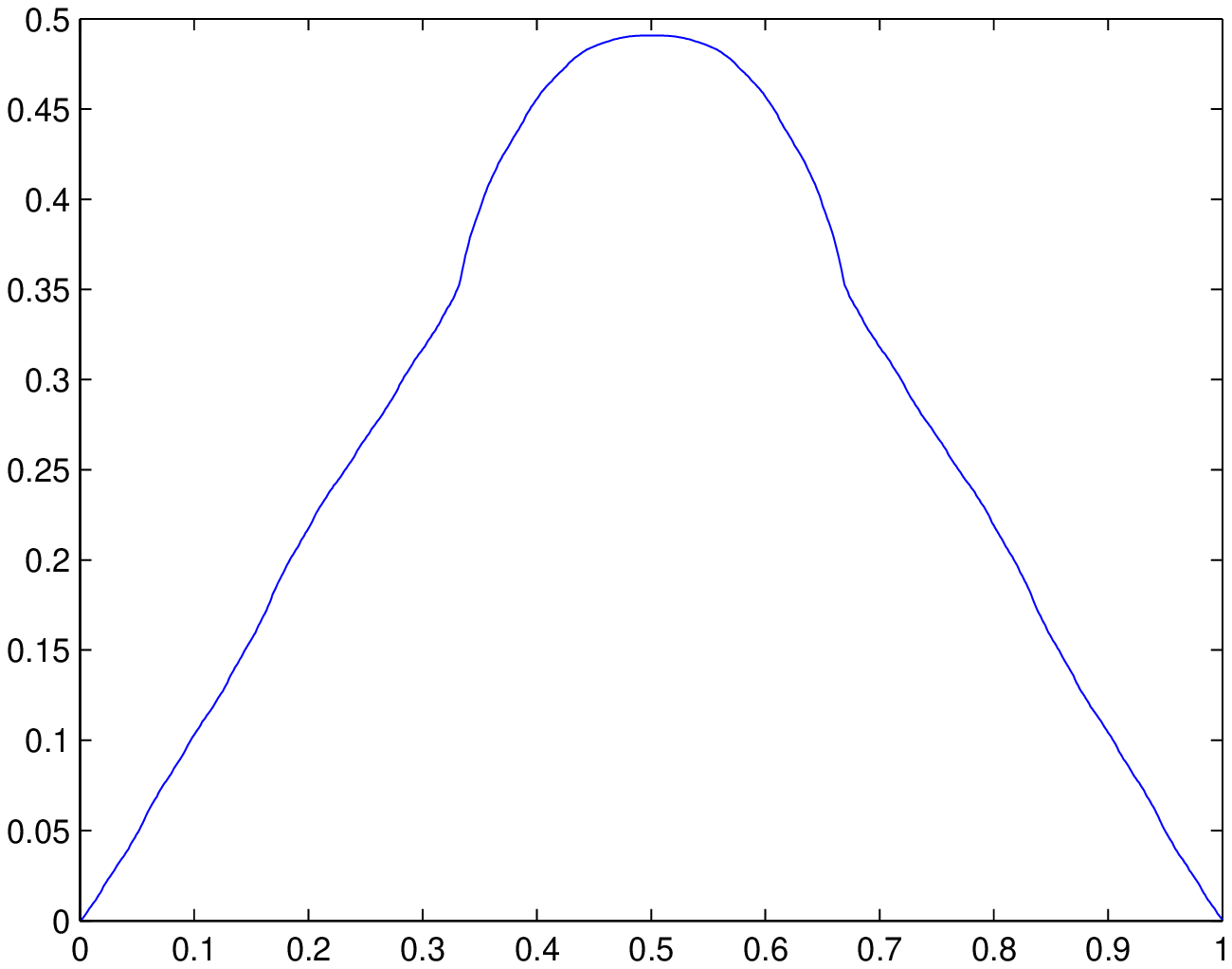}\includegraphics[scale=.3]{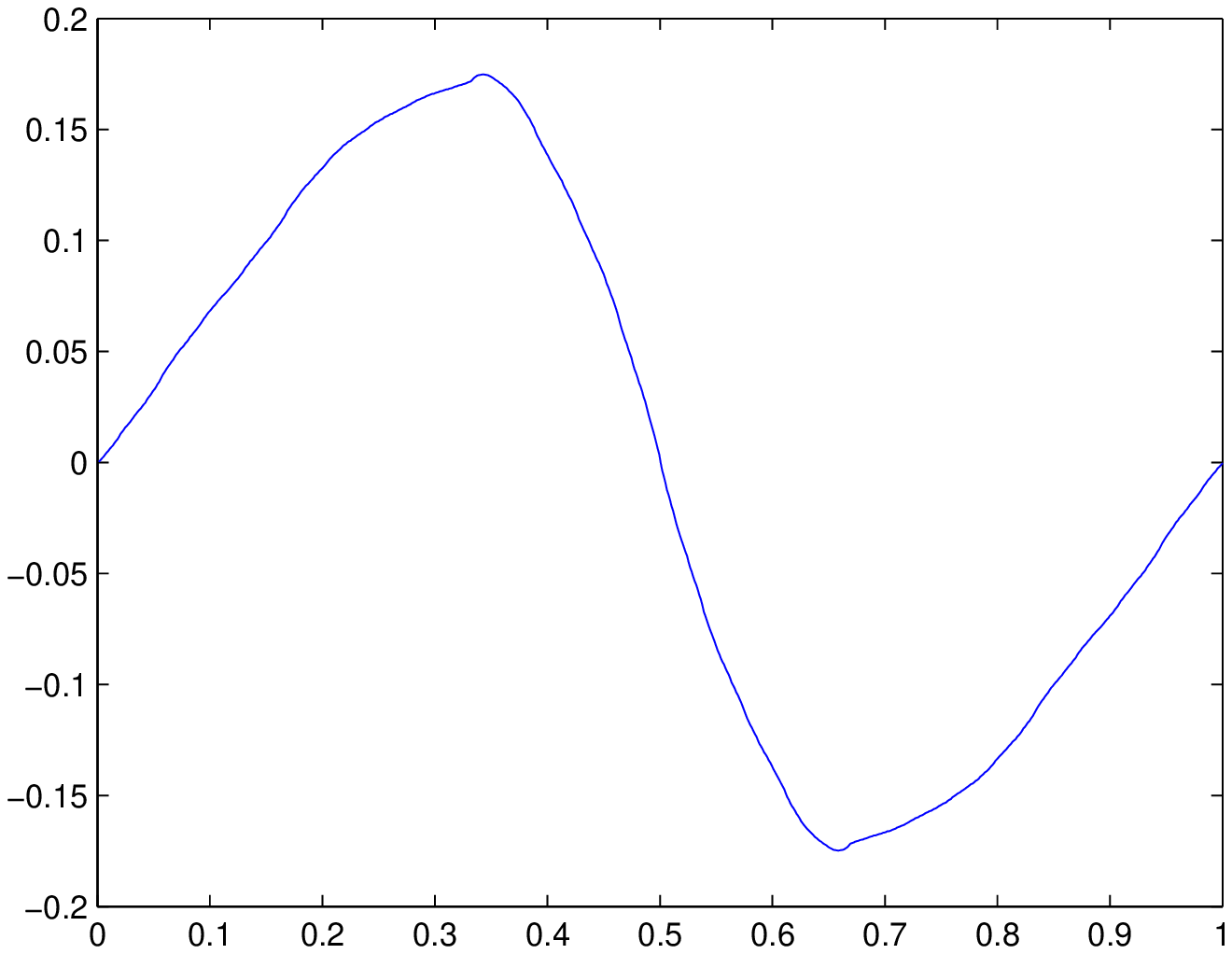}\\
\includegraphics[scale=.3]{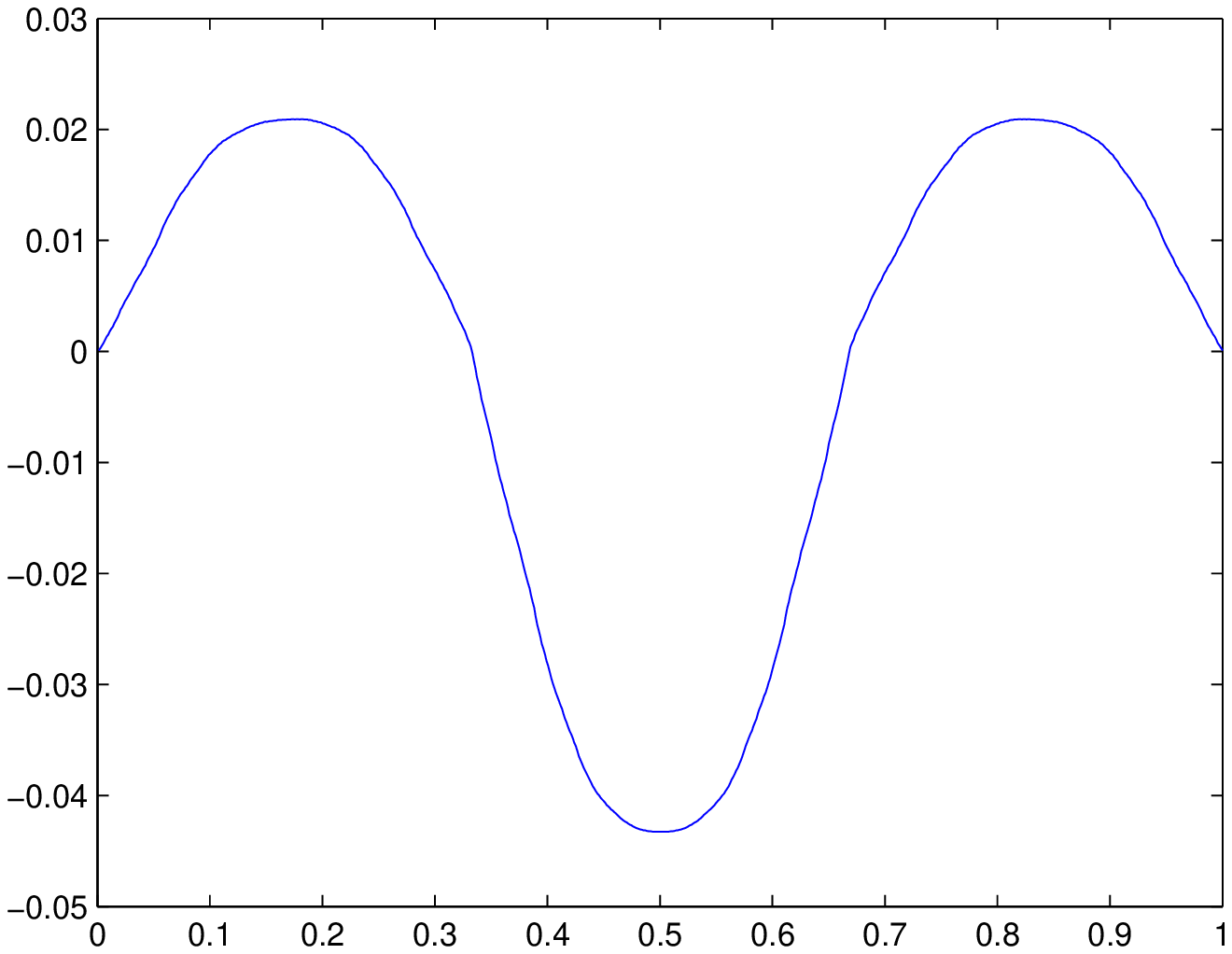}\includegraphics[scale=.3]{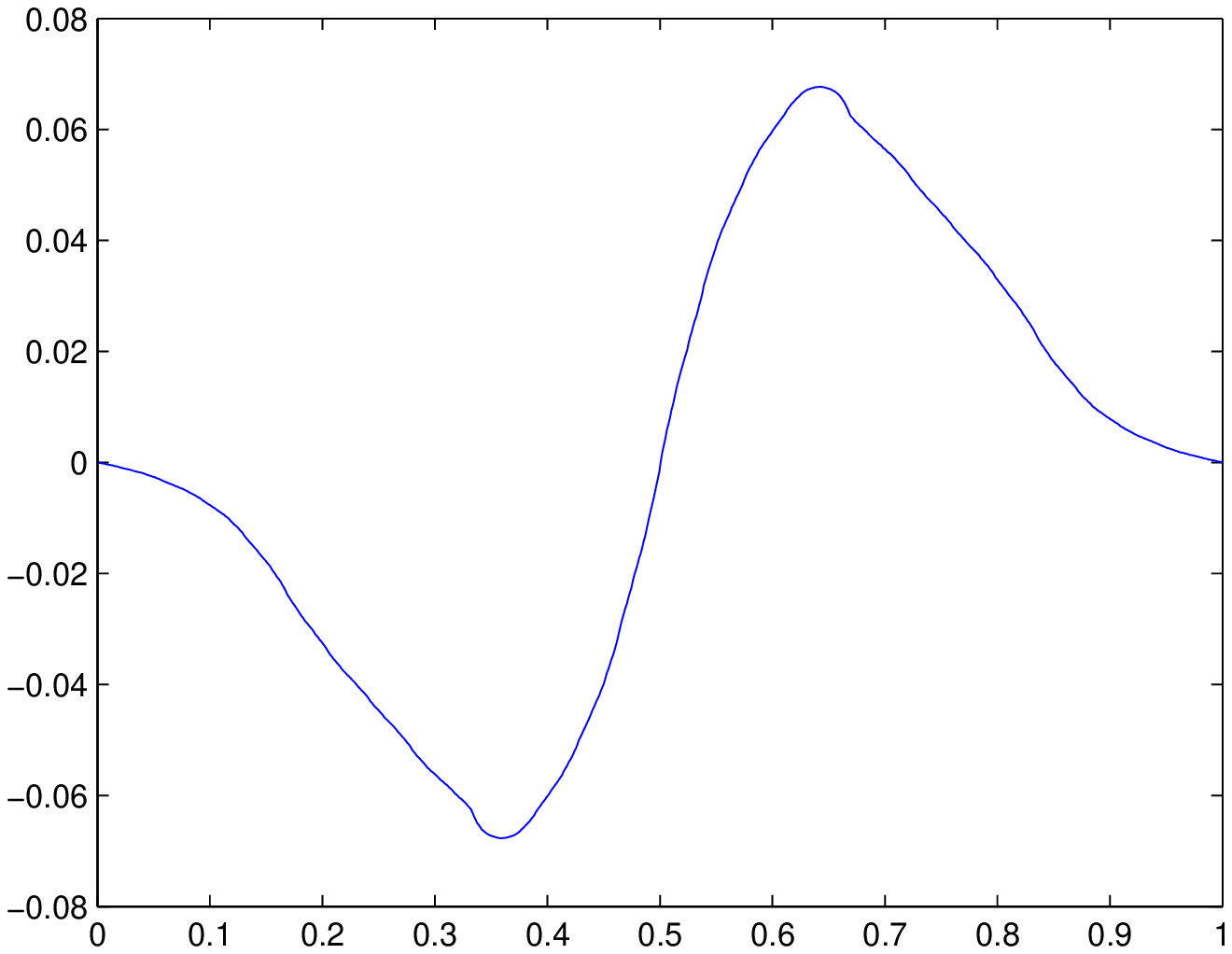}\\
\includegraphics[scale=.3]{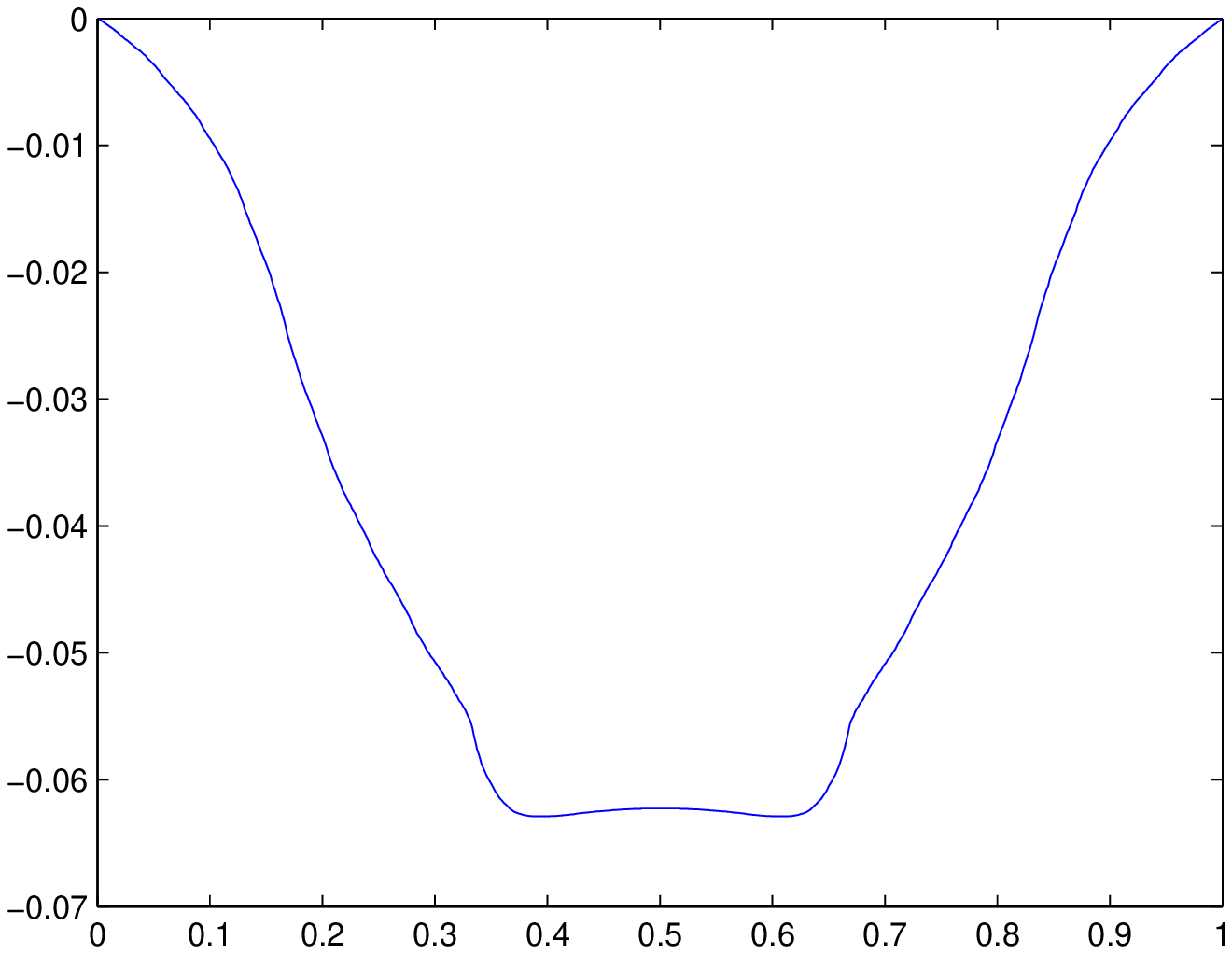}\includegraphics[scale=.3]{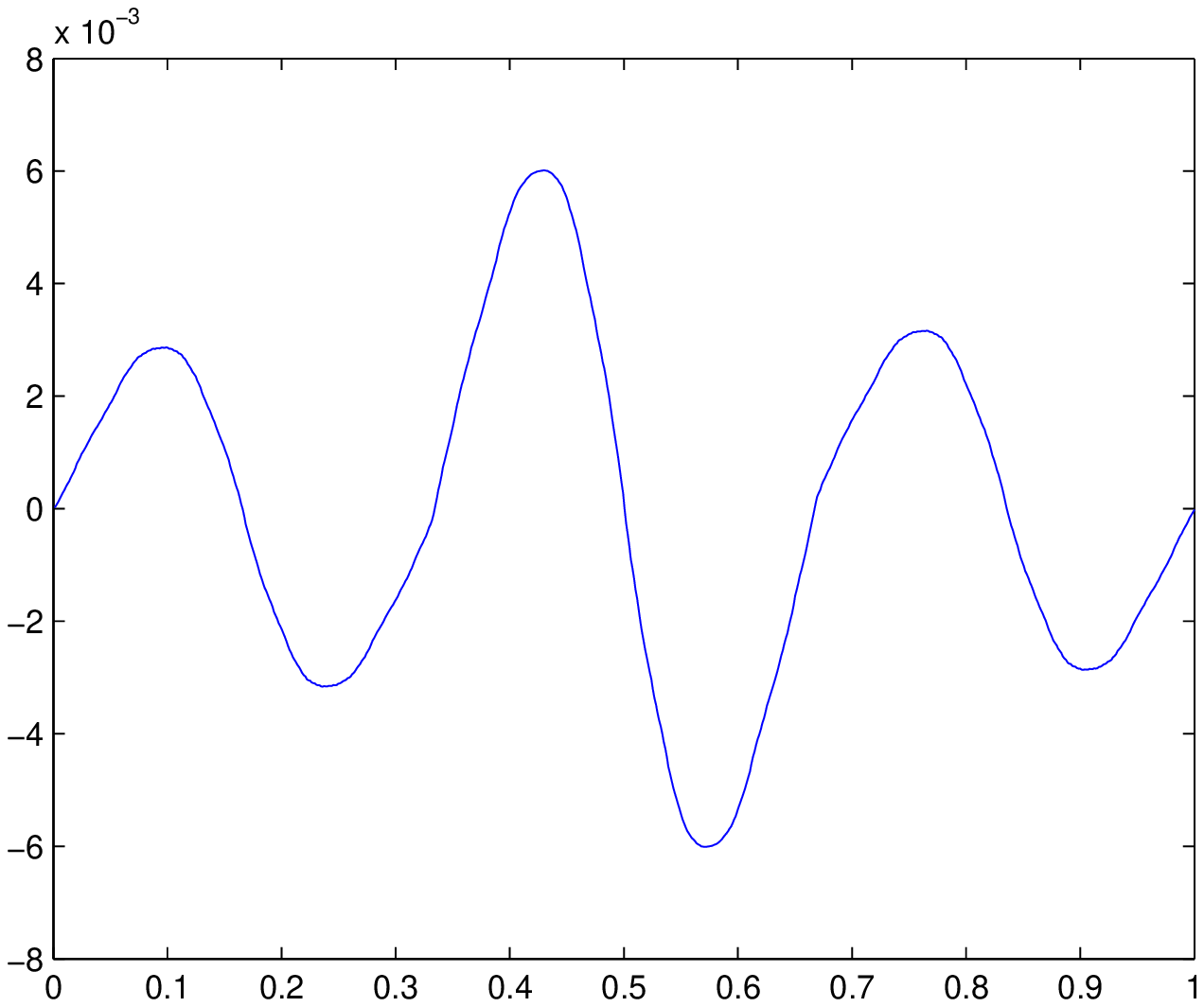}
\end{center}
\caption{Restrictions of the harmonic functions from Figure \ref{sinharmonicfig} along the line one third of the way from the nonzero boundary edge.}\label{sinrestrictionfig}
\end{figure}

For our first example, we consider the boundary data $f(t)=\sin \pi k t$ along one edge of the boundary and zero on the other three edges.  By taking infinite linear combinations of these functions and their three rotations we could obtain any harmonic function.  In Figure \ref{sinharmonicfig} we display the graphs for $k=1,2,3,4,5,6$ (computed at level $m=6$).  In Figure \ref{sinrestrictionfig} we display the graphs of the restrictions of these functions to the line one third of the way to the opposite boundary edge.

\begin{figure}[h]
\begin{center}
\includegraphics[scale=.55]{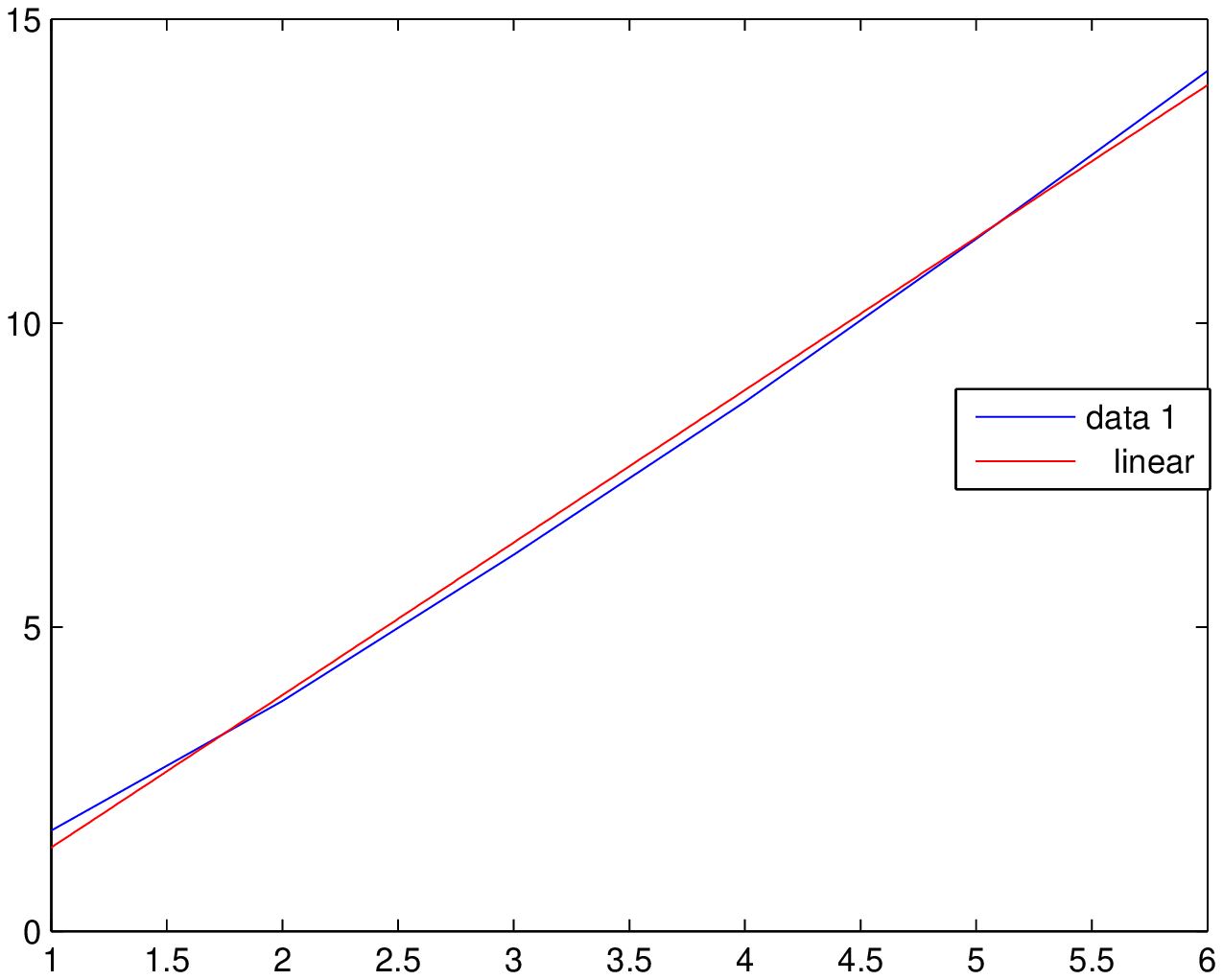}
\end{center}
\caption{Graph of $\mathcal{E}_6(u_k)$ with linear approximation.}\label{harmenergies}
\end{figure}

We do not know whether or not these harmonic functions have finite energy on SC.  By examining the renormalization energies (Table \ref{Table3.1}) for levels $m=1,2,3,4,5,6$ we see that it is reasonable to conjecture that they do have finite energy.  In Figure \ref{harmenergies} we show the graph of $\mathcal{E}_6$ as a function of $k$ together with a linear approximation.  It is difficult to speculate on the dependence for large values of $k$.  Note that under the assumption that these functions have finite energy, it would follow by standard Fourier series arguments that they are orthogonal in the energy inner product for a fixed boundary edge (but not so for rotations).  This would allow us to compute energies for $\sum a_kh_k$ if $h_k$ denotes these harmonic functions for a fixed boundary edge.

\begin{table}[h]
\begin{center}
\begin{tabular}{|r || l | l | l | l | l | l |}
\hline

$m$& $k=1$  & $k=2$ & $k=3$ & $k=4$ & $k=5$ & $k=6$ \\ \hline \hline
1&    0.4012  &  0.2487  &  0.4688  &  0.0622  &  0.0160  &       0 \\ \hline
2&    1.2709  &  2.2158  &  2.5081  &  2.4536  &  2.1673  &  1.7703 \\ \hline
3&    1.5559  &  3.3898  &  5.2038  &  6.8267  &  8.2068  &  9.2851 \\ \hline
4&    1.6286  &  3.6863  &  5.9569  &  8.2735  & 10.6618  & 13.0465  \\ \hline
5&    1.6482  &  3.7626  &  6.1398  &  8.6161  & 11.2343  & 13.9264  \\  \hline
6&    1.6535  &  3.7835  &  6.1890  &  8.7067  & 11.3823  & 14.1488   \\ \hline
\end{tabular}
\caption{Renormalized energies $\mathcal{E}_m(u_k)$ of the harmonic functions given in Figure \ref{sinharmonicfig}.}\label{Table3.1}
\end{center}
\end{table}

For our second example we compute approximations to the Poisson kernel $P(x,t)$ for $x\in$SC and $t\in\partial$SC.  This is the harmonic function in $x$ with boundary values equal to the delta measure at $t$.  This has a singularity as $x$ approaches $t$.  The Poisson kernel allows us to construct any harmonic function from its boundary values via the Poisson integral
\begin{equation}\label{Bob3.4}
h(x)=\int_{\partial\text{SC}} P(x,t)f(t)\,dt.
\end{equation} 

We refer to the point $t$ as the \emph{stimulus point}.  We approximate the delta function in the obvious way, taking the right side of \eqref{Bob3.2} to be $3^m$ on the cell for which $t\in x\cup x^*$ and zero for all other boundary cells (in case $t$ lies at the boundary between two cells $x_1$ and $x_2$ we use $1/2\cdot 3^m$ for each of them).  Our intuition is that the SC Poisson integral should resemble the upper half-plane Poisson kernel
\begin{equation}\label{Bob3.5}
P((x,y),t)=\frac{1}{\pi}\frac{y}{(x-t)^2+y^2} \quad \text{here $x,t\in\mathbb{R}$ and $y>0$}\end{equation}
at all points except the corner points, where it should resemble the quadrant $(x>0,y>0)$ case where the analog of \eqref{Bob3.4} is
\begin{equation}\label{Bob3.6}
h(x,y)=\frac{1}{\pi}\Sum_{\pm}  \int_0^\infty  \frac{4 x y s}{(x^2-y^2\mp s^2)^2+(2xy)^2}f_{\pm}(s)\, ds
\end{equation}
where the boundary values are 
\begin{equation}\label{Bob3.7}
h(x,0)=f_+(x) \quad \text{ and } \quad h(0,y)=f_-(y). 
\end{equation}

\begin{figure}[p]
\begin{center}
\includegraphics[scale=.4]{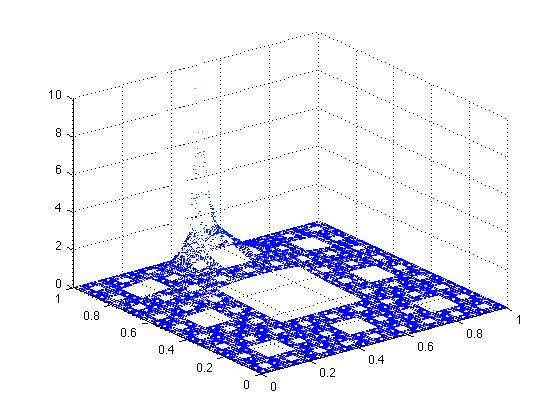}\\ 
\includegraphics[scale=.4]{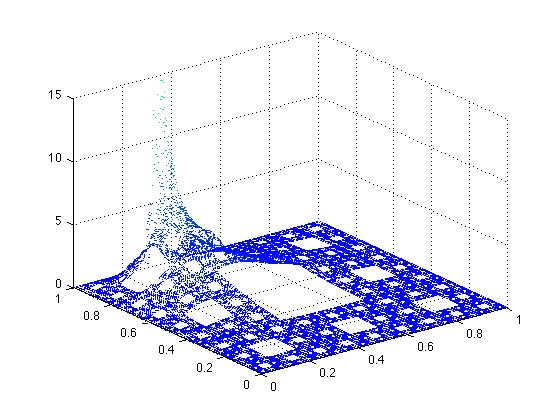}\\
\includegraphics[scale=.4]{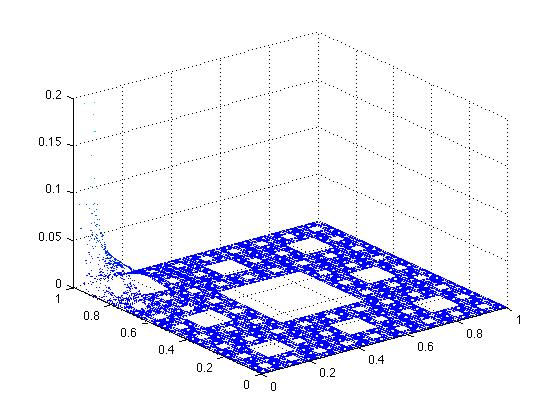}\\
\end{center}
\caption{Graphs of approximations to $P(\cdot,t)$ for stimulus point on the center, junction point, and corner of an edge on the boundary.} \label{fig3.4}
\end{figure}

\begin{figure}[p]
\begin{center}
\includegraphics[scale=.25]{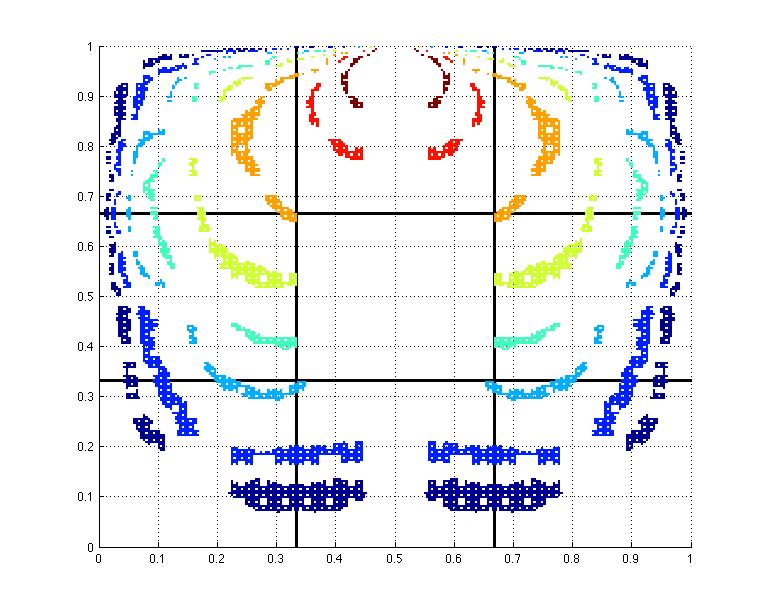}\\ 
\includegraphics[scale=.25]{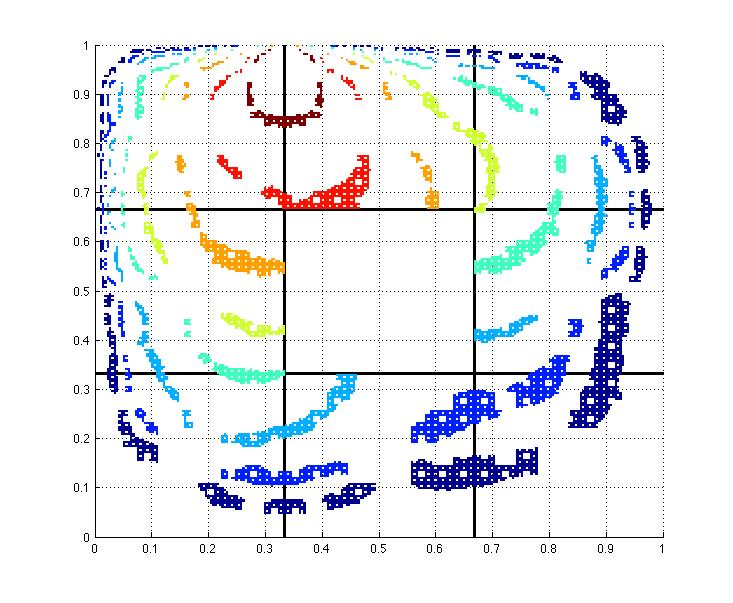}\\
\includegraphics[scale=.25]{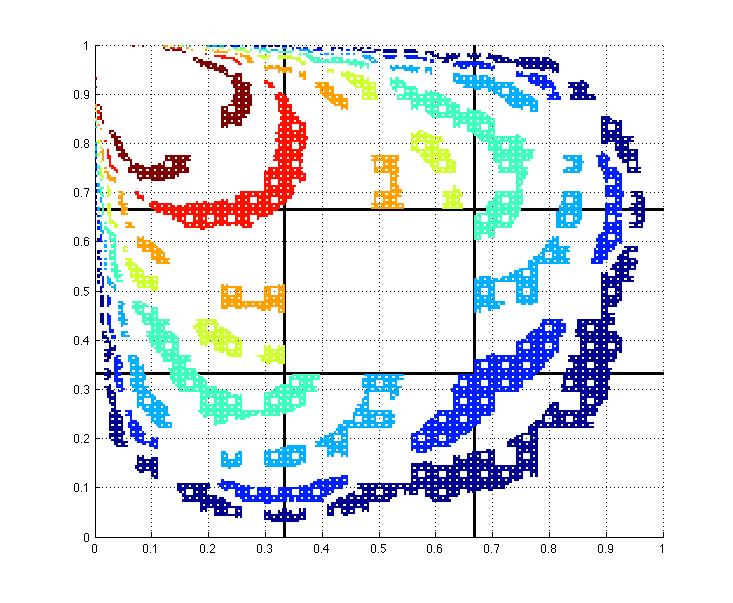}\\
\end{center}
\caption{Thickened level sets to $P(\cdot,t)$ for stimulus point on the center, junction point, and corner of an edge on the boundary.} \label{figrings}
\end{figure}

\begin{figure}[p]
\begin{center}
\includegraphics[scale=.39]{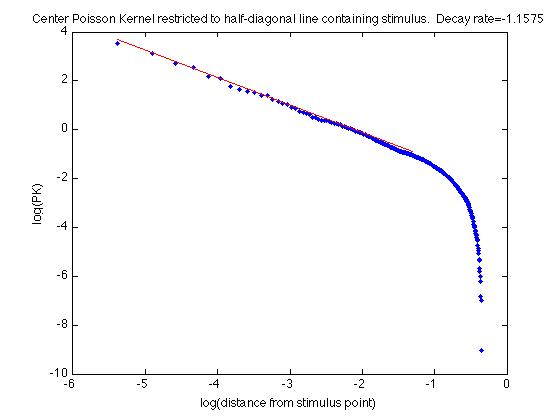}\\ 
\includegraphics[scale=.39]{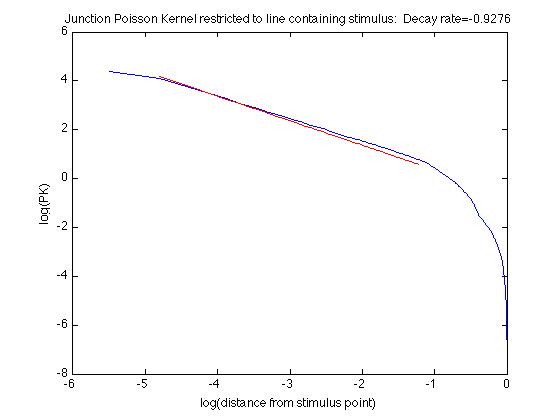}\\
\includegraphics[scale=.39]{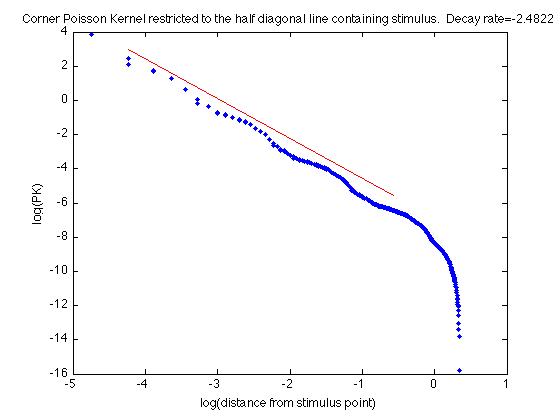}\\
\end{center}
\caption{Log graphs of the restriction of $P(\cdot,t)$ to line segments for stimulus point on the center, junction point, and corner of an edge on the boundary.  The center and corner kernels are along the ``half diagonal" while the junction kernel is along the full perpendicular line segment.} \label{figPKlog}
\end{figure}

The two basic questions we would like to discuss are the following:
\begin{enumerate}[(i)]
\item What do the level sets of $P$ look like (for fixed stimulus point)?  Do they resemble circles tangent to the stimulus point?
\item What is the rate of growth as $x$ approaches the stimulus point?
\end{enumerate}
We display the graphs of approximations to $P(\cdot,t)$ for three choices of stimulus point in Figure \ref{fig3.4}.  The choices are the midpoint of an edge, a point one third of the way along an edge (the intersection of two 1-cells), and a corner point.  In Figure \ref{figrings} we display some thickened approximations to level sets for these choices, giving a pictorial answer to (i).  To study question (ii) we restrict the Poisson kernel to a straight line segment that approaches the stimulus point from inside SC.  In the case of the junction point stimulus point there is a line segment perpendicular to the edge, while for the center and corner points, there is a ``half-diagonal" segment joining a corner point to an opposite center point.  In Figure \ref{figPKlog} we display the log of the restriction of the Poisson kernel to the line segment together with a straight line fitted to the data in the relevant portion of the graph (not too close to the stimulus point where the accuracy of the approximation deteriorates, and not too far from the stimulus point).  We measure the distance to the stimulus point in the Euclidean metric.  We observe a power law blow-up rate, but the power is different for each of the points.  

For our third example we use harmonic functions to compute effective resistances.  The effective resistance metric $R(x,y)$ between points $x,y\in$ SC is defined by
\begin{equation}\label{Bob3.8}
R(x,y)^{-1} =\min\{\mathcal{E}(u): u(x)=0 \text{ and } u(y)=1\}, \end{equation}
and indeed the minimum is attained by a function that is harmonic in the complement of the points $x$ and $y$ and satisfies Neumann boundary conditions.  For our approximations we can use the same formula for $x$ and $y$ cells in SC$_m$ and in fact we can interpret the result as the effective resistance between these cells in SC.

\begin{figure}[p]
\begin{center}
\includegraphics[scale=.55]{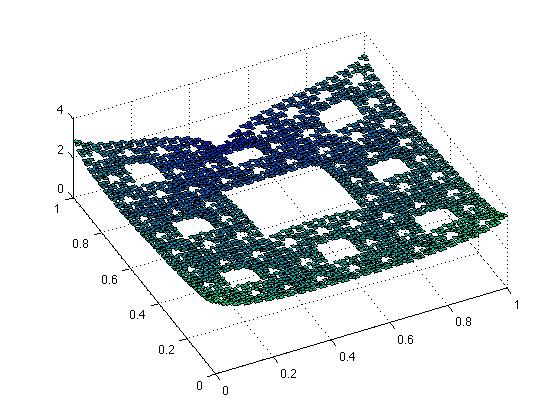}
\end{center}
\caption{Graph of $R(\cdot,y)$ on SC$_4$ where $y$ is the center-most cell on the top boundary edge.}\label{figresistance1111}
\end{figure}

\begin{figure}[p]
\begin{center}
\includegraphics[scale=.65]{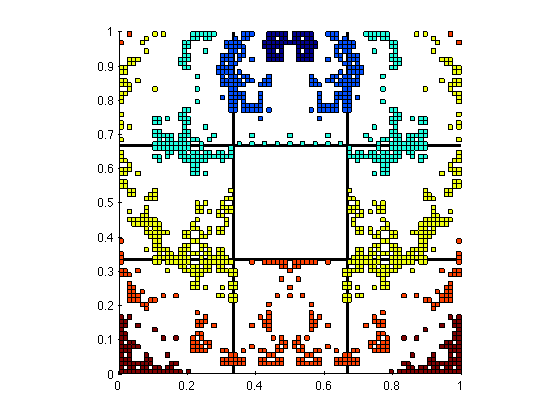}
\end{center}
\caption{Thickened level sets of $R(\cdot,y)$ on SC$_4$ where $y$ is the center-most cell on the top boundary edge.}\label{figresistance1111rings}
\end{figure}

\begin{figure}[h]
\begin{center}
\includegraphics[scale=.5]{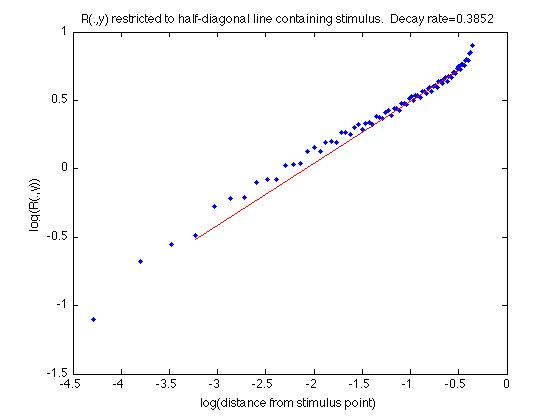}
\end{center}
\caption{Restriction of $R(\cdot,y)$ along the half diagonal on SC$_4$ where $y$ is the center-most cell on the top boundary edge.}\label{figresistance1111halfdiag}
\end{figure}

In Figure \ref{figresistance1111} we show the graph of $R(\cdot,y)$ where $y$ is the center point of a boundary edge.  In Figure \ref{figresistance1111rings} we show thickened level sets of this function.  These are clearly different from the level sets of the Poisson kernel.  In Section 5 we will compare them to level sets of the heat kernel.  In Figure \ref{figresistance1111halfdiag} we graph the restriction to the half-diagonal line segment and estimate the decay rate as $x\to y$.


\section{Spectrum}

In this section we discuss the eigenvalues and eigenfunctions of the Laplacian with either Dirichlet or Neumann boundary conditions, and later other types of periodic boundary conditions.  We note that a different method to approximate the Neumann spectrum was used in \cite{BHS}.  The results obtained using the two methods are consistent.  In our method we solve the eigenvalue problem on SC$_m$
\begin{equation}\label{Bob4.1}
-\Delta_mu(x)=\lambda u(x)
\end{equation}
at all cells in SC$_m$, computing the Laplacian taking into account virtual cells using even or odd extensions
\begin{equation}\label{Bob4.2}
u(x^*)=\pm u(x)
\end{equation}
on virtual cells for Neumann or Dirichlet boundary conditions.  The $D_4$ symmetry group acts on SC and all approximations, so each eigenspace must split according to the irreducible representations of $D_4$ (four 1-dimensional and one 2-dimensional).  Following the notation in \cite{BHS} we denote the 2-dimensional representation by 2, and the 1-dimensional representations by $1++, 1+-, 1-+$ and $1--$.  The first $\pm$ refers to the symmetry $(+)$ or skew-symmetry $(-)$ with respect to the reflections about the diagonals of the square, and the second $\pm$ refers to the symmetry or skew-symmetry with respect to horizontal and vertical reflections of the square.  This implies that there must be eigenspaces of multiplicity two.  In principle there could be higher multiplicities due to coincidences, but we do not observe any in our data, and there is no reason to believe that they occur. 

\begin{figure}[p]
\begin{center}
\includegraphics[scale=.09]{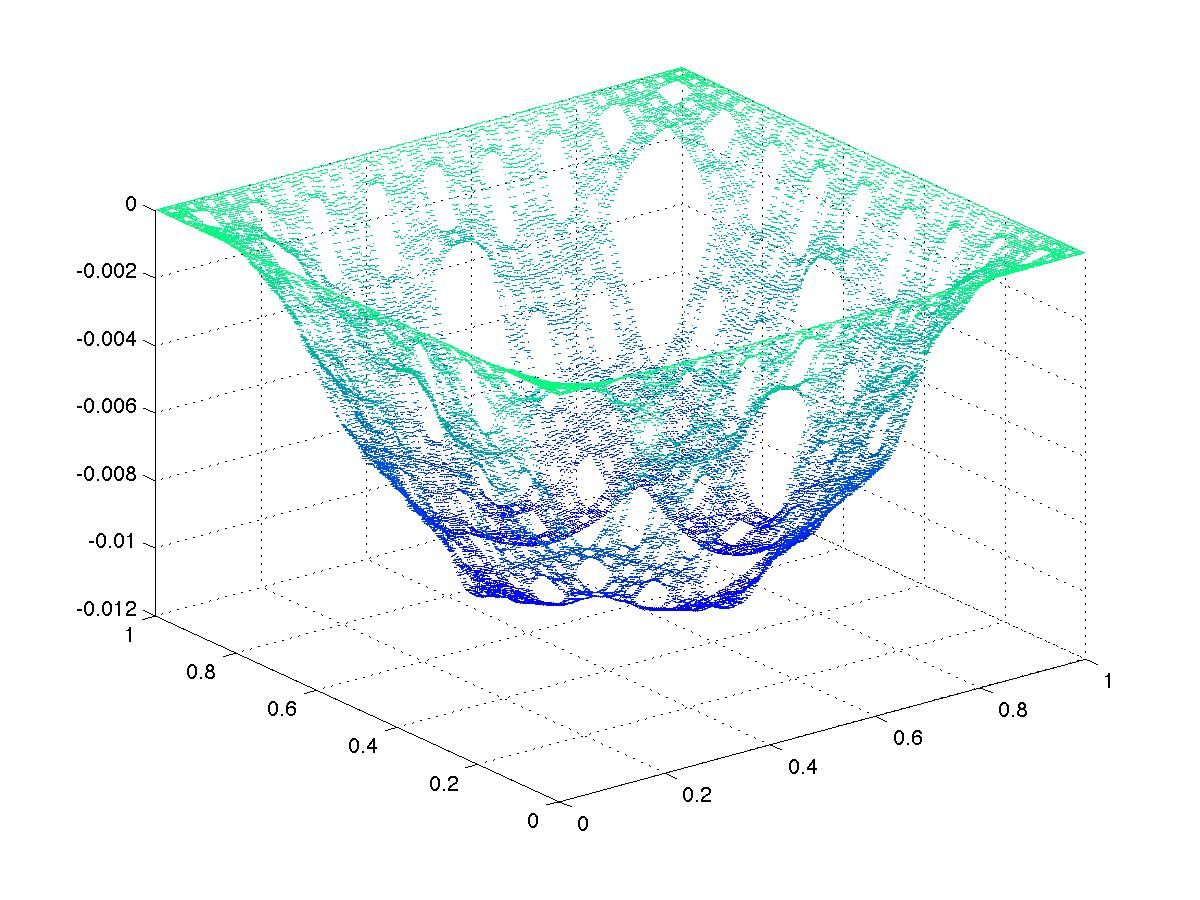}
\includegraphics[scale=.09]{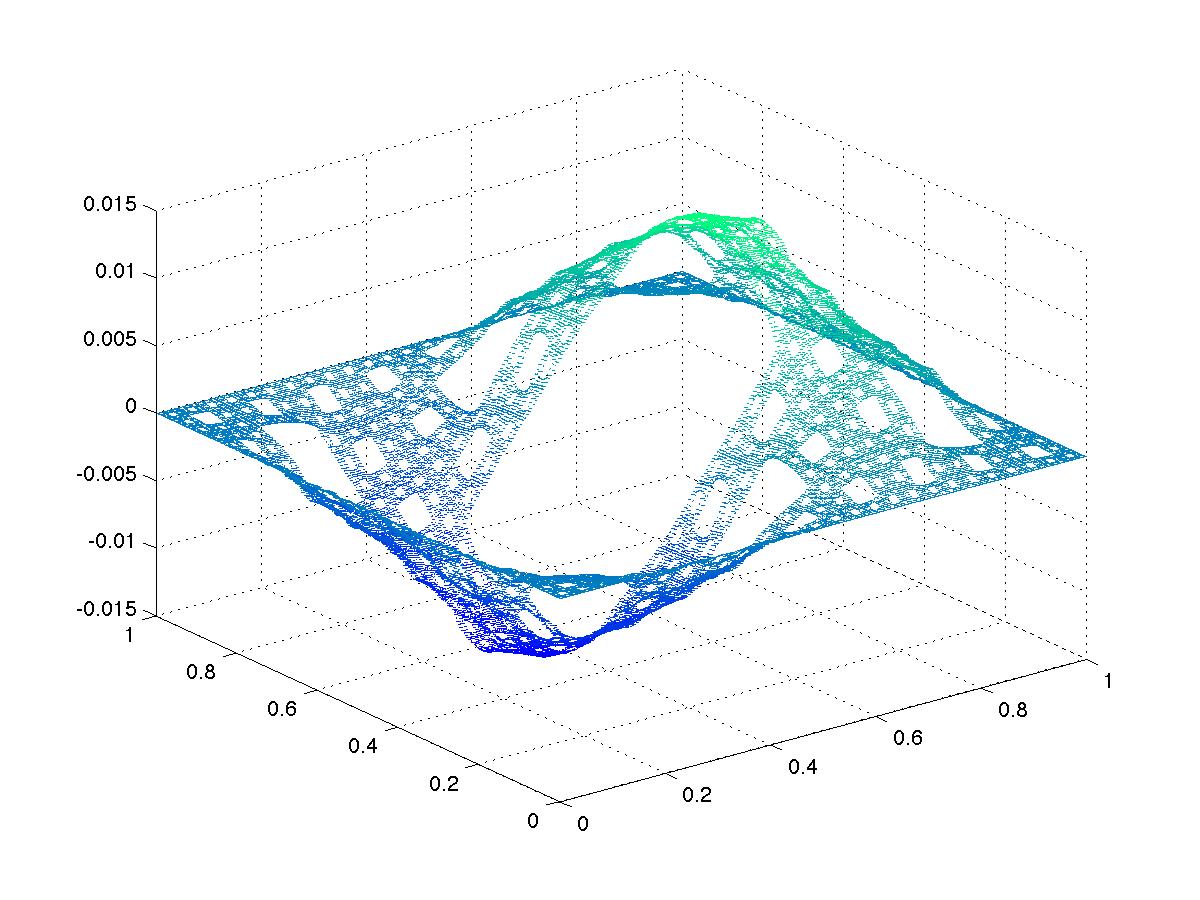}
\includegraphics[scale=.09]{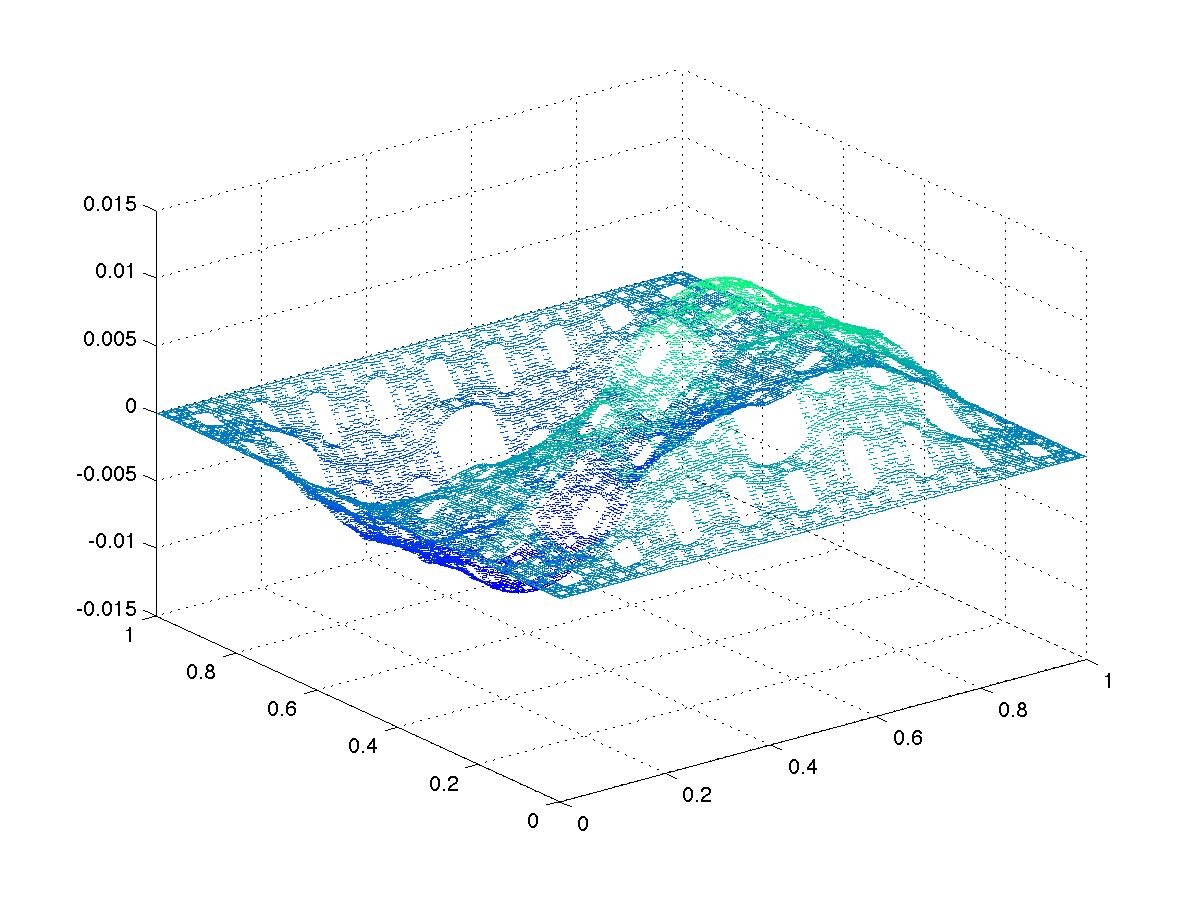}\\
\includegraphics[scale=.1]{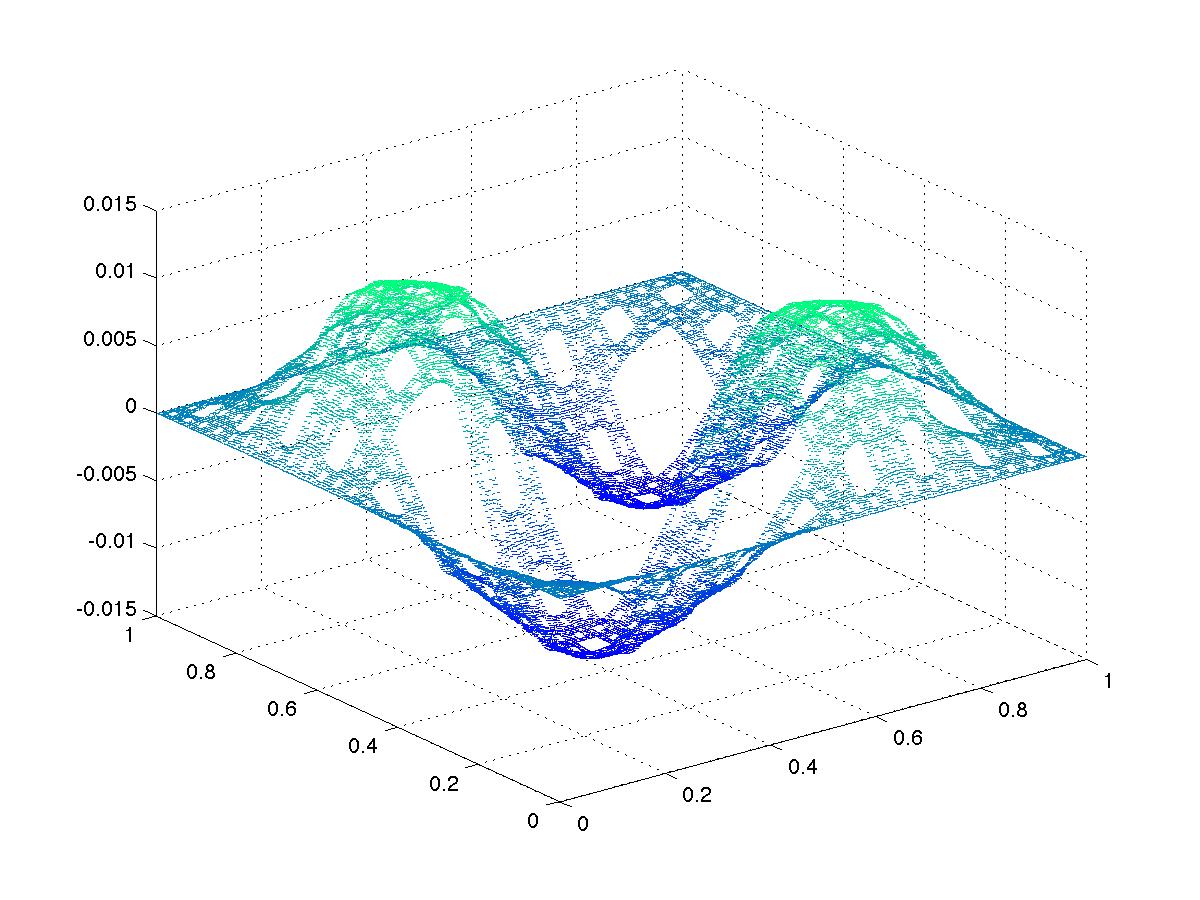}
\includegraphics[scale=.18]{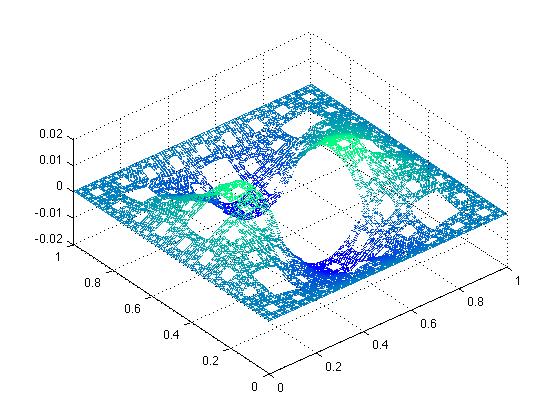}
\includegraphics[scale=.15]{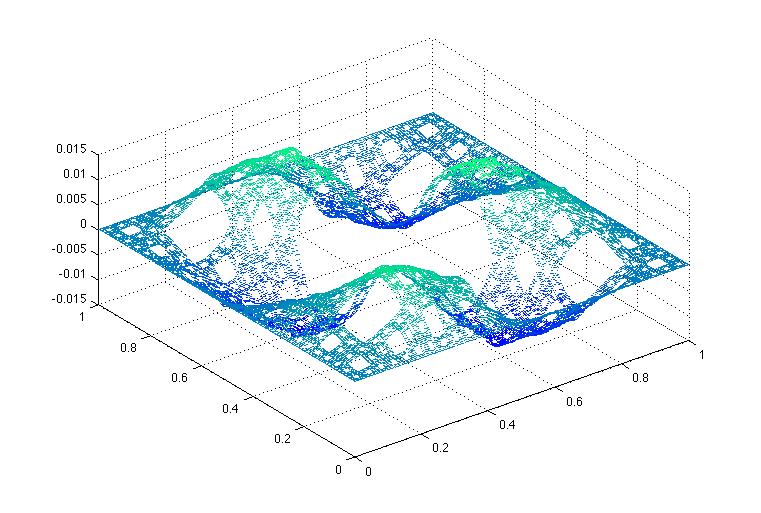}\\
\includegraphics[scale=.15]{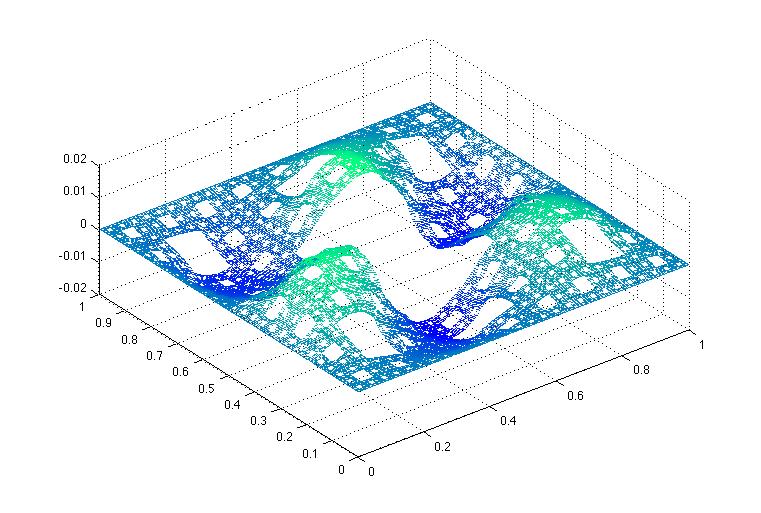}
\includegraphics[scale=.15]{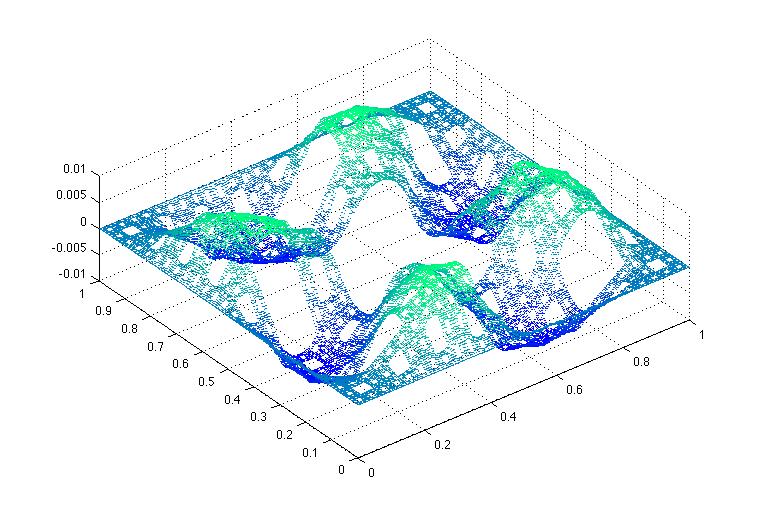}
\includegraphics[scale=.09]{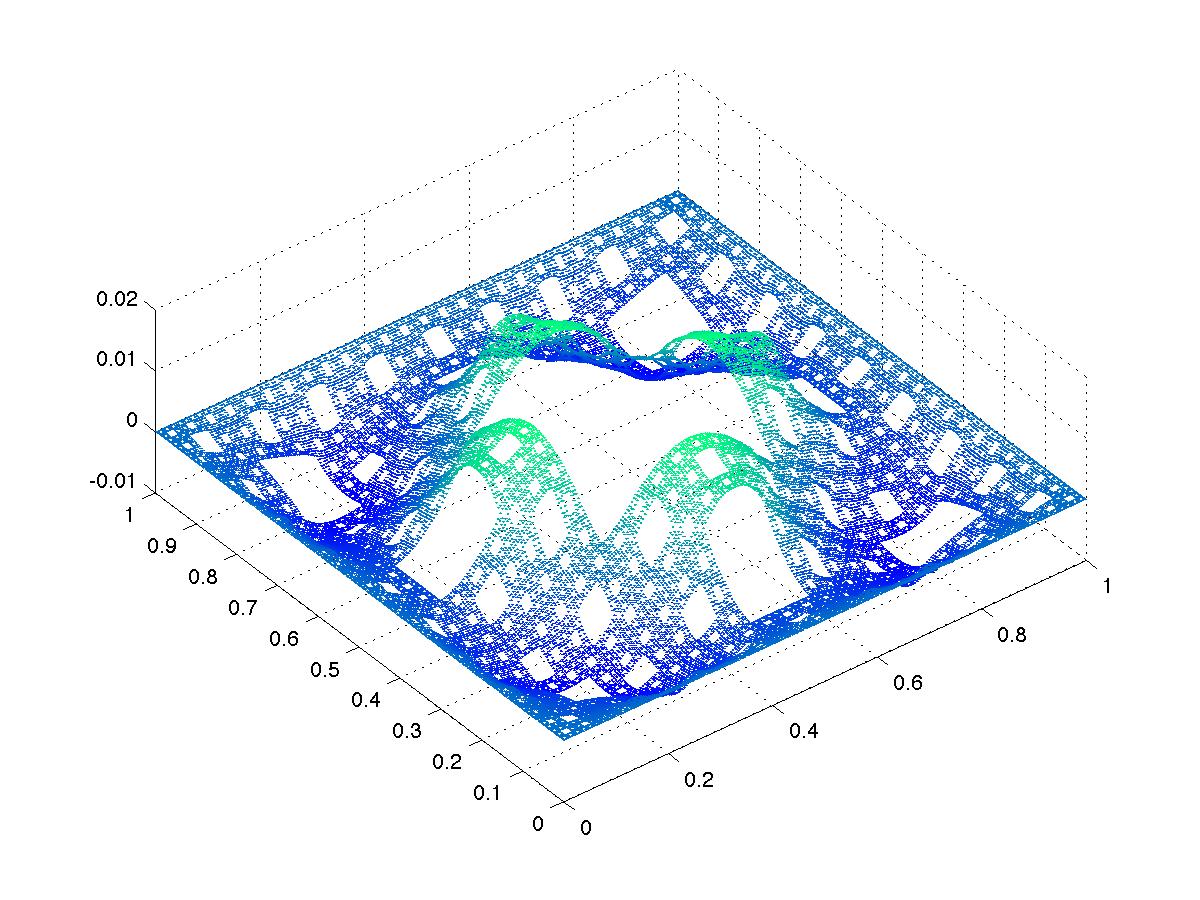}

\end{center}
\caption{The first nine Dirichlet eigenfunctions on SC$_5$}\label{figDeigs}
\end{figure}

\begin{figure}[p]
\begin{center}
\includegraphics[scale=.09]{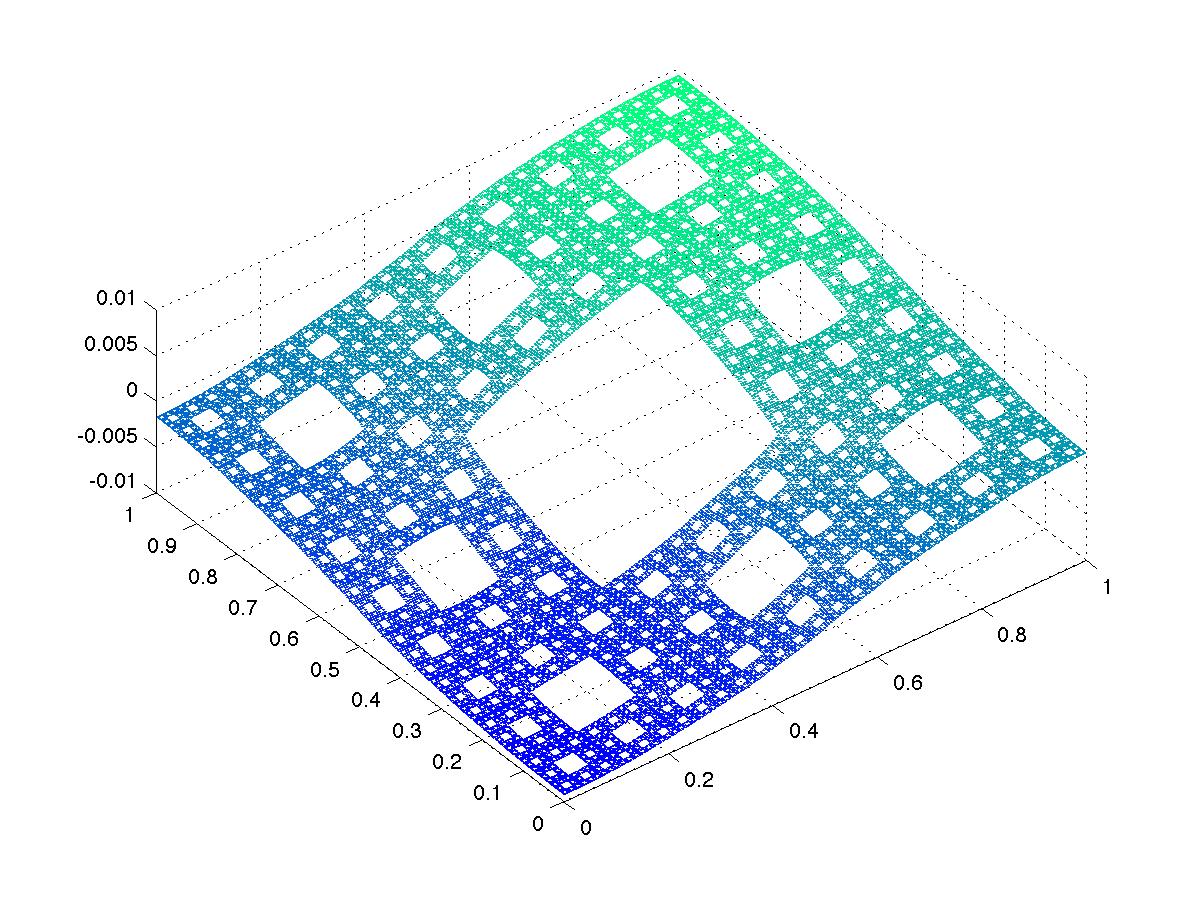}
\includegraphics[scale=.09]{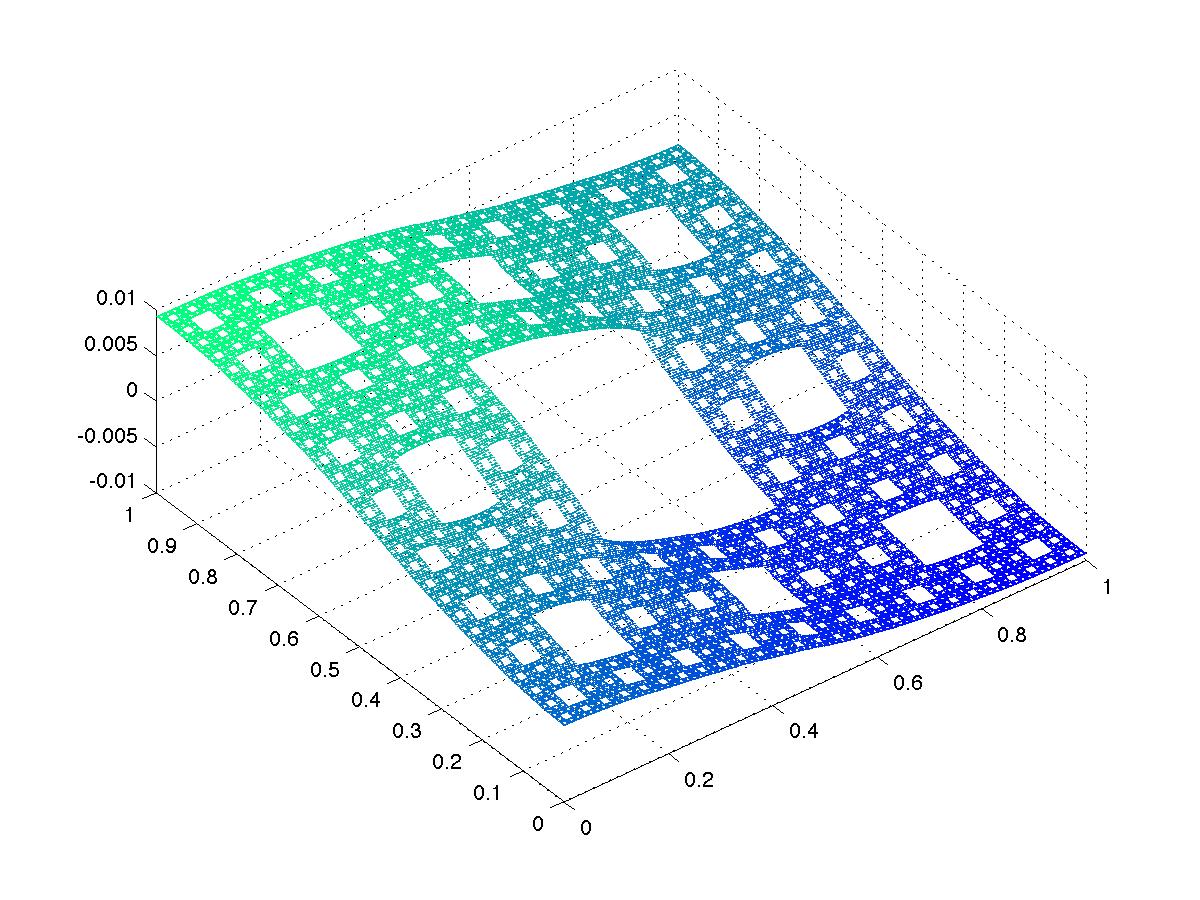}
\includegraphics[scale=.18]{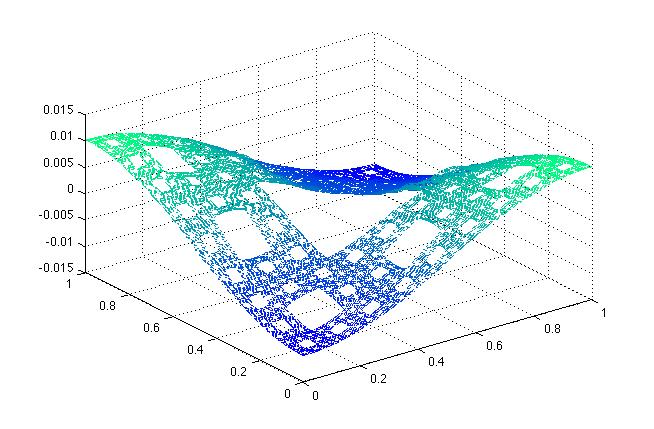}\\
\includegraphics[scale=.09]{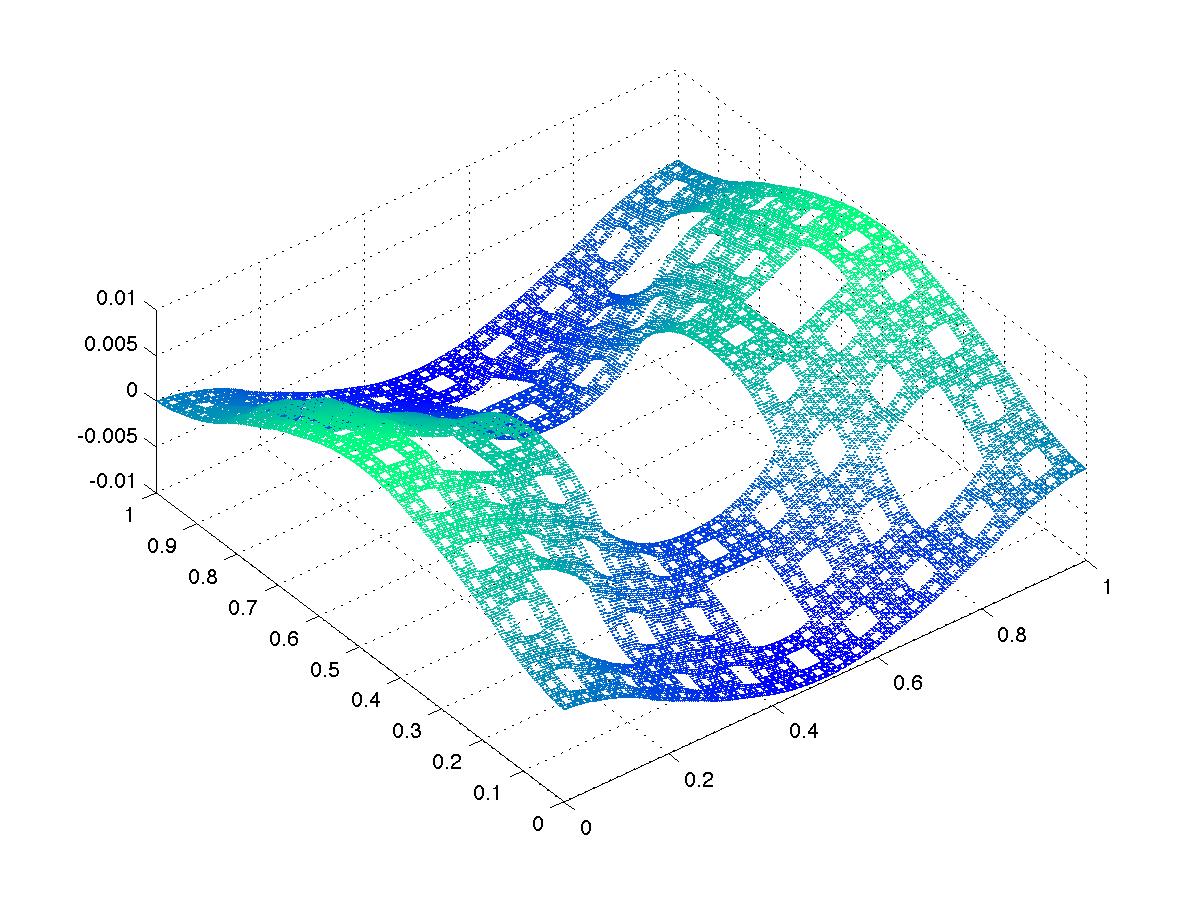}
\includegraphics[scale=.09]{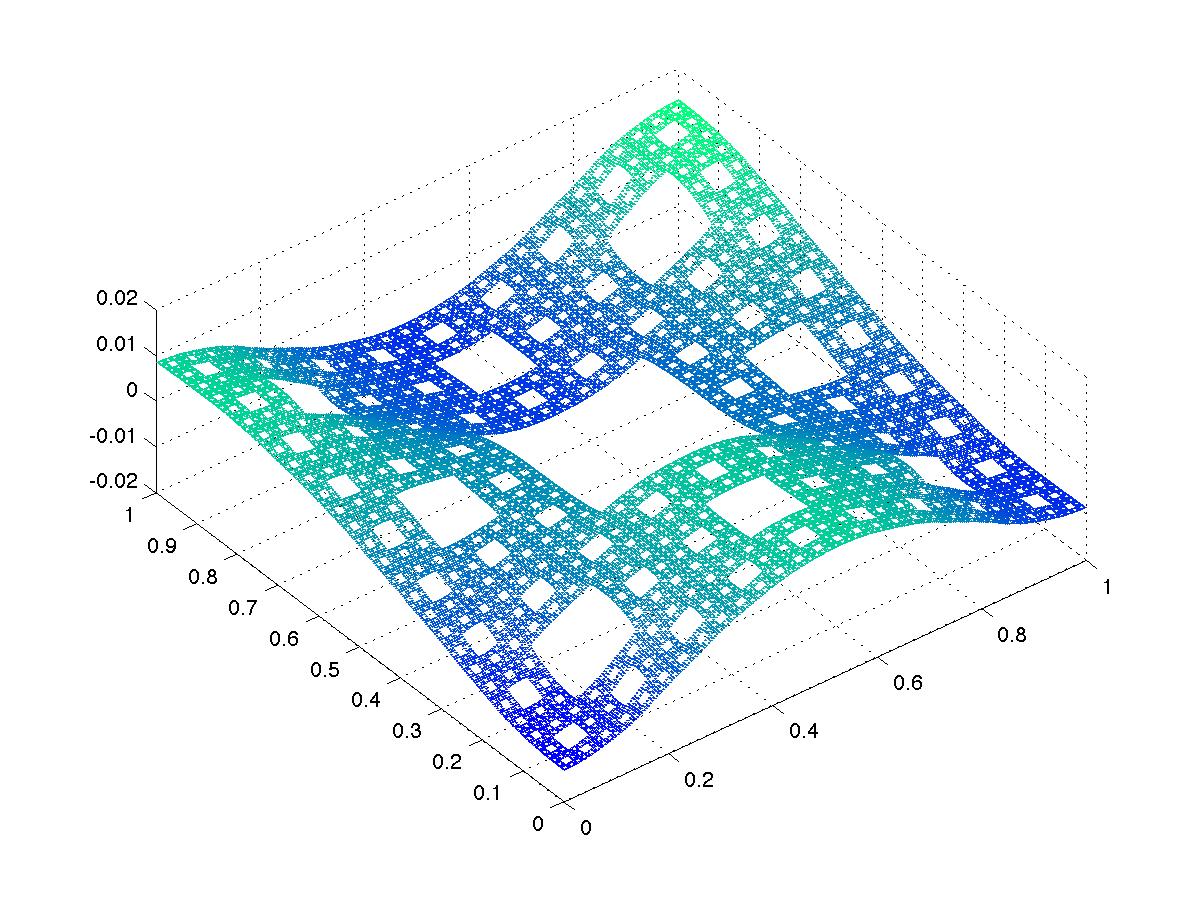}
\includegraphics[scale=.09]{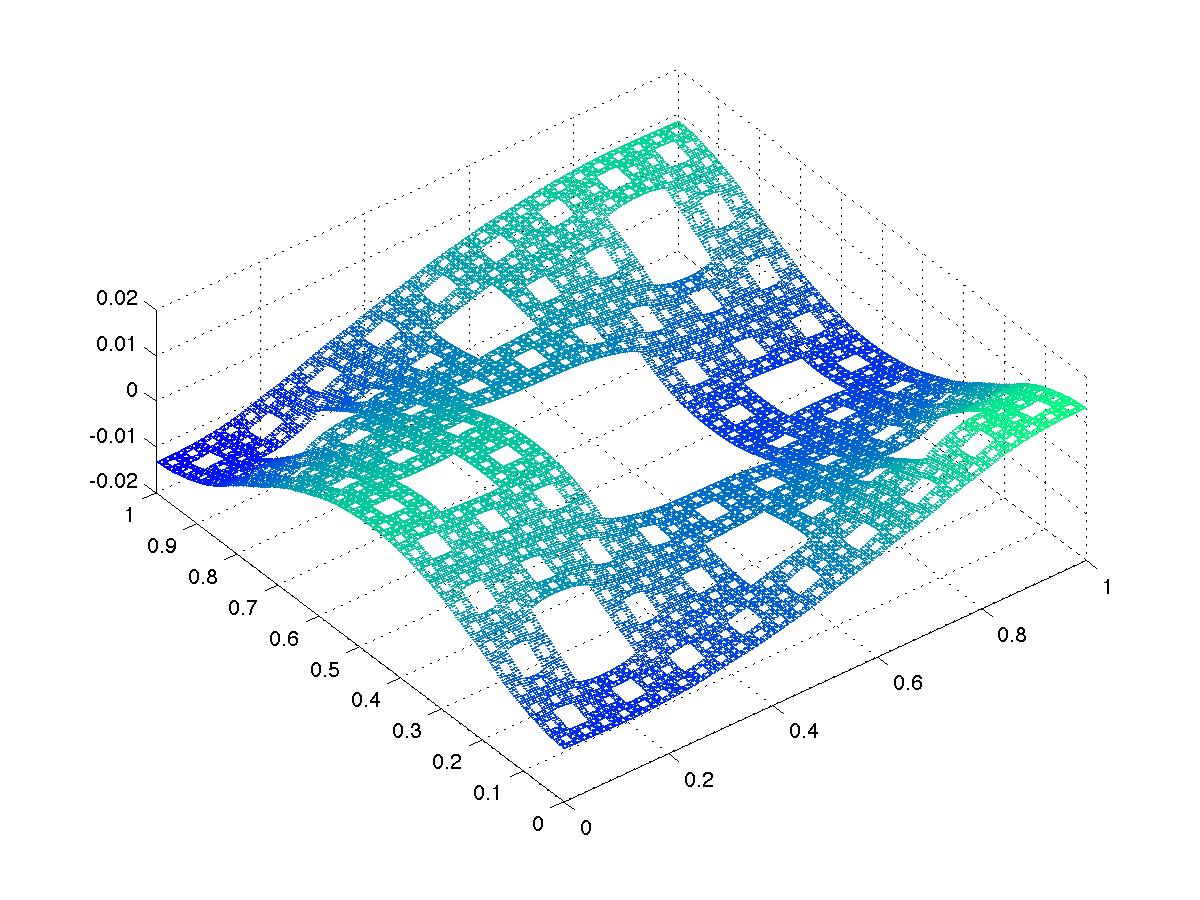}\\
\includegraphics[scale=.1]{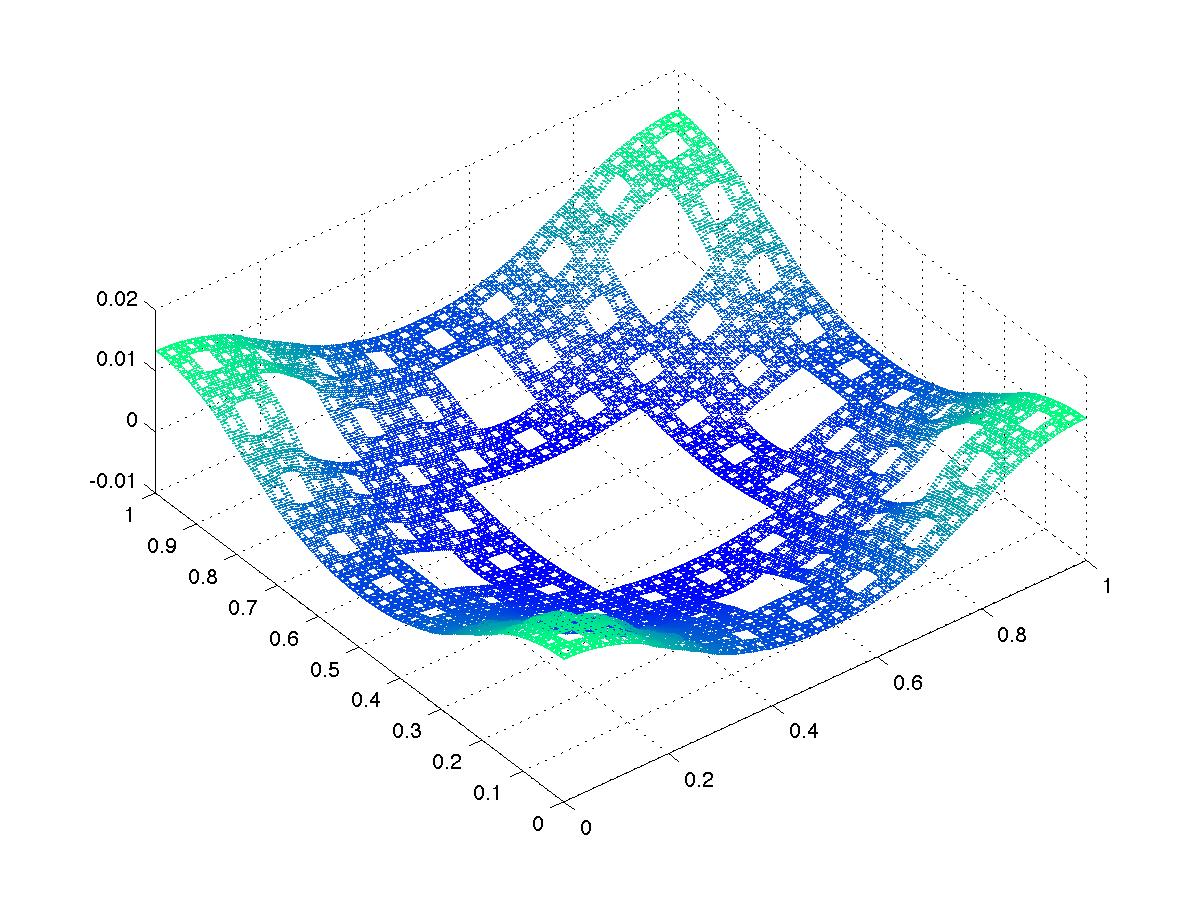}
\includegraphics[scale=.09]{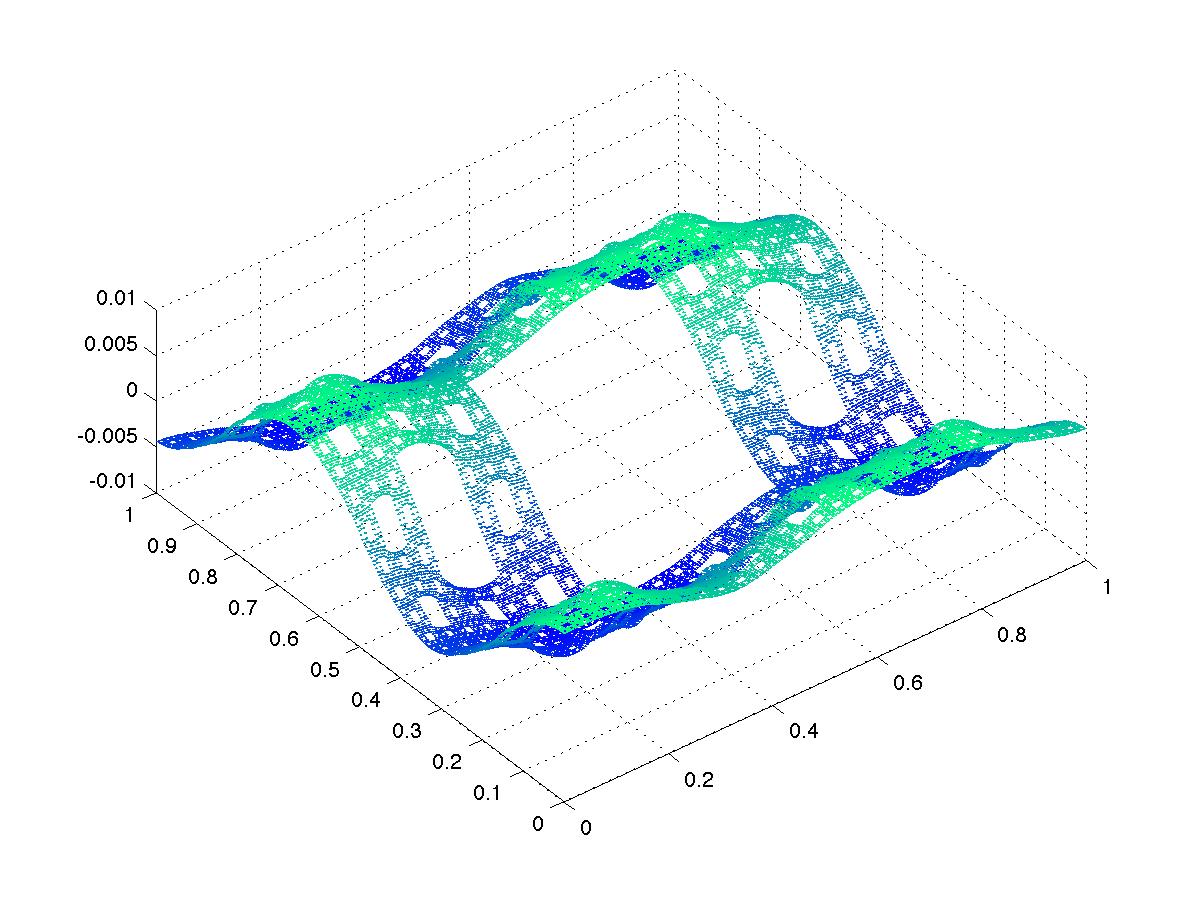}
\includegraphics[scale=.09]{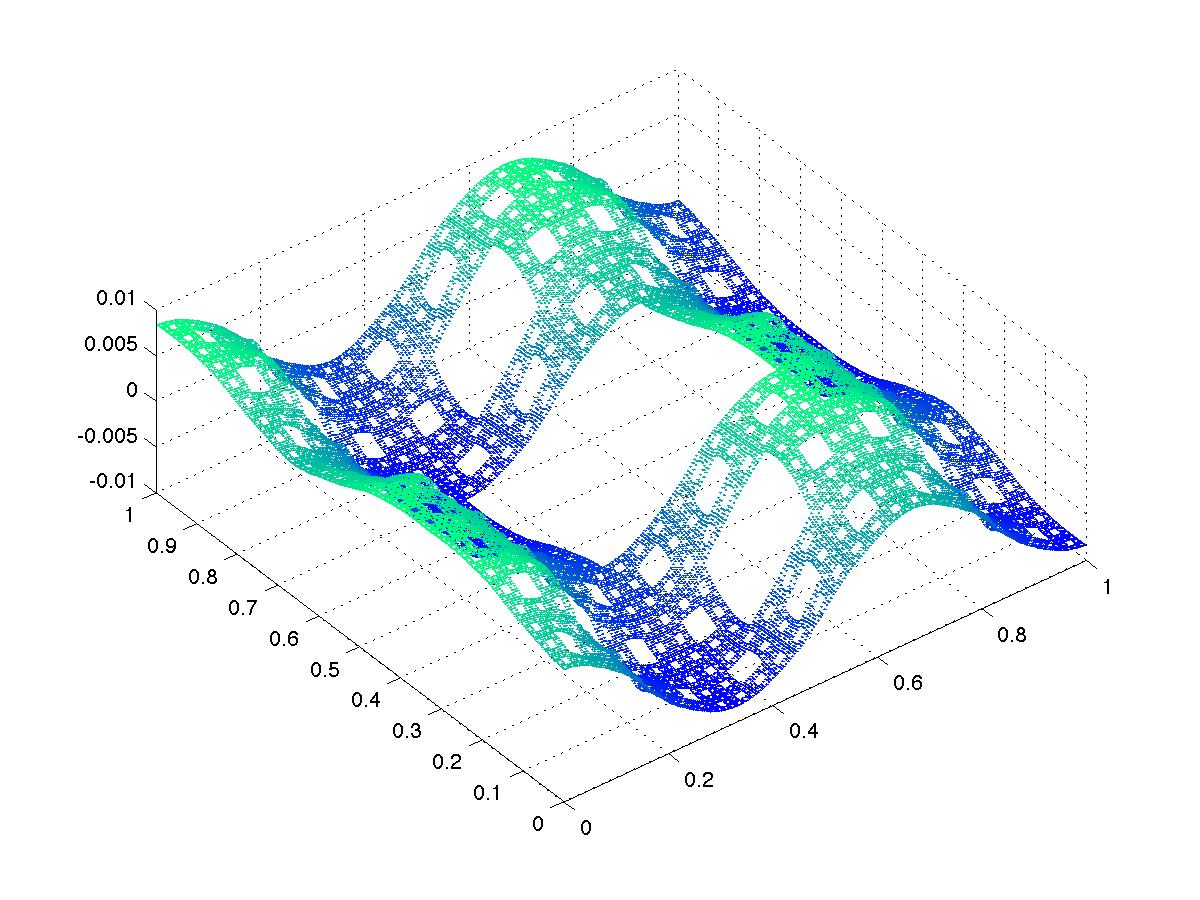}

\end{center}
\caption{The first nine nonconstant Neumann eigenfunctions on SC$_5$}\label{figNeigs}
\end{figure}

As discussed in \cite{BHS}, there is a miniaturization principle: for each eigenfunction $u$ with eigenvalue $\lambda$ there exists an eigenfunction $u'$ with eigenvalue $\rho\lambda$, with $u'$ composed of eight miniaturized versions of $u$ in each 1-cell, depending on the associated representation.  It is easy to recognize the miniaturized eigenfunctions and to use the eigenvalues to estimate the value of $\rho$.  In Figure \ref{figDeigs} we show graphs of typical eigenfunctions with Dirichlet boundary conditions, and in Figure \ref{figNeigs} the same with Neumann boundary conditions.
In Tables \ref{tabdeig} and \ref{tabneig} we show an initial portion of the Dirichlet and Neumann eigenvalues on levels $m=5$ and $6$, along with the representation types of the associated eigenfunctions.  We observe the ``quartet'' structure of each spectrum: if we group together four consecutive eigenvalues then each quartet contains a multiplicity two eigenspace corresponding to the 2-dimensional representation of $D_4$, and two distinct 1-dimensional representation eigenspaces.  (there appear to be some short term exceptions to this structure in the Neumann spectrum, but it is not clear if these represent a true deviation to the quartet structure, or are merely due to inaccuracy in the computation higher up in the spectrum.)

\begin{table}[p]
\begin{center}
\vspace{-2cm}
\begin{tabular}{|r||  c| l| l| c|} \hline
$\lambda_j$ & SC eigenvalue& level $m=5$ & level $m=6$ & Symmetry type \\ \hline \hline
1&	27.6077511113075  	& 0.0002744  &   0.0000274    & 1 + +  \\  \hline
2,3&	39.9082101260222		& 0.0003967   &   0.0000396   & 2  \\  \hline
4&	67.3444577668735		&0.0006694   &   0.0000669   & 1 + -  \\  \hline
5&	83.2634911289046		&0.0008277   &   0.0000827   &  1 - + \\  \hline
6,7&	122.741602145591		&0.0012201   &     0.0001220   & 2  \\  \hline
8&	159.852387587360		&0.0015891   &     0.0001600   & 1 - -  \\  \hline
9&	168.766639100550		&0.0016777   &     0.0001679   & 1 + +  \\  \hline
10,11&190.225270233493	&0.0018910   &   0.0001892   &2   \\  \hline
12&	205.077670619984		&0.0020387   &   0.0002042   &1 - +  \\  \hline
13&	251.810358450044		&0.0025032   &   0.0002504   &1 + -  \\  \hline
14&	278.906546773232		&0.0027726   &   0.0002774   &  1 + + \\  \hline
15,16&278.933835316832	&0.0027729   &   0.0002774   &  2 \\  \hline
17,18&305.367892038249	&0.0030357   &   0.0003017   &  2 \\  \hline
19&	323.883828126434		&0.0032197   &   0.0003233   & 1 + +  \\  \hline
20&	333.851304283982		&0.0033188   &   0.0003323   & 1 - -  \\  \hline
21&	345.835817148736		&0.0034380   &   0.0003408   & 1 + - \\  \hline
22&	359.328736486837		&0.0035721   &   0.0003583   & 1 + +   \\  \hline
23,24&362.527180325156	&0.0036039   &   0.0003591   & 2  \\  \hline
25&	374.717241516490		&0.0037251   &   0.0003714   & 1 + -  \\  \hline
26,27&402.586401338684	&0.0040021   &   0.0003927   &2   \\  \hline
28&	455.500281595392		&0.0045282   &   0.0004527   & 1 + +  \\  \hline
29&	462.626312181879		&0.0045990   &   0.0004663   & 1 - -  \\  \hline
30&	488.817301901377		&0.0048594   &   0.0004866   & 1 - +  \\  \hline
31,32&507.723951331926	&0.0050473   &   0.0005049   &2   \\  \hline
33,34&556.611933010238	&0.0055333   &   0.0005467   &2   \\  \hline
35&	576.031531252910		&0.0057264   &   0.0005733   & 1 + +  \\  \hline
36&	581.772410001422		&0.0057835   &   0.0005787   &1 + +   \\  \hline
37&	582.121682504993		&0.0057869   &   0.0005788   &1+ -   \\  \hline
38,39&592.639482105066	&0.0058915   &   0.0005802   & 2  \\  \hline
40&	611.949121939020		&0.0060835   &   0.0006092   & 1 - - \\  \hline
41&	          640.105293846754		&	0.0063567	& 0.0006358   & 1 +- \\  \hline		
42,43&	677.088392255753		&	0.0067251	&0.0006726   & 2 \\  \hline
44&	681.41621596676			&	0.0067691	&0.0006769   & 1 - - \\  \hline
45,46&708.53358170651			&	0.0070382	&0.0007038   & 2 \\  \hline
47&	722.461456690457			&	0.0071761	&0.0007177   & 1 - - \\  \hline
48&	745.824838954798			&	0.0074080	&0.0007409   & 1 - -\\  \hline
49&	777.354706737803			&	0.0077231	&0.0007722   & 1 + + \\  \hline
50&	782.601255409013			&	0.0077766	&0.0007774   & 1 + + \\  \hline
51,52&819.682344081039		&	0.0081444	&0.0008142   & 2 \\  \hline
53,54&862.870732092531		&	0.0085714	&0.0008571   & 2 \\  \hline
55&	864.146522755131			&	0.0085848	&0.0008584   & 1 + - \\  \hline
56&	877.058239744236			&	0.0087057	&0.0008712   & 1 ++ \\  \hline  
57,58&917.009721010054		&	0.0091012	&0.0009109   & 2 \\  \hline
59&	936.979846819245			&	0.0093055	&0.0009308   & 1 - - \\  \hline
60&	959.636776984823			&	0.0095249	&0.0009533   & 1 - +\\  \hline
\end{tabular}
\caption{The first 60 renormalized Dirichlet eigenvalues including the unrenormalized approximations for levels $m=5$ and $6$ along with the $D_4$ symmetry type of the associated eigenfunction} \label{tabdeig}
\end{center}
\end{table}

\begin{table}[p]
\begin{center}
\vspace{-2cm}
\begin{tabular}{|r|| c| l| l| c|} \hline
$\lambda_j$ & SC eigenvalue& level $m=5$ & level $m=6$ & Symmetry type \\ \hline \hline
1 &  0&0    &    0  &  constant \\  \hline
2,3 &	6.70561233677260	&  0.0000666   &    0.0000066   & 2 \\  \hline
4 &	17.7873066921127	& 0.0001768   &    0.0000176  &  1+ -\\  \hline
5 &	33.0503189296770	&  0.0003285   &    0.0000328  &  1 - +\\  \hline
6,7 &	44.1452040931981	& 0.0004388   &    0.0000438  &  2\\  \hline
8 &	46.1613387561034	& 0.0004588   &    0.0000458  &  1 + +\\  \hline
9,10 &67.1332757001913& 0.0006671   &    0.0000666  & 2 \\  \hline
11 &	72.9910330202890	& 0.0007254   &    0.0000724  &  1 + -\\  \hline
12 & 	76.5459449443382	&0.0007607   &    0.0000760  &  1 - -\\  \hline
13 &	91.2139397192494	& 0.0009066   &    0.0000905  & 1 + + \\  \hline
14 &	109.211435146286	& 0.0010854   &    0.0001084   & 2  \\  \hline
16 & 	112.309816147566	&0.0011160   &    0.0001115  &  1 - + \\  \hline
17,18 &160.534164976961& 0.0015954   &    0.0001594  & 2  \\  \hline
19 &	164.944259240120	& 0.0016393   &    0.0001637  & 1 - + \\  \hline
20 &	178.083831423592	& 0.0017703   &    0.0001768  &  1 + -\\  \hline
21 &	182.514354031062	& 0.0018139   &    0.0001812  &  1 + +\\  \hline 
22 & 	218.422016530391	&0.0021701   &    0.0002168  & 1 - - \\  \hline
23,24 &219.762176750716& 0.0021843   &    0.0002182  &  2\\  \hline 
25,26 &255.585276540520& 0.0025395   &    0.0002537  &  2\\  \hline
27 &	274.051345703083	& 0.0027237   &    0.0002721  &  1 + +\\  \hline
28 &	289.636409851151	& 0.0028784   &    0.0002875  &  1 + -\\  \hline
29 &	307.546192379348	& 0.0030560   &    0.0003053  &  1 - -\\  \hline
30 &	330.850994543218	&0.0032856   &    0.0003285  &  1 - +\\  \hline
31,32 & 339.697975499827	&0.0033753   &    0.0003373  &  2\\  \hline
33 &	368.889301910548	& 0.0036657   &    0.0003662  &  1 + -\\  \hline
34,35 &	387.765600423861	& 0.0038516   &    0.0003850  &  2\\  \hline
36 &	422.672579153705	& 0.0041989   &    0.0004196  &  1 + +\\  \hline
37 &	437.037035109719	& 0.0043418   &    0.0004339  &  1 - +\\  \hline
38,39 &441.945885521613& 0.0043910   &    0.0004388  &  2\\  \hline
40 &	450.427989161186	& 0.0044729   &    0.0004472  &  1 - -\\  \hline
41 &     462.106411740673   &    0.0045898    &  0.0004588& 1++ \\  \hline
42,43 &  475.895366084487 &   0.0047269&   0.0004725&	2	\\  \hline
44 &           486.753395979367    &   0.0048352   &   0.0004833&1+-\\  \hline
45&          506.813197803626     &  0.0050346    &  0.0005032& 1++  \\  \hline
46,47&      518.365005140405  &     0.0051477 &     0.0005147&2	\\  \hline
48&          526.708786555689     &  0.0052313   &   0.0005229& 1--\\  \hline
49&          558.331449520164     &  0.0055449     & 0.0005543&1+-\\  \hline
50&      569.973846496736         &  0.0056662    &  0.0005665& 1+-\\  \hline
51&  570.615629740279		&0.0056662	&0.0005665&  1+-\\  \hline
52&570.690152093376		&0.0056733	&0.0005673&1 -+\\  \hline
53&644.327571161679		&0.0064053	&0.0006408&1-+\\  \hline
54&649.874198213270		&0.0064871	&0.0006463&1++\\  \hline
55,56&652.554315294009	&0.0064871	&0.0006488&2\\  \hline
57,58&670.494059872279	&0.0066655	&0.0006671&2\\  \hline
59&671.509205341636		&0.0066756	&0.0006671&1++\\  \hline
60&674.251348906568		&0.0067028	&0.0006679&1++\\  \hline

\end{tabular}
\caption{The first 60 renormalized Neumann eigenvalues including the unrenormalized approximations for levels $m=5$ and $6$ along with the $D_4$ symmetry type of the associated eigenfunction} \label{tabneig}
\end{center}
\end{table}

\begin{figure}[p]
\begin{center}
\includegraphics[scale=.55]{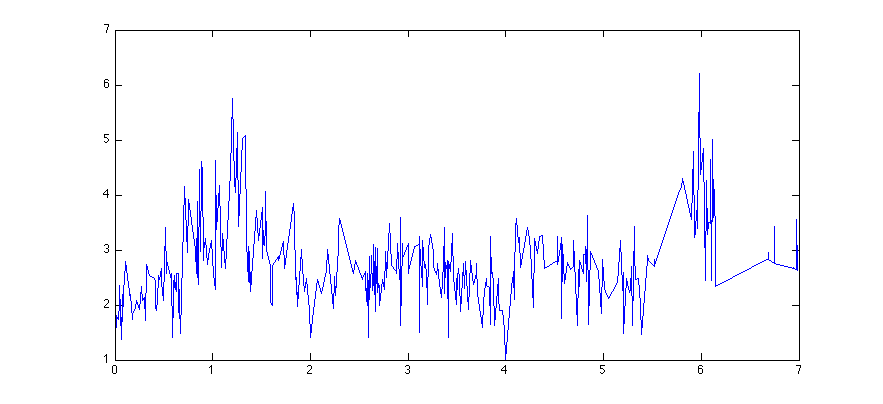}\\
\includegraphics[scale=.55]{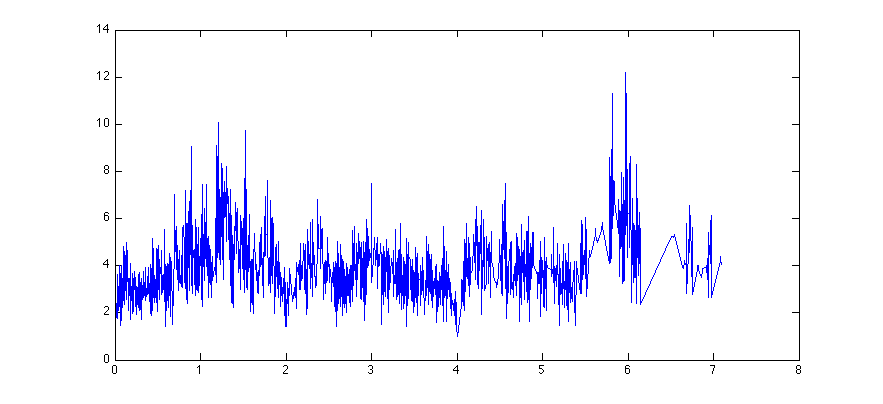}\\
\includegraphics[scale=.5]{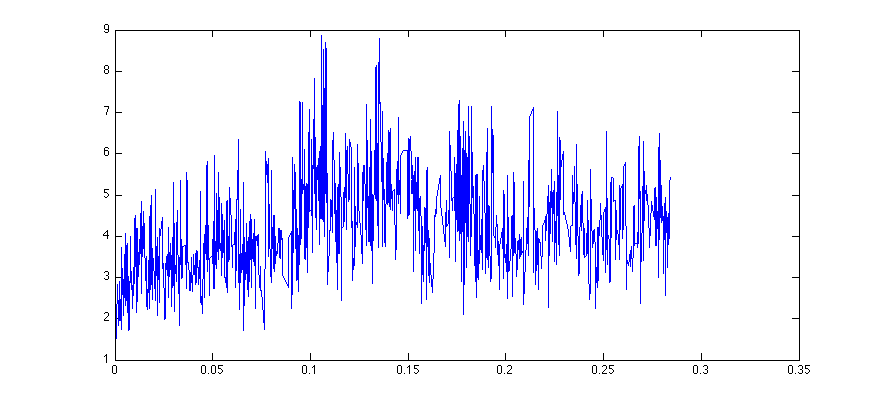}
\end{center}
\caption{The values of $\max |\varphi_j|$ plotted as a function of $\lambda_j$ for levels 3, 4, and 5 respectively.  The plot for level 5 is incomplete as we are only able to compute the first 1500 eigenvalues but it still appears to exhibit the same behavior.} \label{figmaxeig}
\end{figure}

We computed the $L^\infty$ norm of the eigenfunctions $\varphi_j$ (normalized so $\|\varphi_j\|_2=1)$.  Figure \ref{figmaxeig} shows the values plotted as a function of the eigenvalue.  The data is consistent with the uniform boundedness of $\|\varphi_j\|_\infty$ as $j\to\infty$, although we could not realistically expect to detect slow growth rates such as $\|\varphi_j\|_\infty= O(\log \lambda_j).$  Of course uniform boundedness hold for the ordinary Laplacian on the square, but is in sharp contrast with the power law growth on general Riemannian manifolds and on SC (localized eigenfunctions provide obvious examples).

\begin{figure}[p]
\begin{center}
\includegraphics[scale=.5]{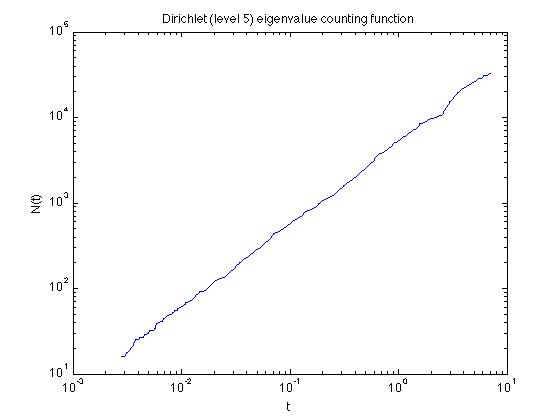}
\end{center}
\caption{Eigenvalue counting function for the level 5 Dirichlet eigenvalues on a log-log scale.}\label{figNd}
\end{figure}

\begin{figure}[p]
\begin{center}
\includegraphics[scale=.5]{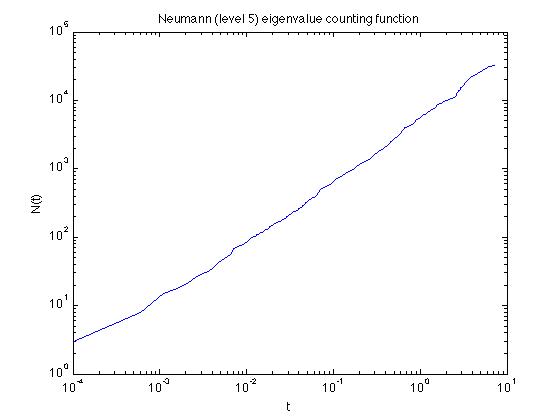}
\end{center}
\caption{Eigenvalue counting function for the level 5 Neumann eigenvalues on a log-log scale.}\label{figNn}
\end{figure}

In Figures \ref{figNd} and \ref{figNn} we graph the eigenvalue counting functions
\begin{equation}\label{4.3}
N(t)=\#\{\lambda_j : \lambda_j \leq t \}
\end{equation}
on a log-log scale, confirming the predicted slope of
\begin{equation} \label{4.4}
\alpha = \frac{\log 8}{\log \rho} \approx 0.9026.
\end{equation}


\begin{figure}[p]
\begin{center}
\includegraphics[scale=.4]{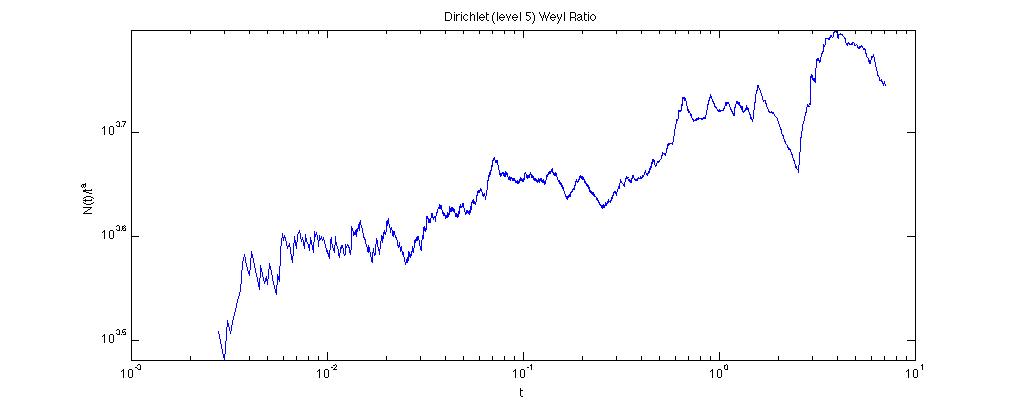}
\end{center}
\caption{Weyl ratio for the level 5 Dirichlet eigenvalues.}\label{figDWeyl}
\end{figure}

\begin{figure}[p]
\begin{center}
\includegraphics[scale=.35]{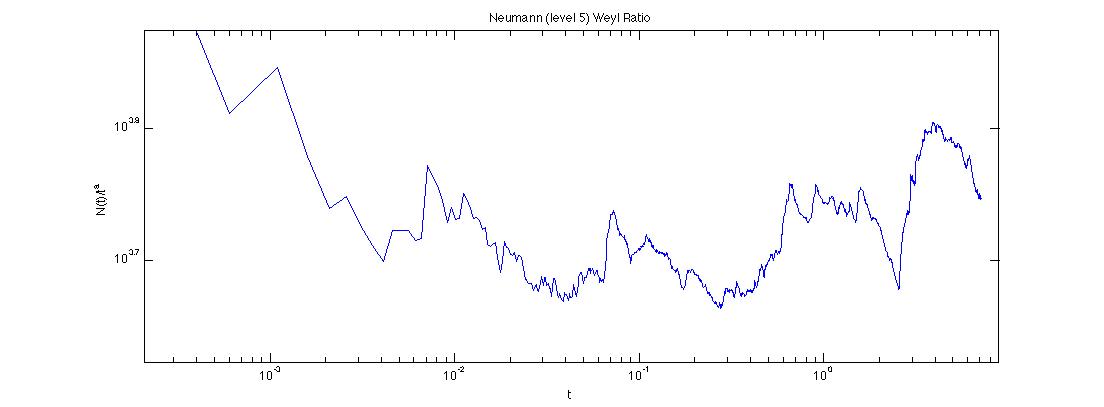}
\end{center}
\caption{Weyl ratio for the level 5 Neumann eigenvalues.}\label{figNWeyl}
\end{figure}

In Figures \ref{figDWeyl} and \ref{figNWeyl} we graph the Weyl ratios
\begin{equation}\label{4.5}
W(t)=\frac{N(t)}{t^\alpha}.
\end{equation}


It is believed that asymptotically $W(t)$ should approach a multiplicatively periodic function (period $\rho$) bounded and bounded away from zero (the same function for both boundary conditions).  Our data is inconclusive on this conjecture.

We are also interested in the difference between the two counting functions.  By general principles, the Neumann counting function $N^{(N)}(t)$ will be greater than the Dirichlet counting functions  $N^{(D)}(t)$, and we expect
\begin{equation}\label{4.6}
N^{(N)}(t)-N^{(D)}(t) \sim t^\beta
\end{equation}
for some $\beta$ that measures the relative sizes of $\partial$SC and SC.  A simple guess is 
\begin{equation}\label{4.7}
\beta=\frac{\log 3}{\log \rho} \approx 0.4769.
\end{equation}

\begin{figure}[p]
\begin{center}
\includegraphics[scale=.5]{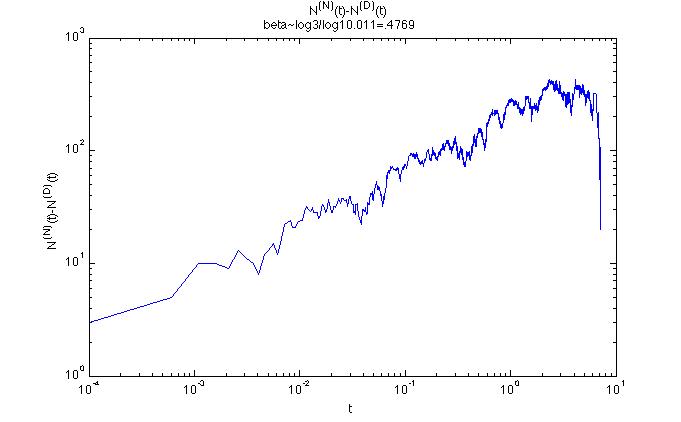}
\end{center}
\caption{The difference of the eigenfunction counting functions $N^{(N)}(t)-N^{(D)}(t)$ at level 5.}\label{figNn-Nd}
\end{figure}

\begin{figure}[p]
\begin{center}
\includegraphics[scale=.4]{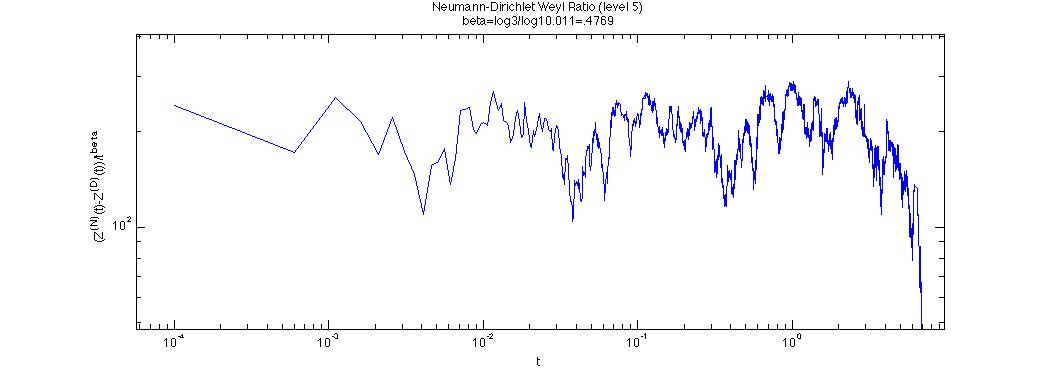}
\end{center}
\caption{Weyl ratio for $N^{(N)}(t)-N^{(D)}(t)$ given in \eqref{4.7}.}\label{figNn-NdWeyl}
\end{figure}

In Figure \ref{figNn-Nd} we graph the difference and in Figure \ref{figNn-NdWeyl} the associated Weyl ratio
\begin{equation}\label{4.7}
\frac{N^{(N)}(t)-N^{(D)}(t)}{t^\beta}.
\end{equation}
Again this is presumed to be asymptotic to a multiplicatively periodic function.


In our data there is no evidence for large gaps in the spectrum, but there is evidence for spectral clustering (distinct eigenvalues that are close together, see \cite{CSW}).  The spectral asymptotics predicts that the separation between distinct eigenvalues should go to infinity on average.  Nevertheless, in the Dirichlet spectrum we have eigenvalue pair \#53-54 with value 860.58 and \#55 with values 861.92, we have four eigenvalues \#118-121 in the range 2005.8 to 2015.6, and four eigenvalues \#152-155 in the range 2786.7 to 2787.8.

\begin{figure}[h]
\begin{center}
\includegraphics[scale=.3]{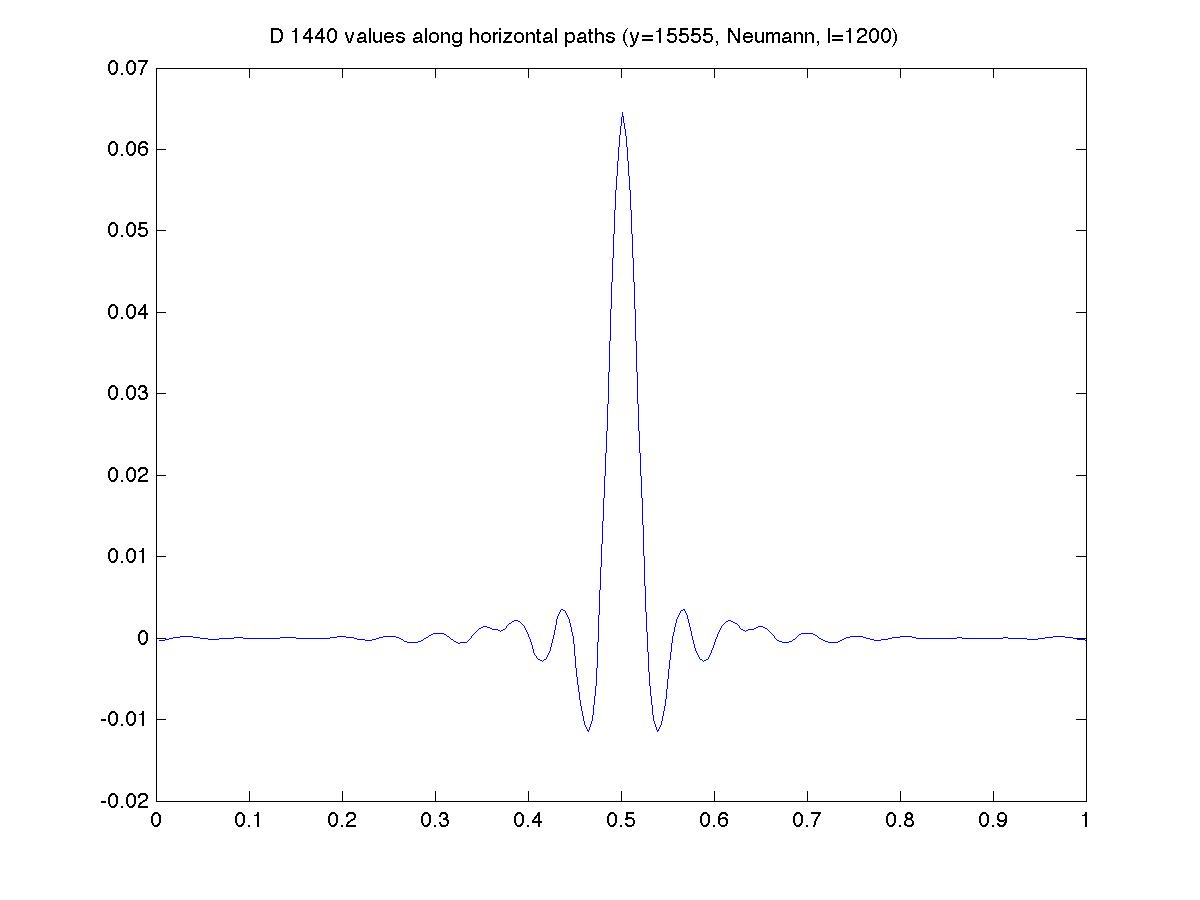}
\end{center}
\caption{Dirichlet kernel $D_N(x,y)$ on SC$_5$ using the first 1200 Neumann eigenvalues where $y$ is the cell '1555' (the center cell directly above the level-1 hole) and $x$ varies along the horizontal line containing $y$.}\label{figDn}
\end{figure}

Using our eigenfunction computations we were able to construct Dirichlet kernels
\begin{equation}\label{4.8}
D_N(x,y)=\Sum_{j=1}^N \varphi_j(x)\varphi_j(y) 
\end{equation}
where $\{\varphi_j\}$ are the orthonormal eigenfunctions.  To the extent that $D_N(\cdot,y)$ resembles an approximate identity about the point $y$, this will tell us about the convergence of eigenfunction expansions.  It was observed in \cite{OSS} for the SG fractal that choosing $N$ so that the sum goes up to a spectral gap produces a strong approximate identity appearance, and later in \cite{S2} it was shown for SG and related fractals that convergence of eigenfunction expansions for these special partial sums is much better than for ordinary Fourier series (continuous functions have uniformly convergent special partial sums).  As already mentioned, the SC spectra do not have such large spectral gaps, but even if we choose $N$ to reach one of the more moderate gaps, the Dirichlet kernels do not show any approximate identity appearance.  In Figure \ref{figDn} we show the graph of $D_N(x,y)$ for $x$ restricted to a line segment in SC passing through the point $y$.  The slow decay away from $x=y$ is reminiscent of the Dirichlet kernels for ordinary Fourier series. 

\begin{figure}[p]
\begin{center}
\includegraphics[scale=.08]{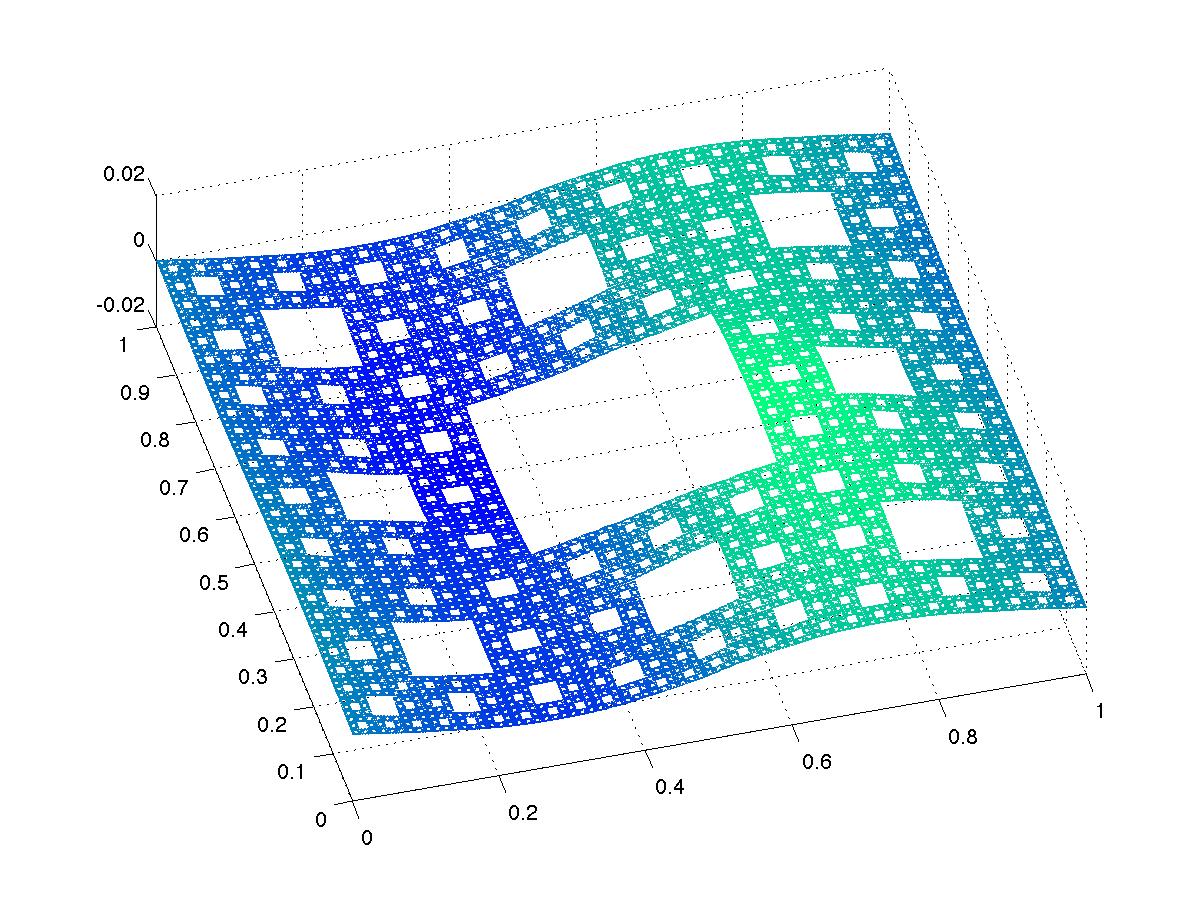}
\includegraphics[scale=.08]{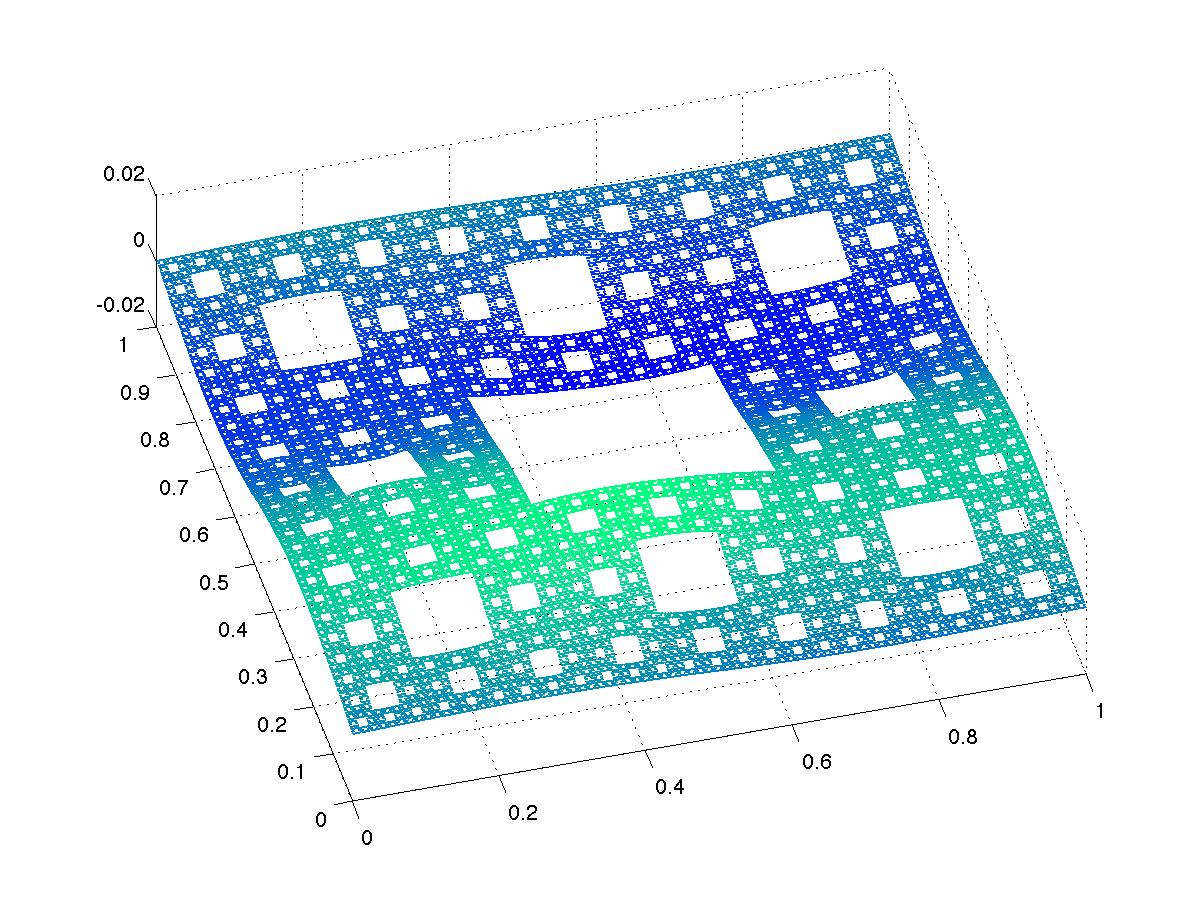}
\includegraphics[scale=.08]{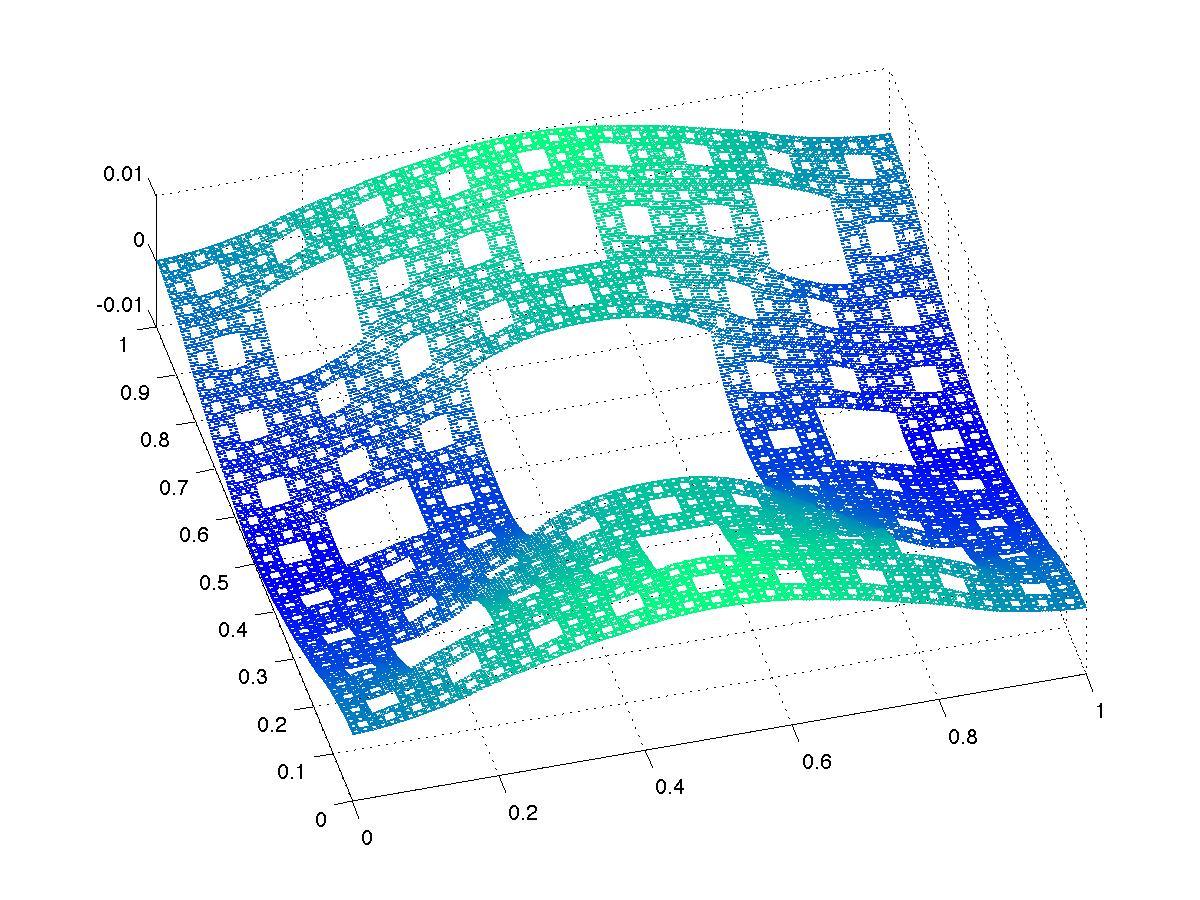}\\
\includegraphics[scale=.08]{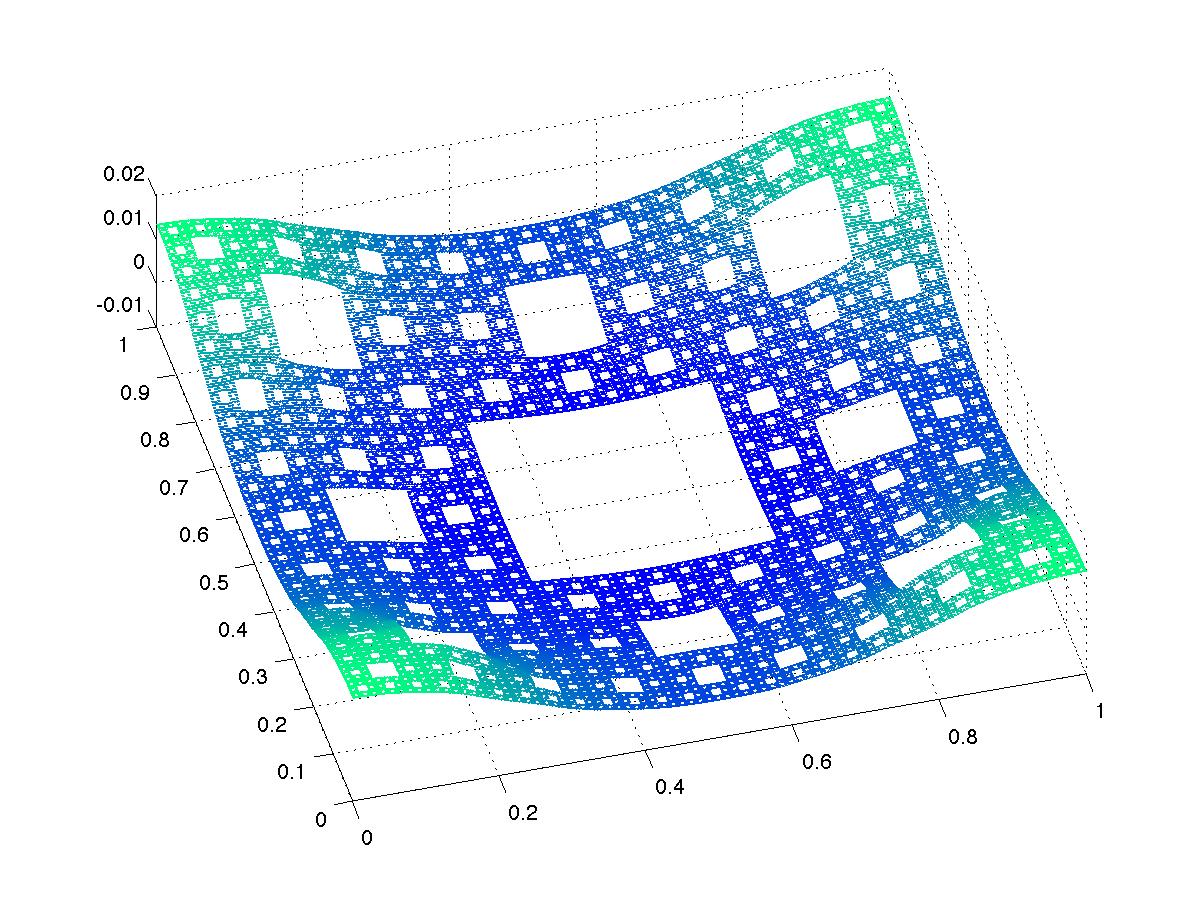}
\includegraphics[scale=.08]{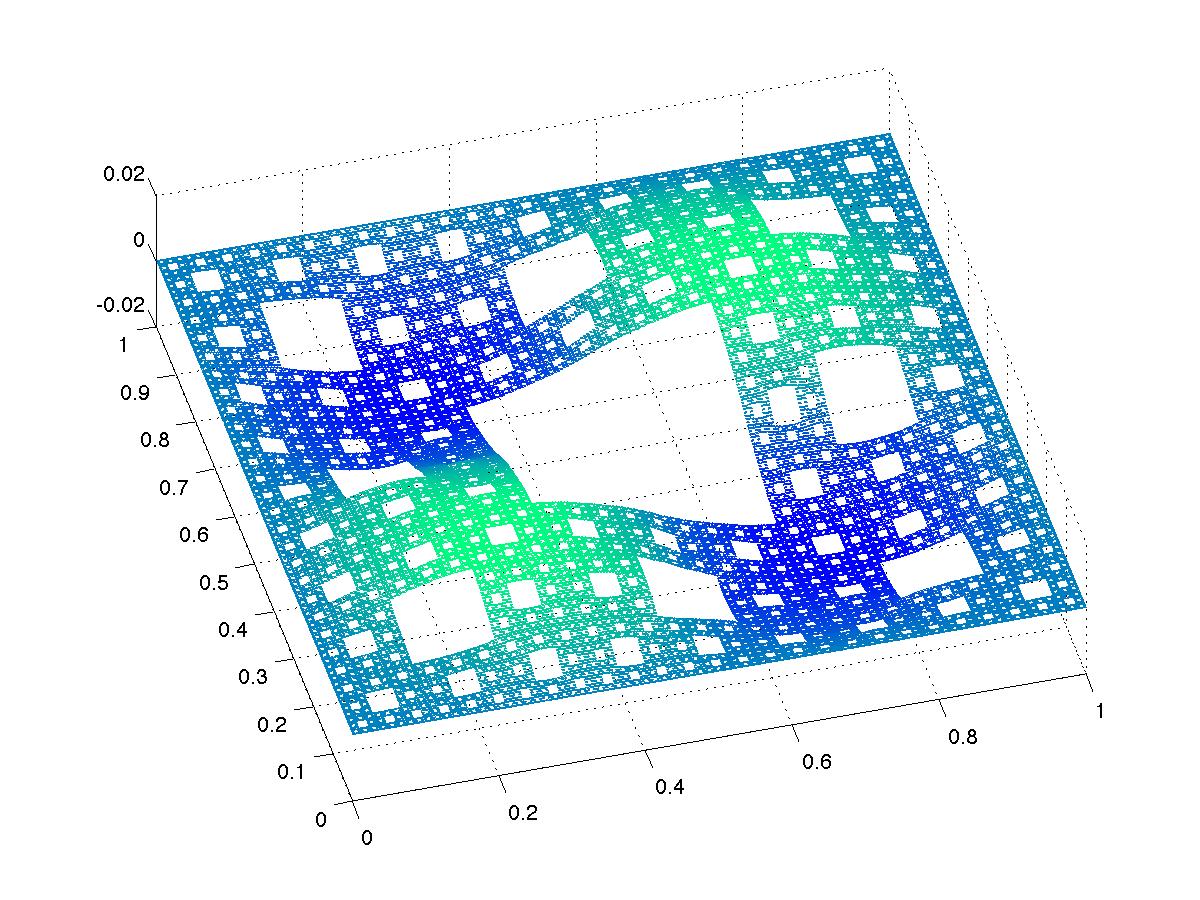}
\includegraphics[scale=.08]{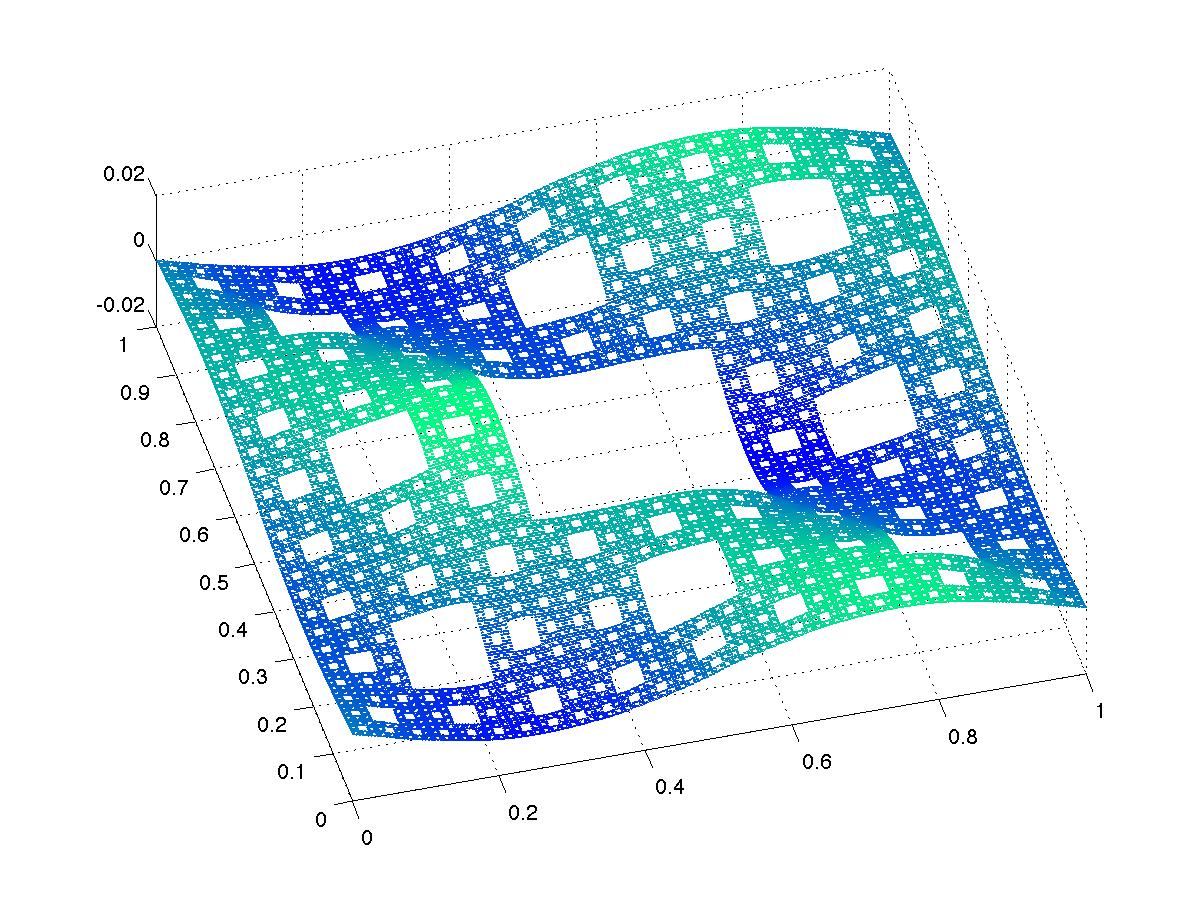}\\
\includegraphics[scale=.08]{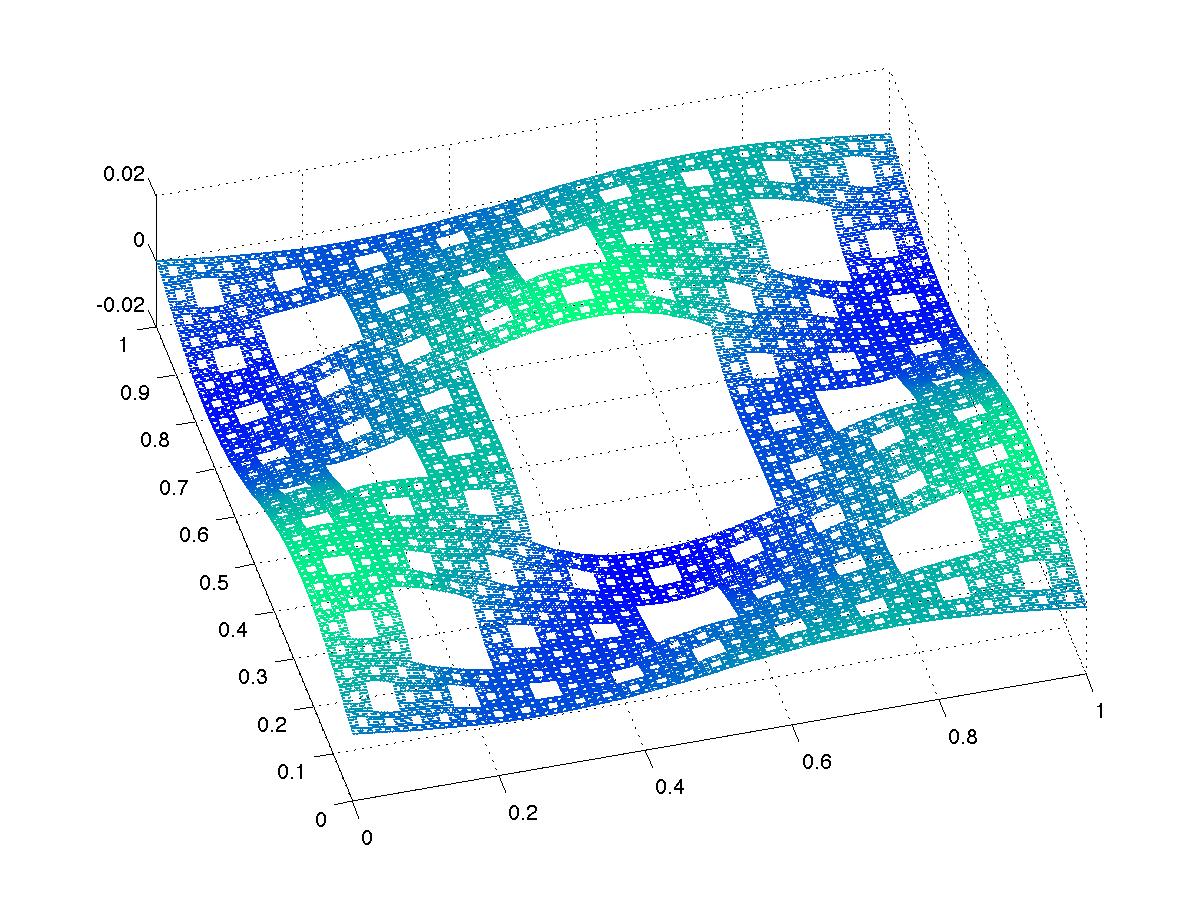}
\includegraphics[scale=.08]{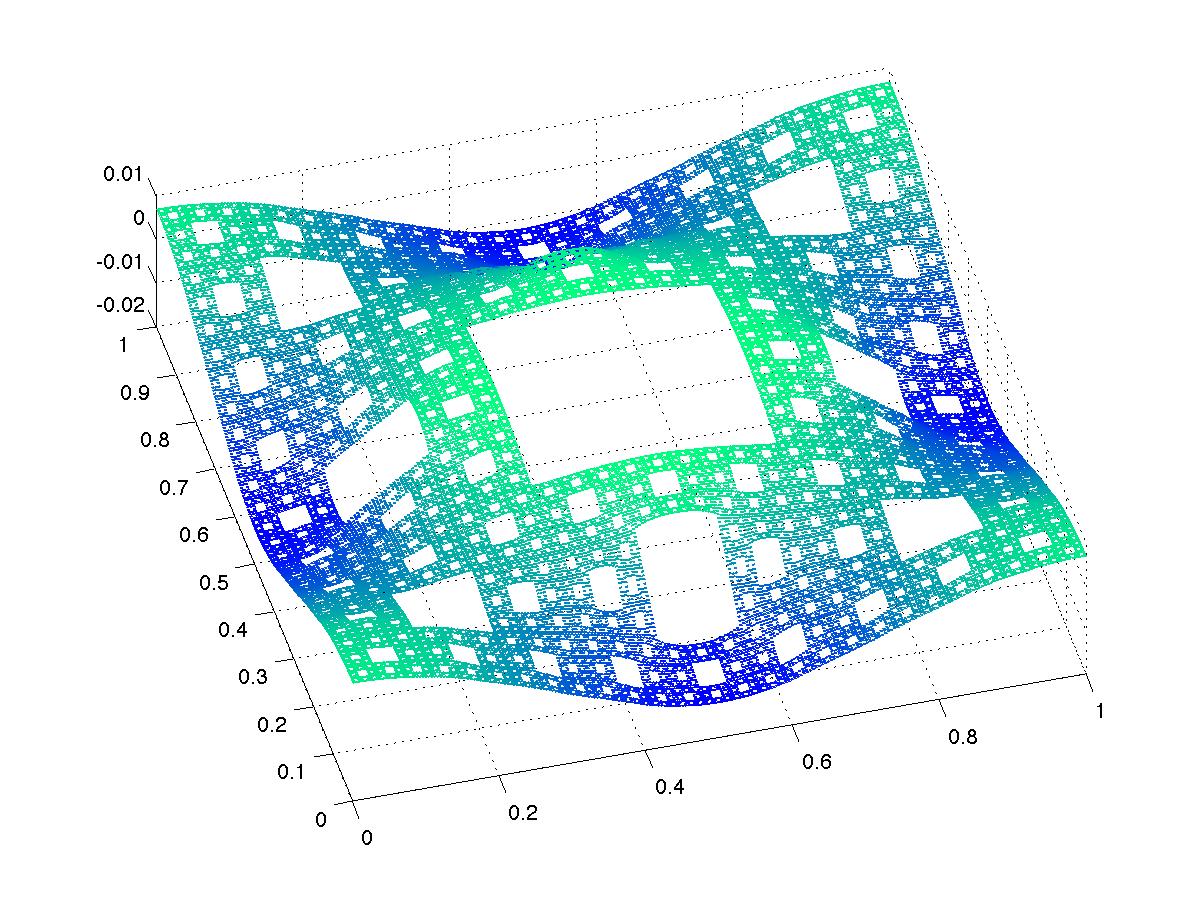}
\includegraphics[scale=.08]{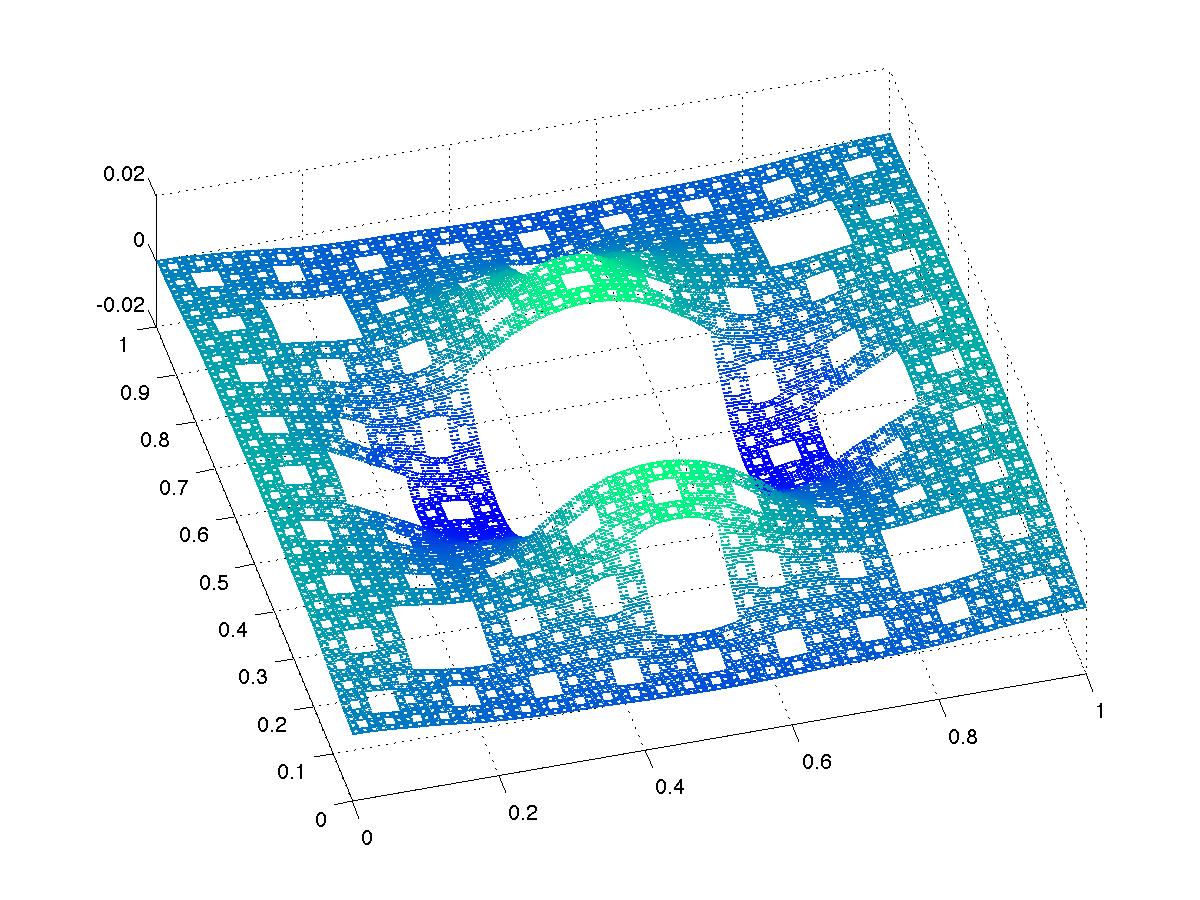}
\end{center}
\caption{The first nine nonconstant eigenfunctions on SC$_5$ with torus boundary conditions}\label{figtoruseigs}
\end{figure}

\begin{figure}[p]
\begin{center}
\includegraphics[scale=.08]{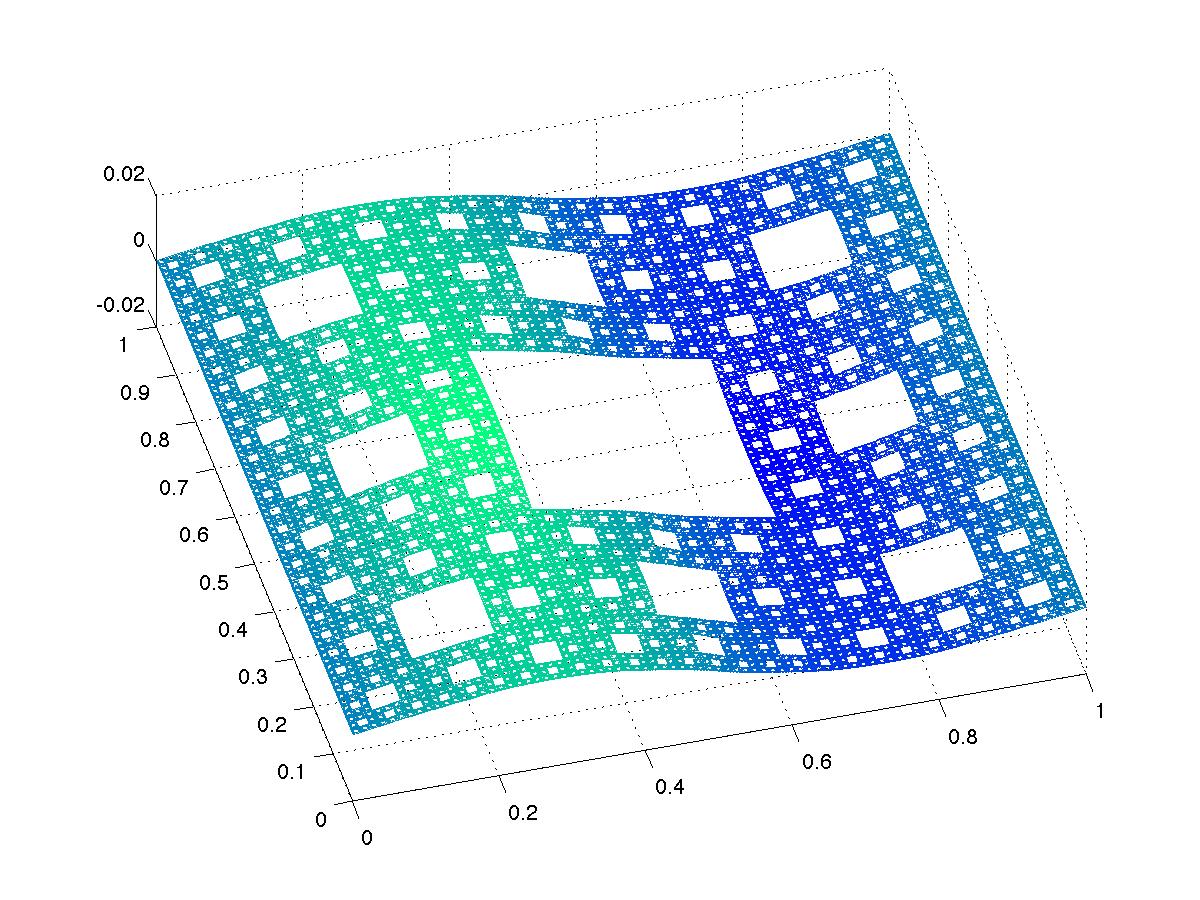}
\includegraphics[scale=.08]{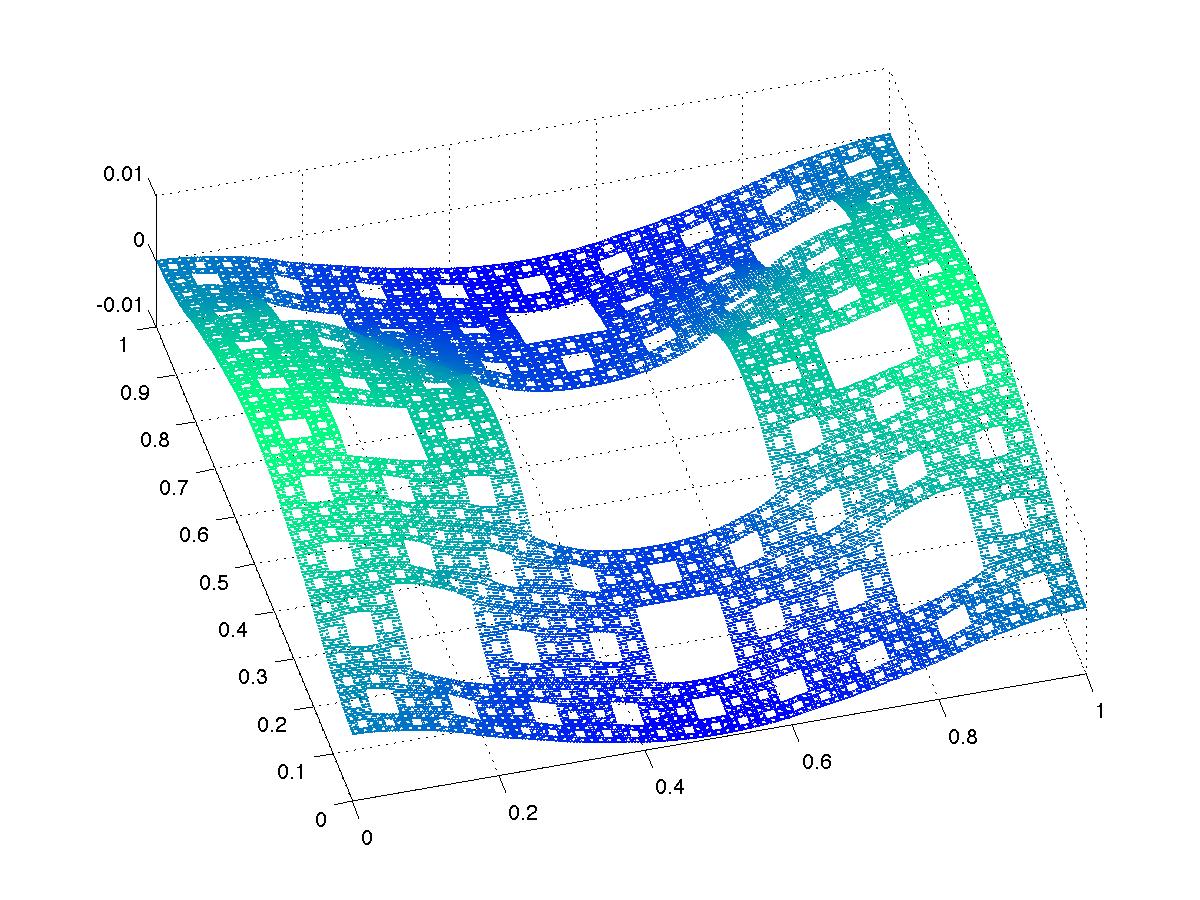}
\includegraphics[scale=.08]{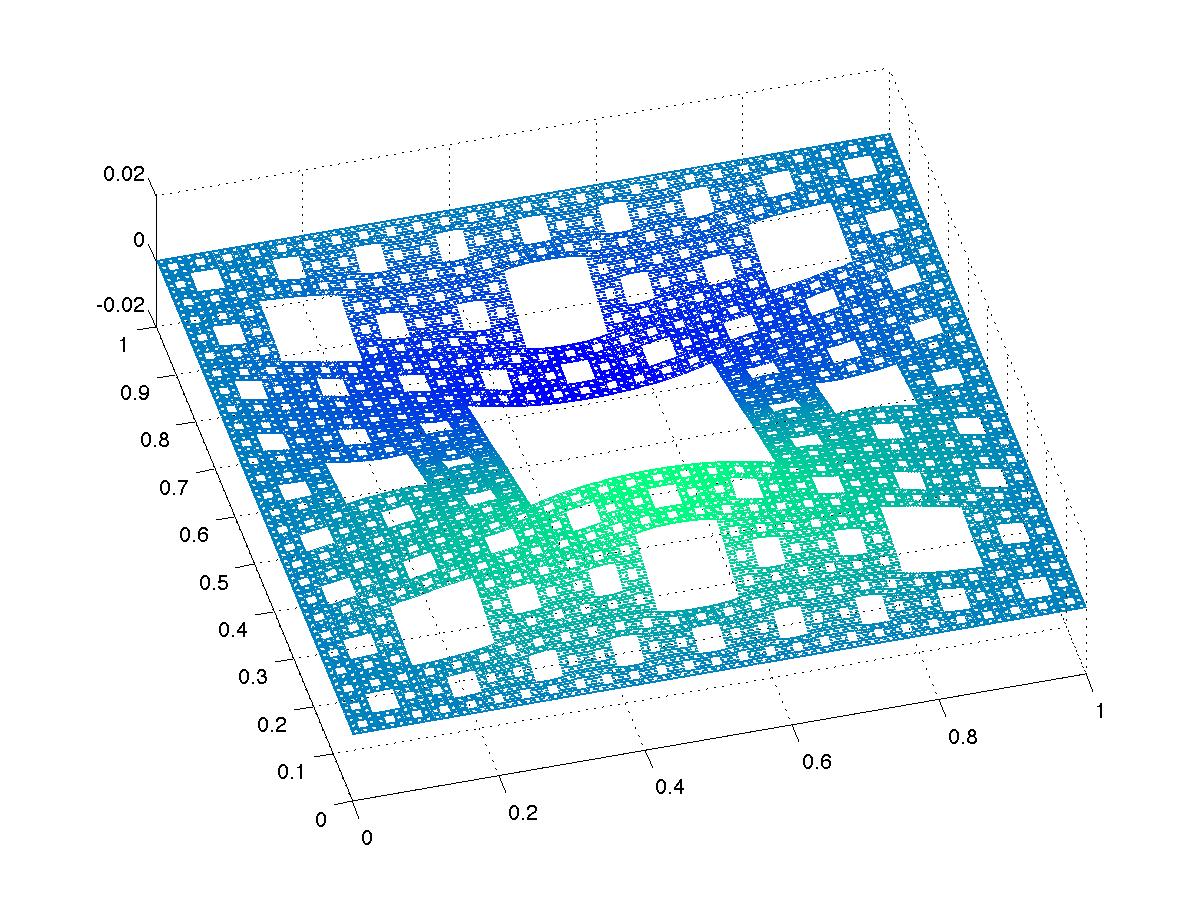}\\
\includegraphics[scale=.08]{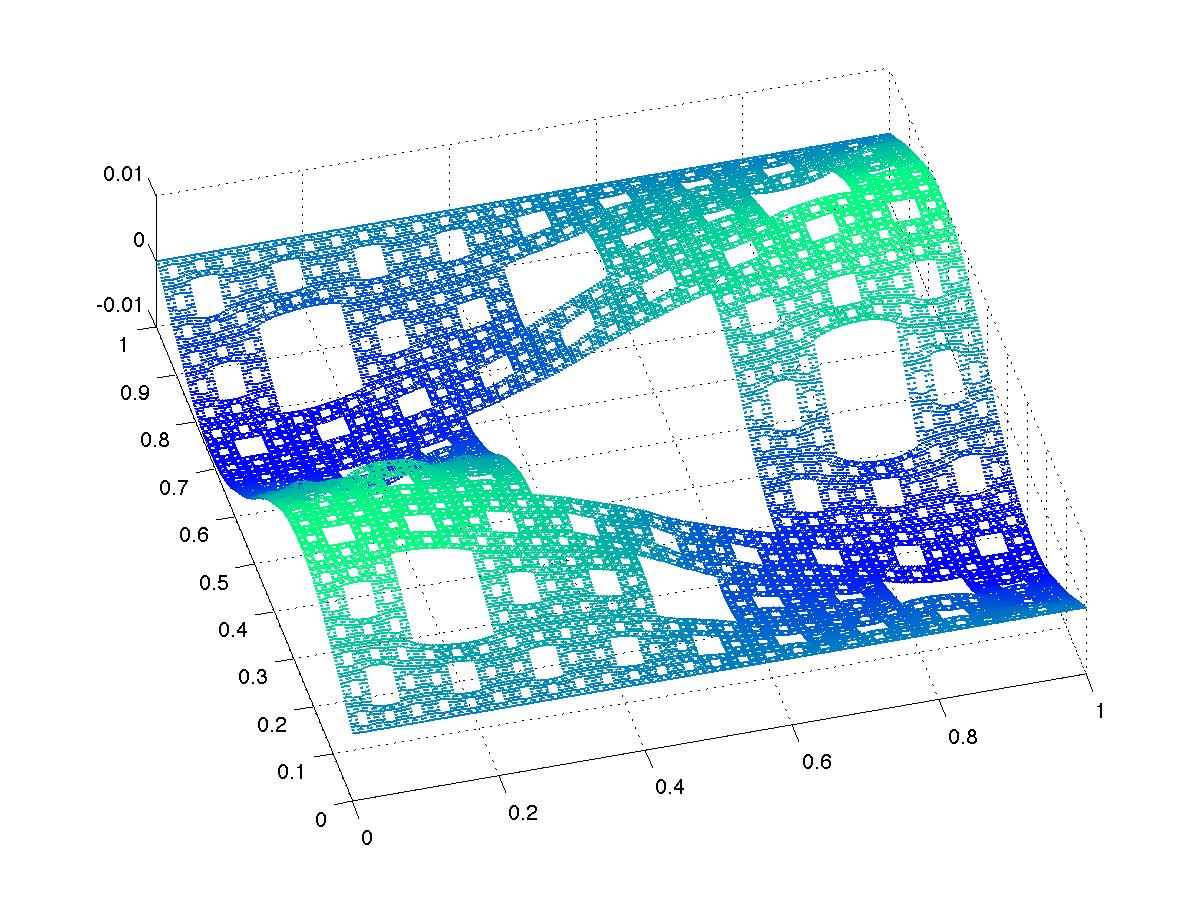}
\includegraphics[scale=.08]{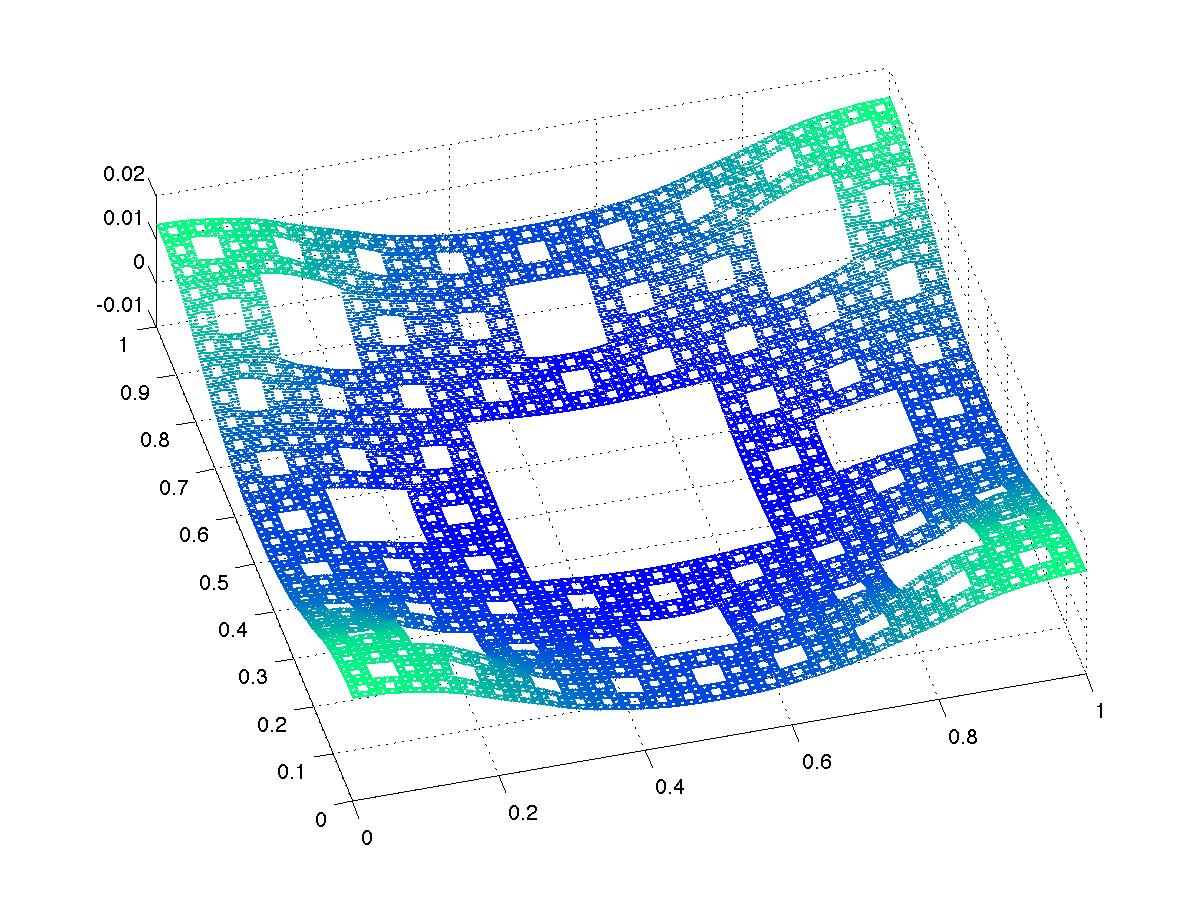}
\includegraphics[scale=.08]{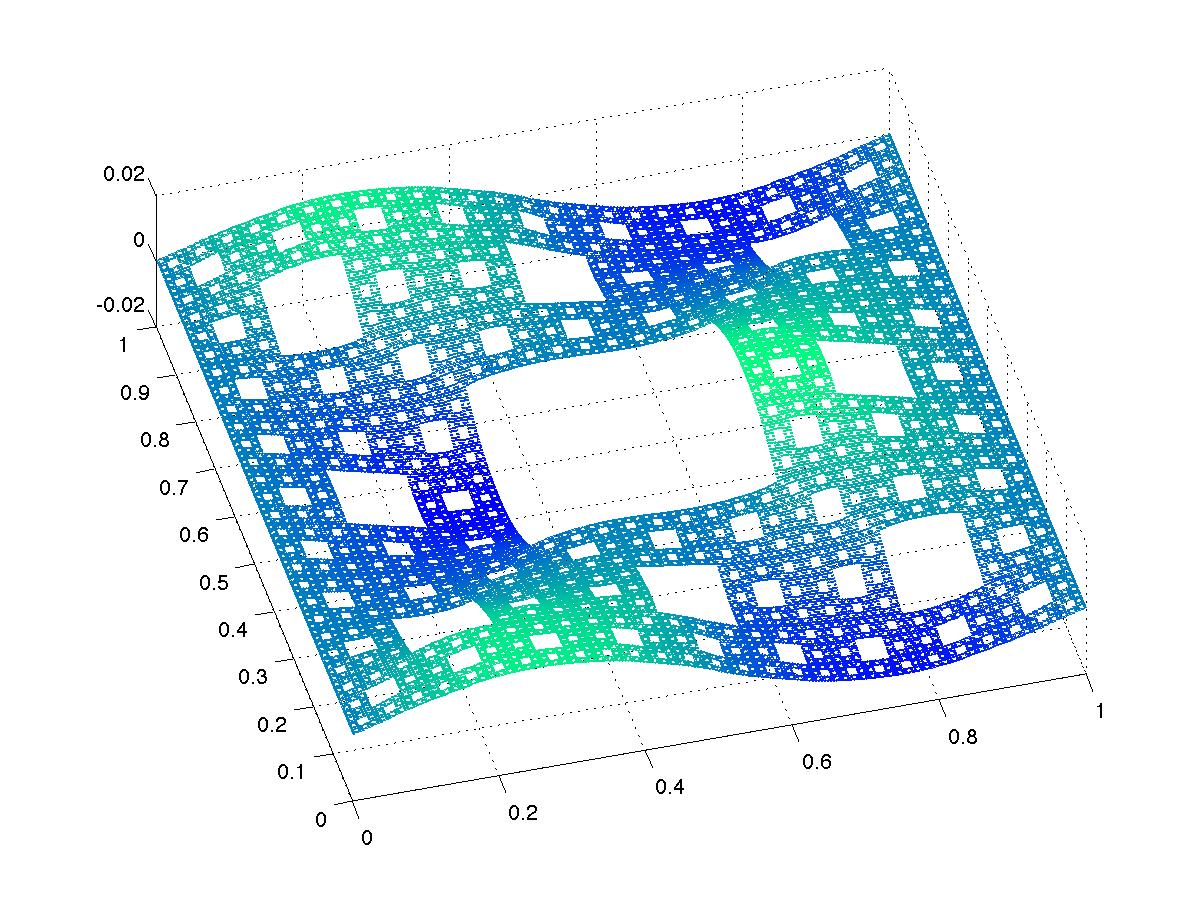}\\
\includegraphics[scale=.08]{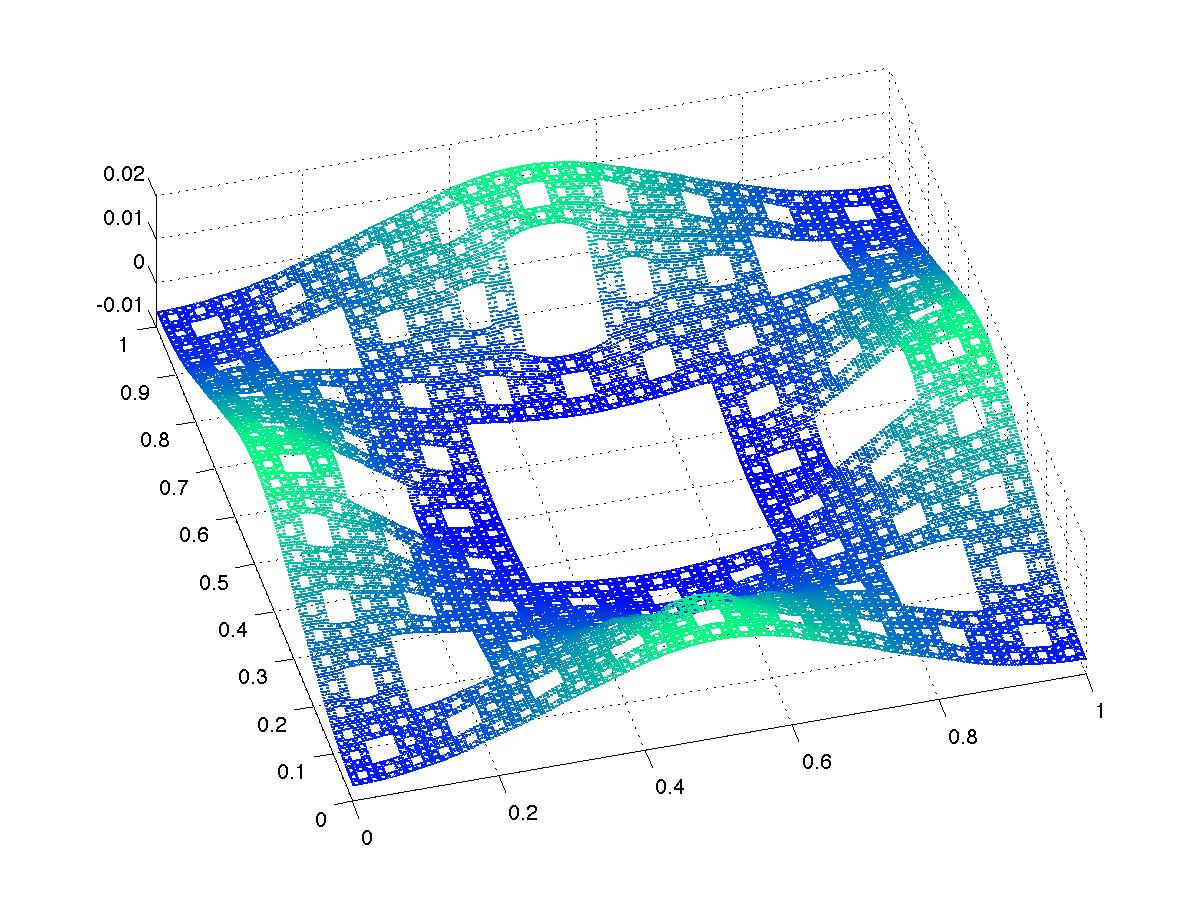}
\includegraphics[scale=.08]{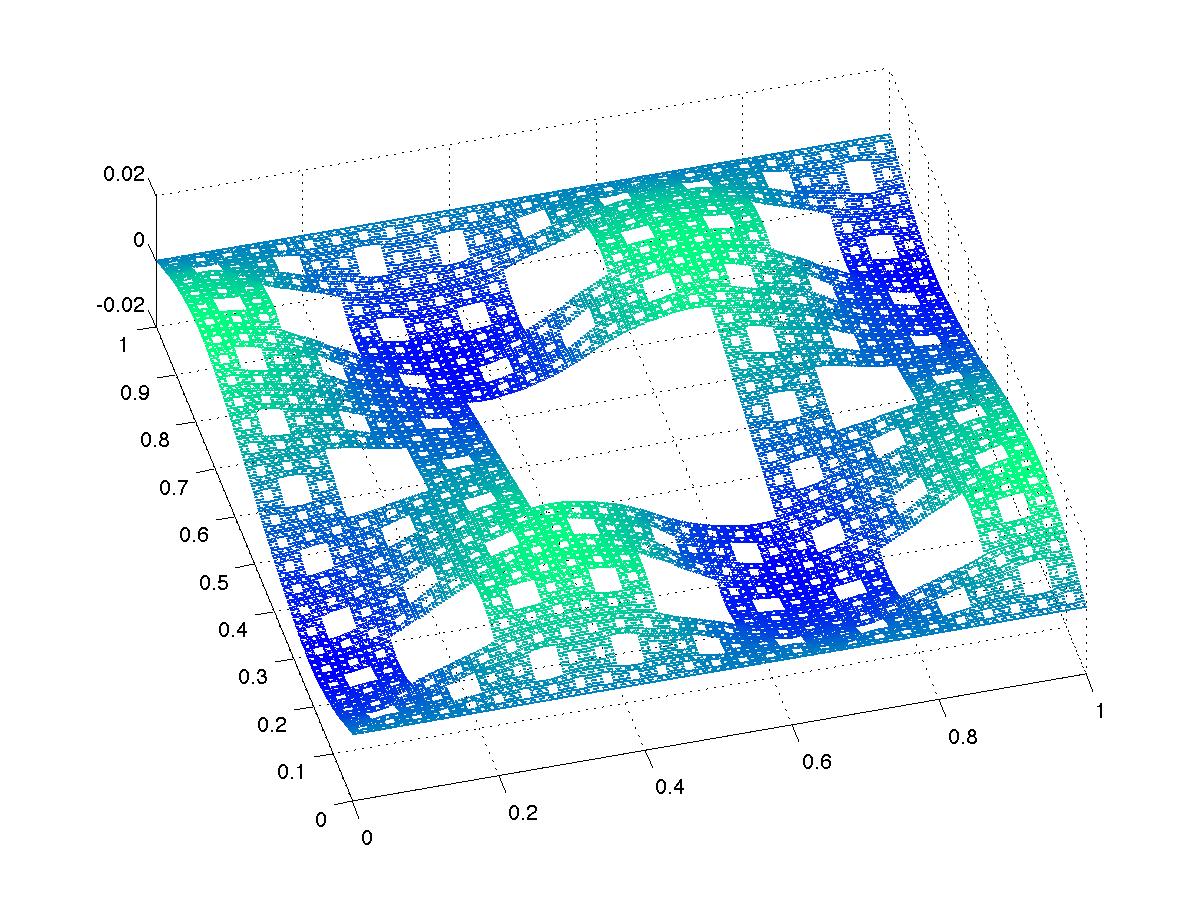}
\includegraphics[scale=.08]{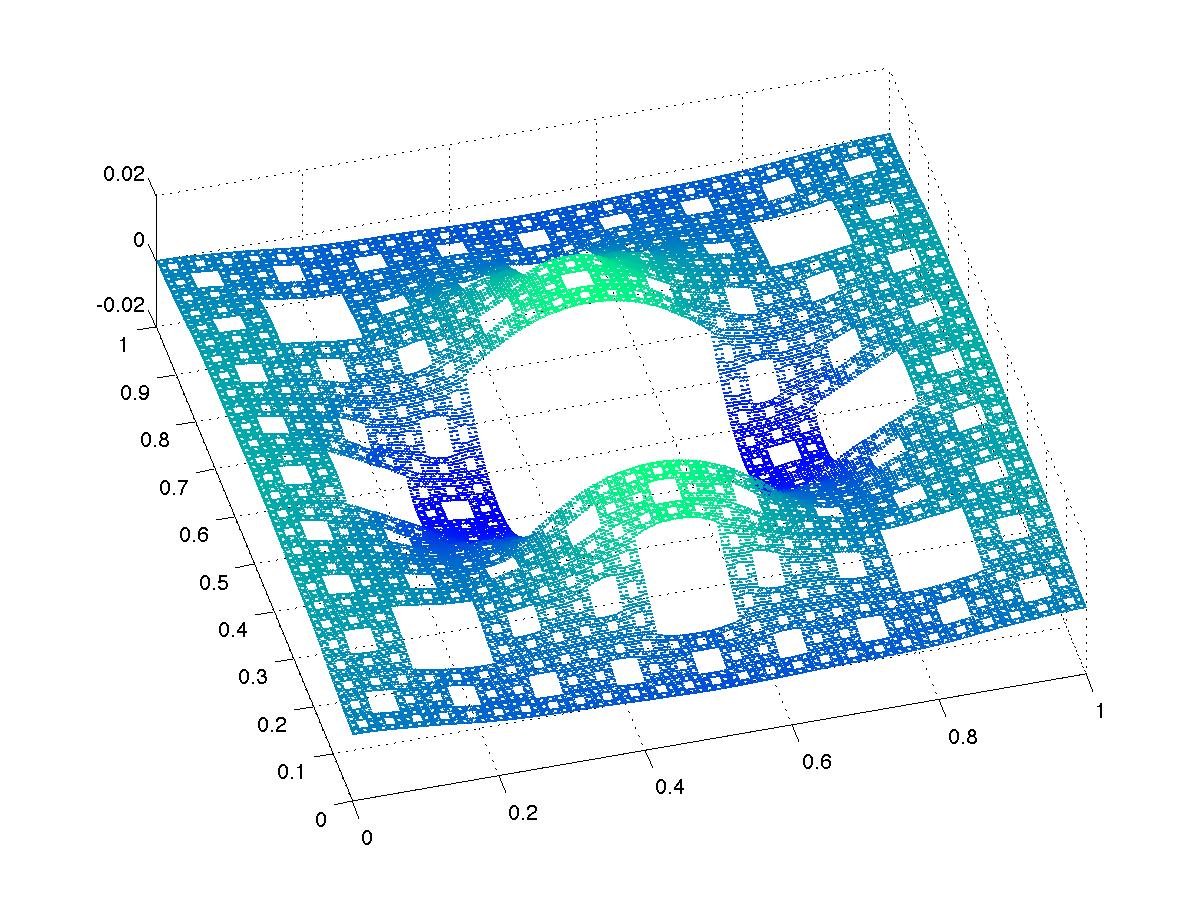}
\end{center}
\caption{The first nine nonconstant eigenfunctions on SC$_5$ with Klein bottle boundary conditions}\label{figkleineigs}
\end{figure}

\begin{figure}[p]
\begin{center}
\includegraphics[scale=.08]{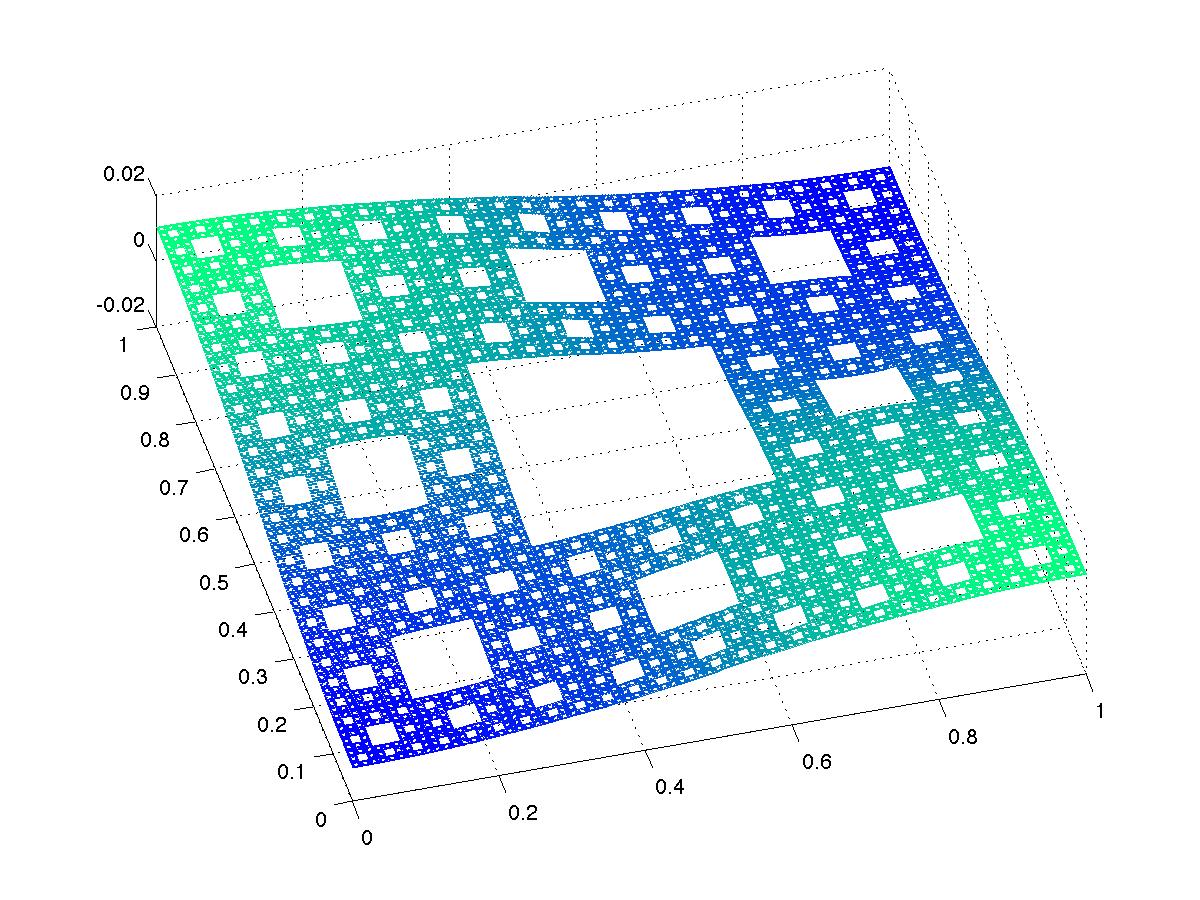}
\includegraphics[scale=.08]{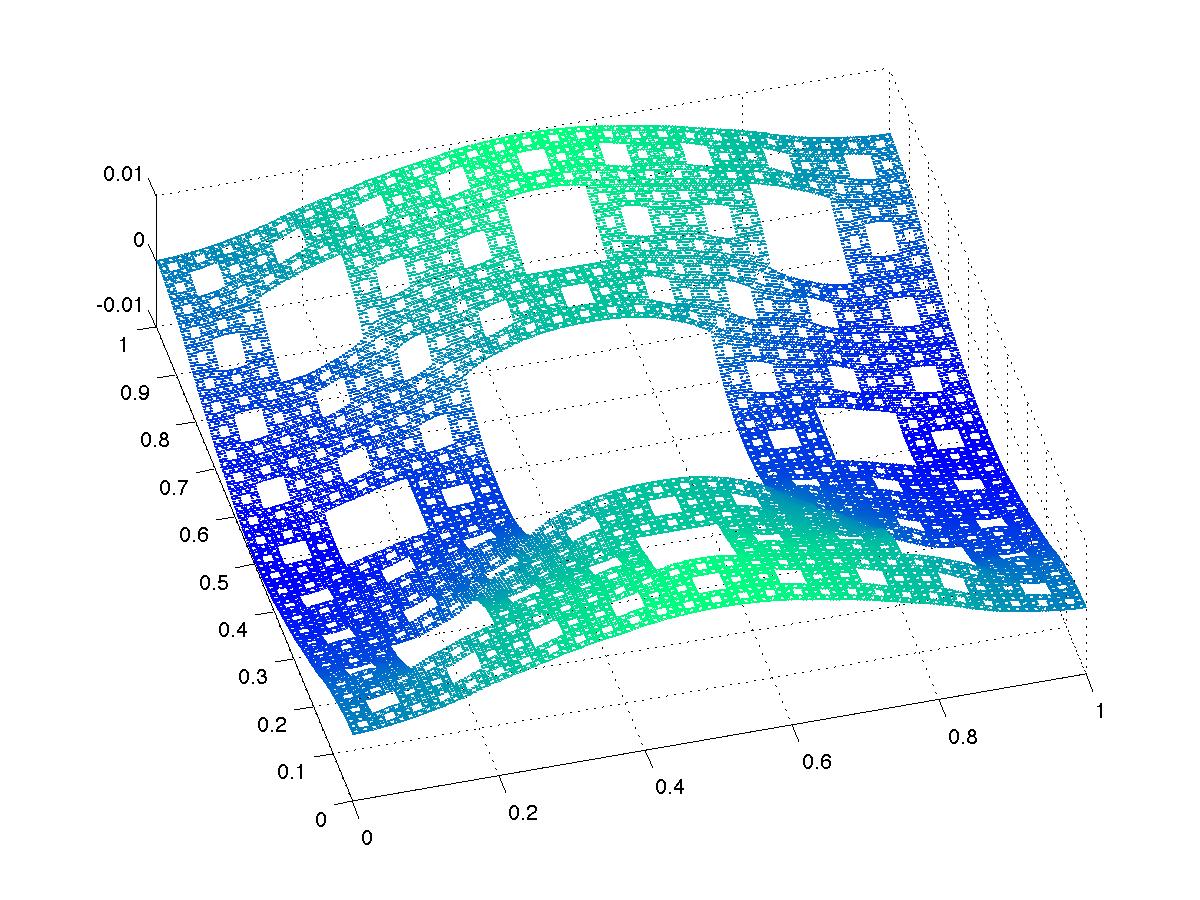}
\includegraphics[scale=.08]{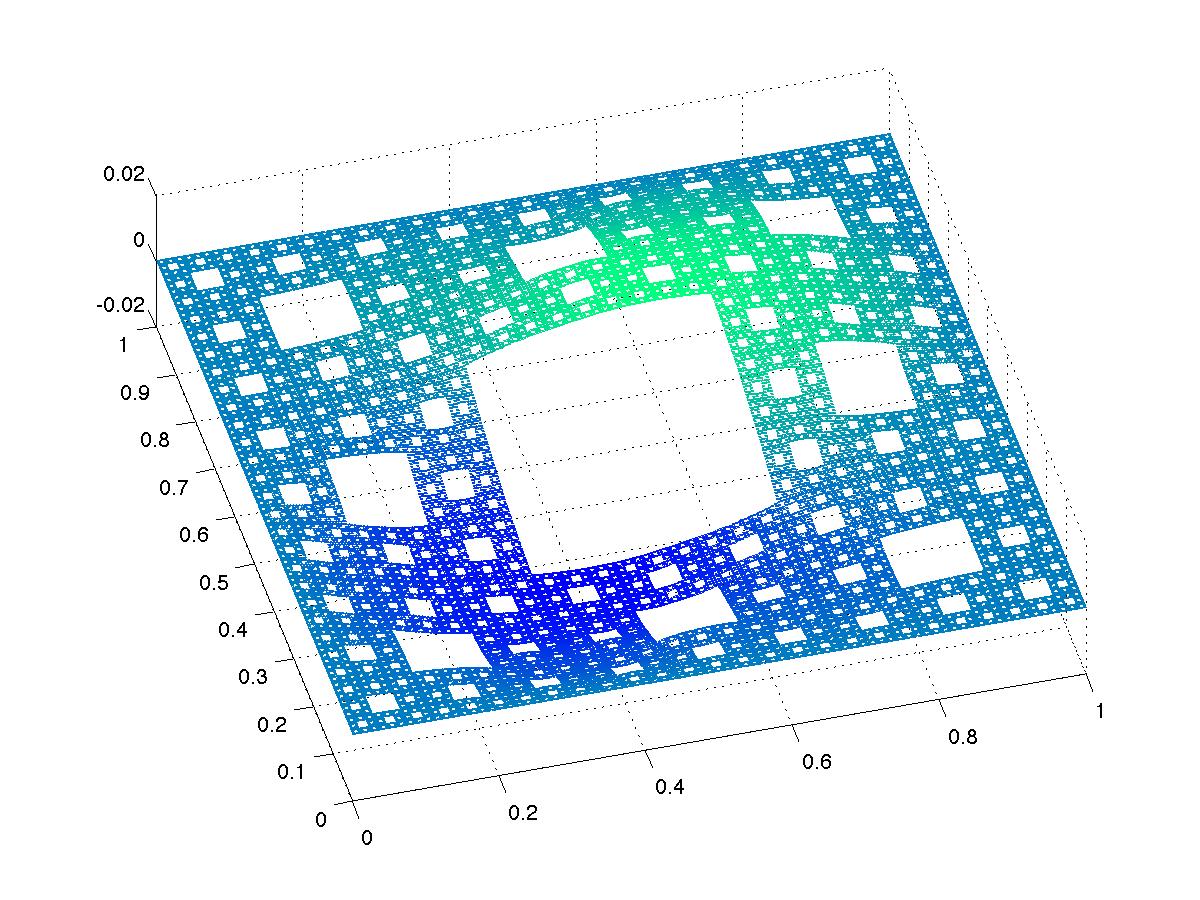}\\
\includegraphics[scale=.08]{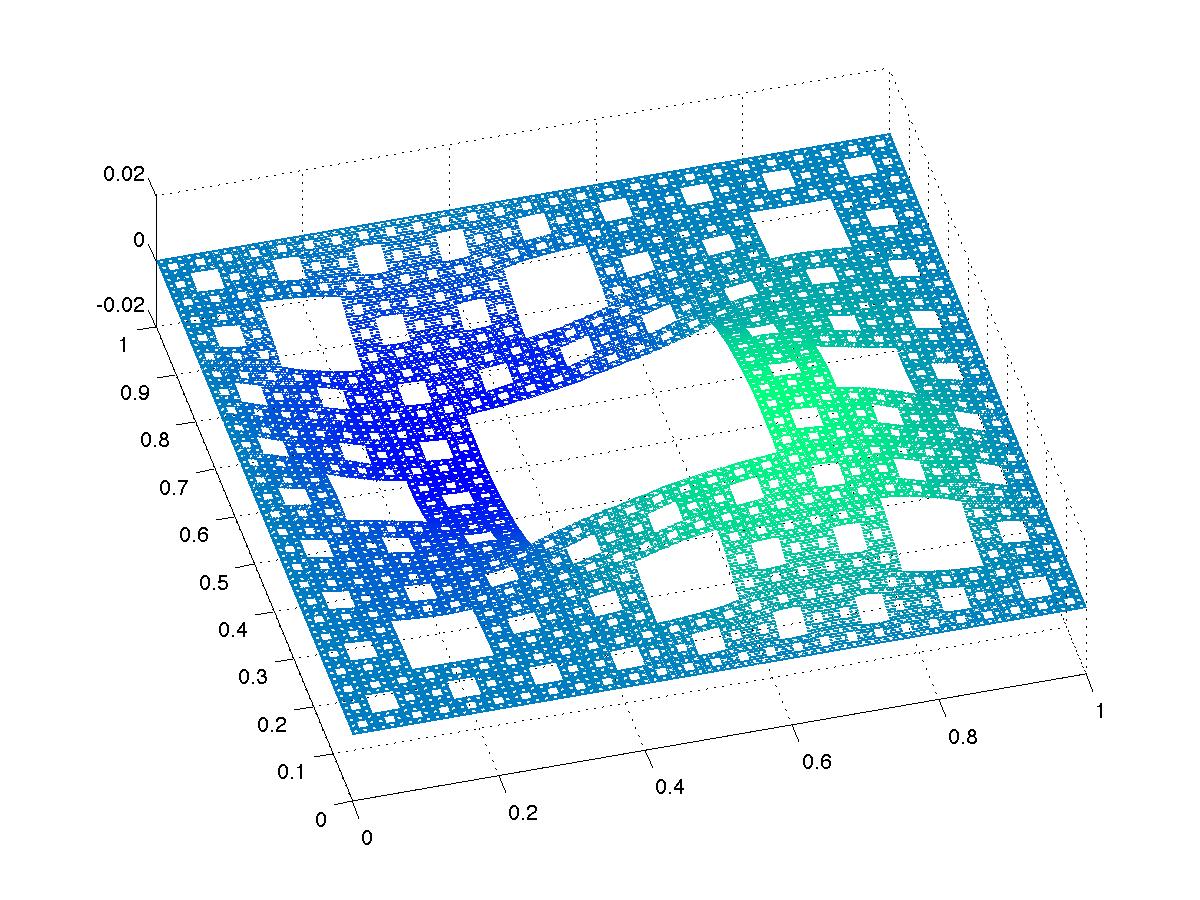}
\includegraphics[scale=.08]{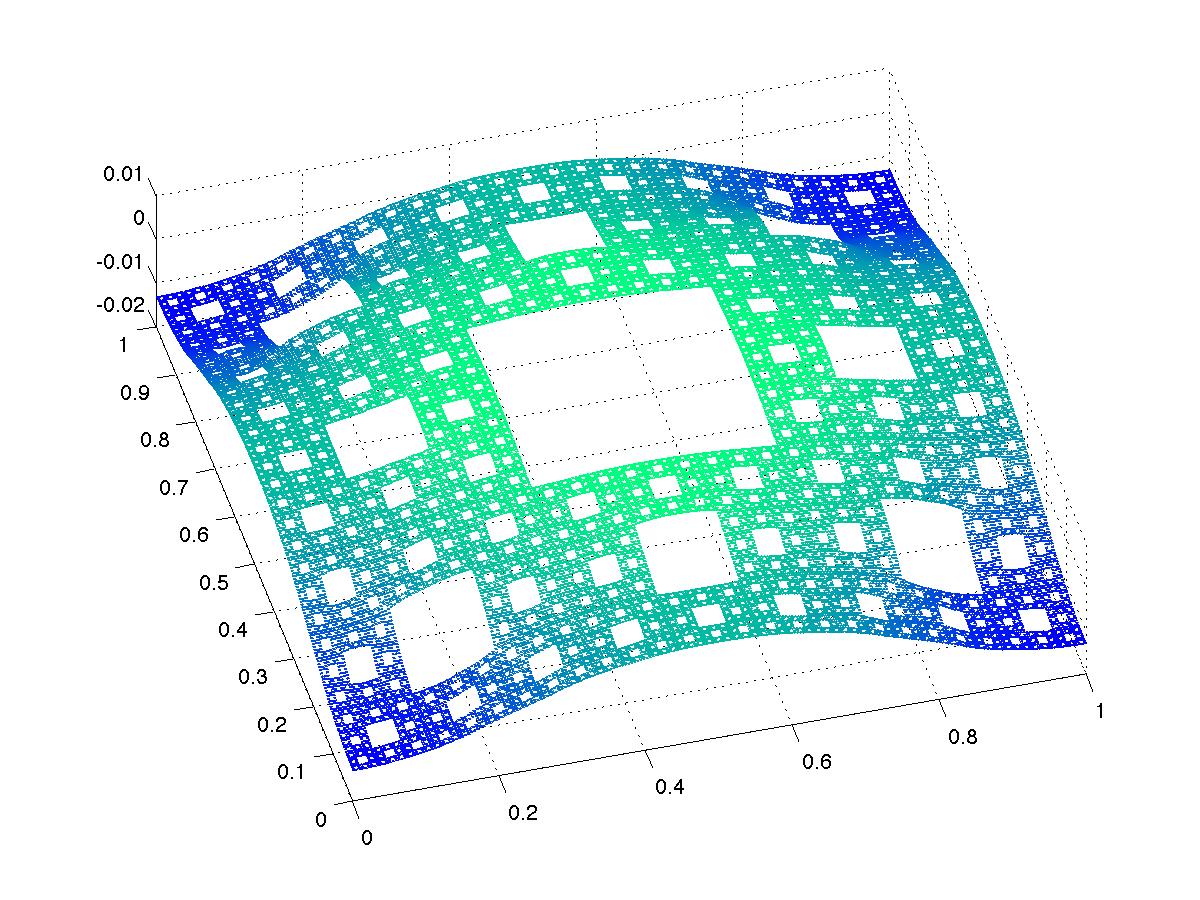}
\includegraphics[scale=.08]{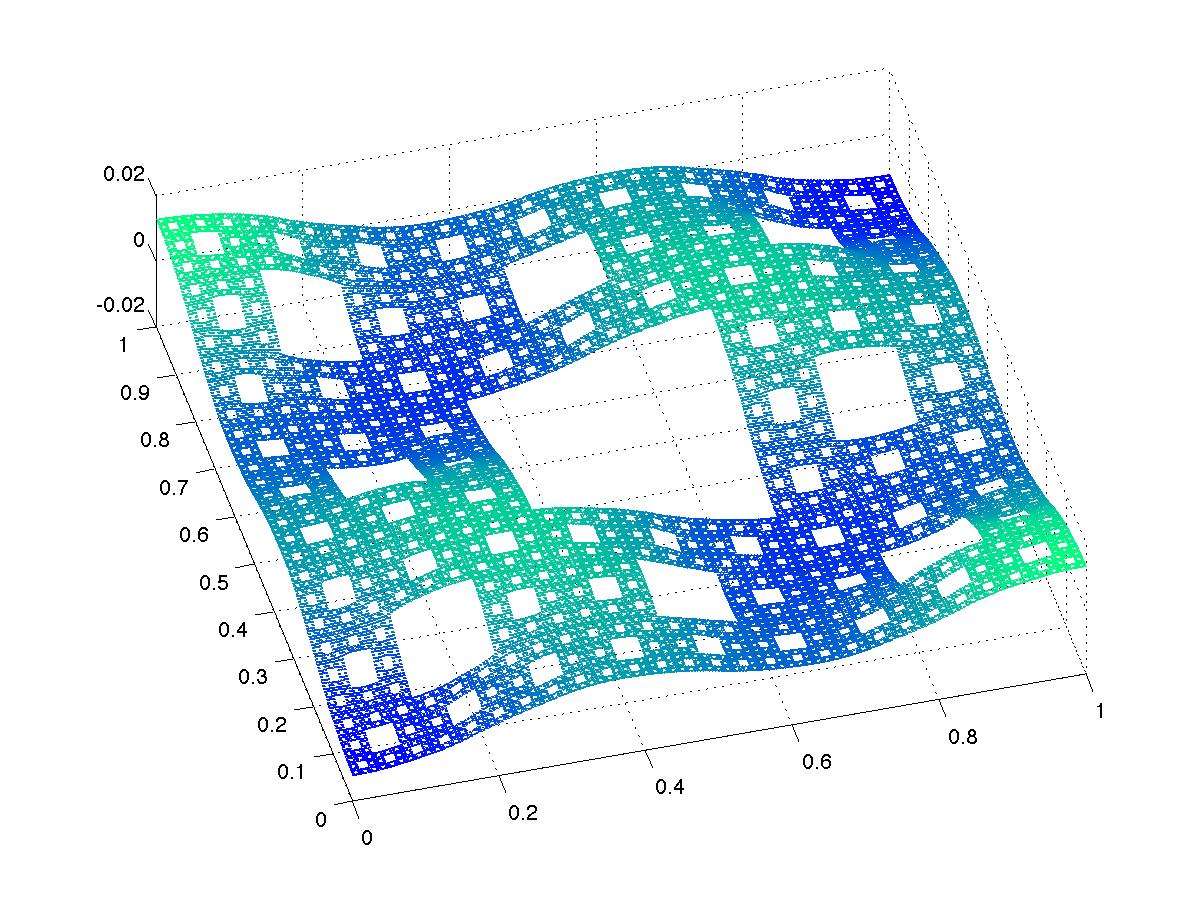}\\
\includegraphics[scale=.08]{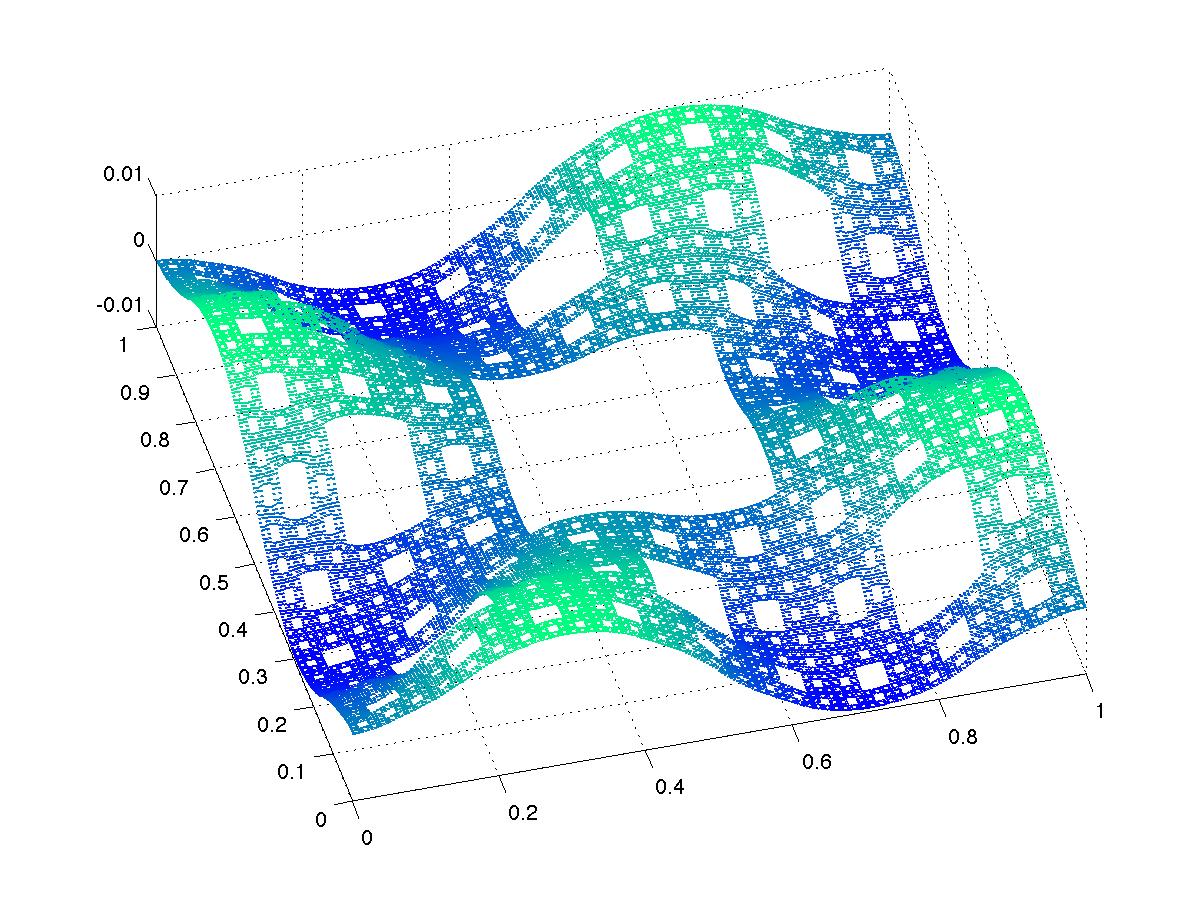}
\includegraphics[scale=.08]{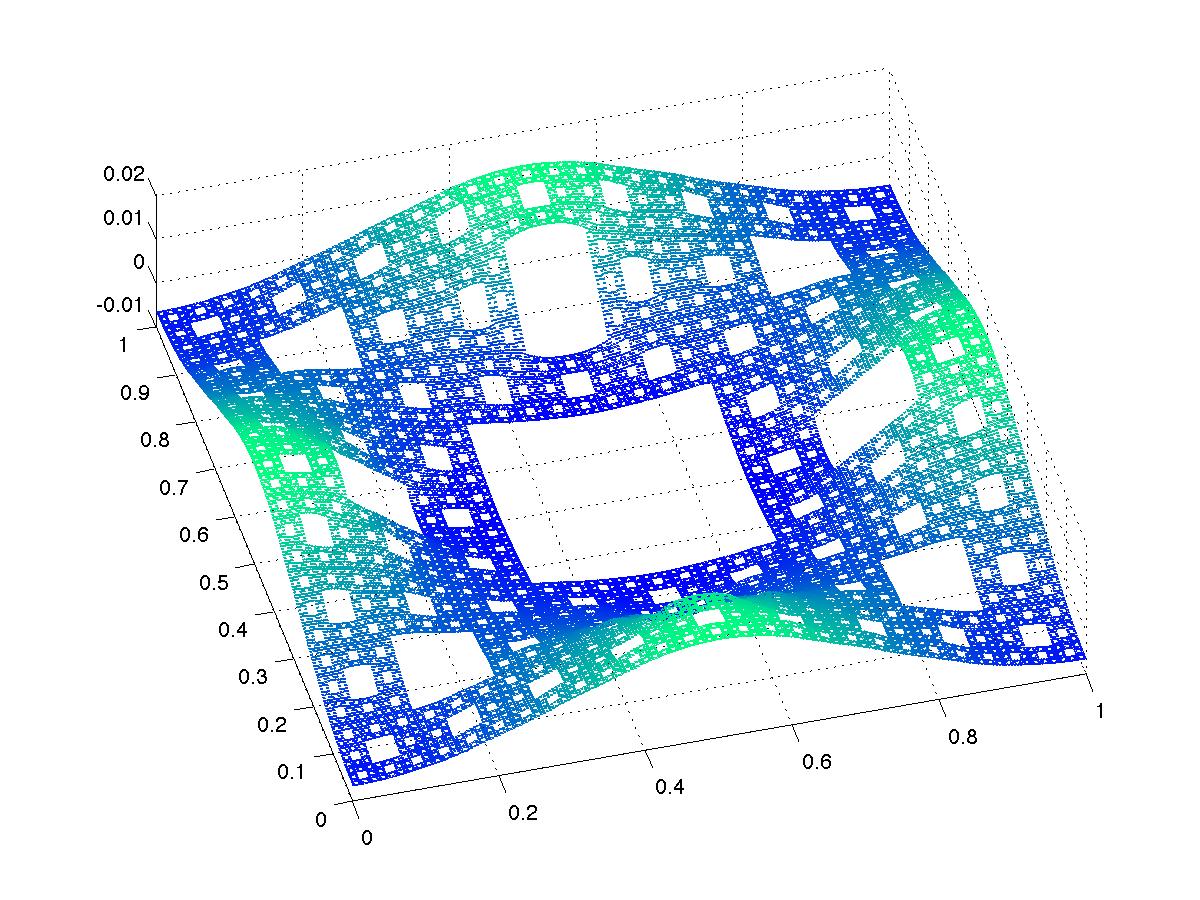}
\includegraphics[scale=.08]{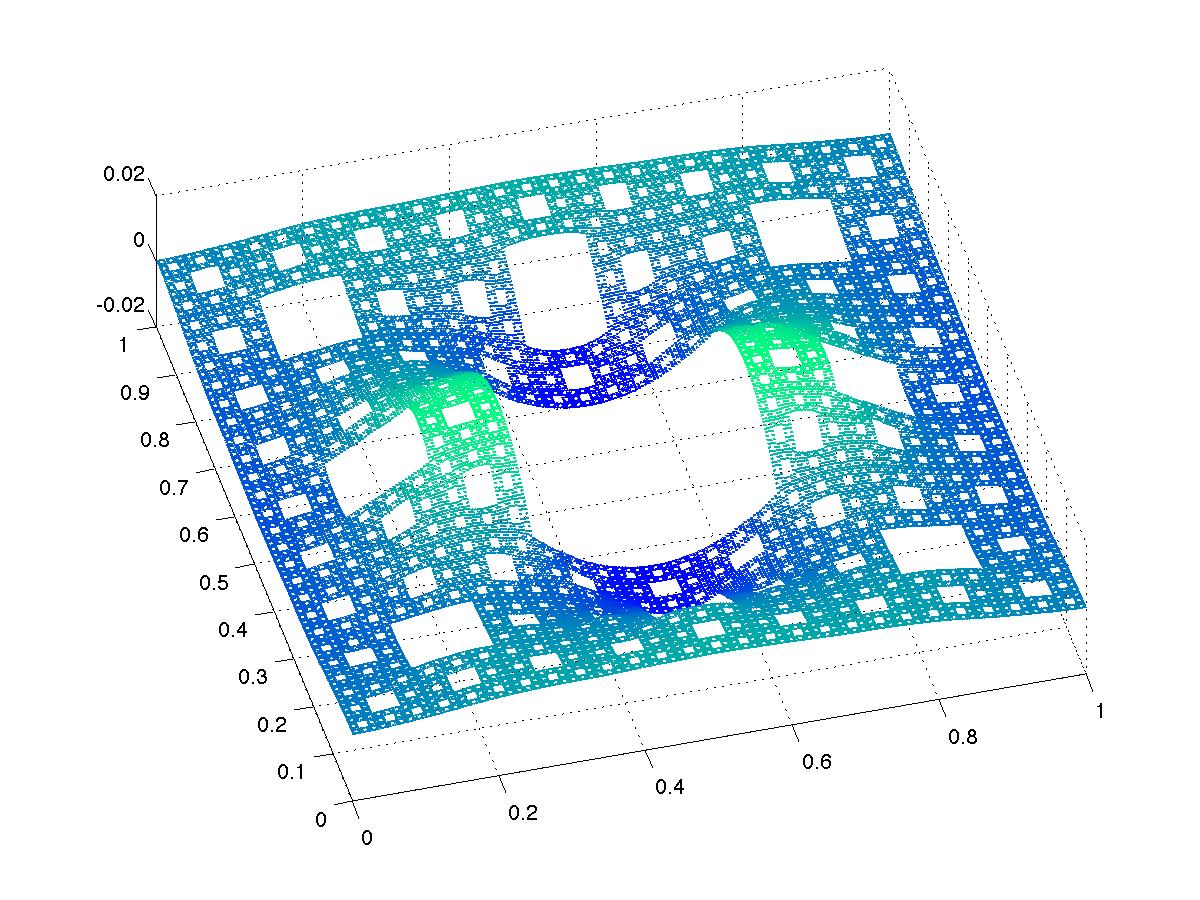}
\end{center}
\caption{The first nine nonconstant eigenfunctions on SC$_5$ with projective plane boundary conditions}\label{figprojeigs}
\end{figure}

We have also examined eigenfunctions and eigenvalues for periodic type boundary conditions corresponding to the construction of the torus, Klein bottle, and projective space from the square.  SC with these identifications may be regarded as \emph{fractafolds} based on SC in the spirit of \cite{S1}.  Note that in the case of the projective space we create two singularities at the corner points of the square, identified in pairs.
We show some of the results in Figures \ref{figtoruseigs}-\ref{figprojeigs} and Tables \ref{tabtoruseig}-\ref{tabprojeig}.

\begin{table}[p]
\begin{center}
\begin{tabular}{|r||  c| c| c| } \hline
$\lambda_j$ & Eigenvalue& level $m=4$ & level $m=5$  \\ \hline \hline
1&0	&2.10360108348969e-15   &   -2.14570386198176e-17     \\  \hline
2& 32.2114556413112	&0.00320568102788810   &   0.000320348757759280     \\  \hline
3&	32.2117876737192&0.00320568102788826   &   0.000320352059881687     \\  \hline
4&33.0335325255338	&0.00328561228543238   &   0.000328524461197705     \\  \hline
5&46.1378042666663	&0.00458982380491410   &   0.000458848815997367      \\  \hline
6&67.3159845178793	&0.00670022323157356   &   0.000669469652591202      \\  \hline
7&70.3055712815539	&0.00699438207618352   &   0.000699201634176303     \\  \hline
8&70.3059096613708	&0.00699438207618378   &   0.000699204999424840     \\  \hline
9&91.1670279725552	&0.00906849340248511   &   0.000906672597625728     \\  \hline
10&112.219464252958	&0.0111377385936996   &   0.00111604288766578      \\  \hline
11&153.725854579008	&0.0152688242028740   &   0.00152883145357489     \\  \hline
12&153.726931440876	&0.0152688242028747   &   0.00152884216316112     \\  \hline
13&159.788304787400	&0.0158639440470136   &   0.00158912361841407     \\  \hline
14&164.837696052388	&0.0163805216075794   &   0.00163934072866179     \\  \hline
15&182.394416891886	&0.0181239623631475   &   0.00181394549579459      \\  \hline
\end{tabular}
\caption{The first 15 eigenvalues for levels $m=4$ and $5$ with torus boundary conditions} \label{tabtoruseig}
\end{center}
\end{table}

\begin{table}[p]
\begin{center}
\begin{tabular}{|r|| c|  c| c| } \hline
$\lambda_j$ & Eigenvalue &level $m=4$ & level $m=5$  \\ \hline \hline
1&	0	&-3.02058725254462e-14   &   -2.19737287176415e-17      \\  \hline
2&	32.2117498844683	&0.00320568102788416   &   0.000320351684060745      \\  \hline
3&	33.032810672492	&0.00328561228541770   &   0.000328517282238523      \\  \hline
4&	39.8922202086433	&0.00397107095374325   &   0.000396735351869938      \\  \hline
5&	43.2714800226914	&0.00430597895848081   &   0.000430342702485531      \\  \hline
6&	46.1377316931061	&0.00458982380490713   &   0.000458848094240172      \\  \hline
7&	70.3057087137930	&0.00699438207617901   &   0.000699203000964800      \\  \hline
8&	91.1664483134311	&0.00906849340247104   &   0.000906666832810801      \\  \hline
9&	103.298679176177	&0.0102746017084947   &   0.00102732406510132      \\  \hline
10&	112.219004510499	&0.0111377385936637   &   0.00111603831544379      \\  \hline
11&	122.693174548126	&0.0121921338975400   &   0.00122020583266118      \\  \hline
12&	153.726542595912	&0.0152688242028903   &   0.00152883829602755      \\  \hline
13&	162.162017612389	&0.0161071330316981   &   0.00161273062218409      \\  \hline
14&	164.841195409480	&0.0163805216076038   &   0.00163937553040158      \\  \hline
15&	182.388331131098	&0.0181239623631782   &   0.00181388497180182      \\  \hline
\end{tabular}
\caption{The first 15 eigenvalues for levels $m=4$ and $5$ with Klein bottle boundary conditions} \label{tabkleineig}
\end{center}
\end{table}

\begin{table}[p]
\begin{center}
\begin{tabular}{|r||  c|  c| c| } \hline
$\lambda_j$ & Eigenvalue & level $m=4$ & level $m=5$  \\ \hline \hline
1&	0	&2.15702836851335e-15   &   -2.13740293909888e-17      \\  \hline
2&	17.7803169580100	&0.00177037771649474   &   0.000176828471010163      \\  \hline
3&	33.0329798073433	&0.00328561228543052   &   0.000328518964315239      \\  \hline
4&	39.8922635742169	&0.00397107095374375   &   0.000396735783148419      \\  \hline
5&	39.8924984277846	&0.00397107095374659   &   0.000396738118809666      \\  \hline
6&	46.1377416688106	&0.00458982380491383   &   0.000458848193450360      \\  \hline
7&	72.9450515387370	&0.00724990726688279   &   0.000725451743172754      \\  \hline
8&	76.4963443558820	&0.00760210249889050   &   0.000760769993148170      \\  \hline
9&	91.1663022922193	&0.00906849340248375   &   0.000906665380603414      \\  \hline
10&	112.220120732730	&0.0111377385937001   &   0.00111604941647595      \\  \hline
11&	122.691748138782	&0.0121921338975364   &   0.00122019164676202      \\  \hline
12&	122.692643256546	&0.0121921338975388   &   0.00122020054887023      \\  \hline
13&	164.843452415657	&0.0163805216075811   &   0.00163939797673661      \\  \hline
14&	178.014270428489	&0.0177288779201235   &   0.00177038414625554      \\  \hline
15&	182.393894269051	&0.0181239623631491   &   0.00181394029821588      \\  \hline
\end{tabular}
\caption{The first 15 eigenvalues for levels $m=4$ and $5$ with projective plane boundary conditions} \label{tabprojeig}
\end{center}
\end{table}

There are some interesting coincidences in the Neumann spectrum and these spectra.  For simplicity we denote the four cases by N, T, KB, and PS.  The full $D_4$ symmetry group acts on N, T, and PS, but not KB, and indeed there appear to be no eigenspaces of dimension greater than one for KB.  The one-dimensional eigenspaces for N and PS are identical.  Exactly half the one-dimensional eigenspaces for N appear in the T and KB spectra, namely those that are symmetric under horizontal and vertical reflections (labels $1++$ and $1-+$ in Table \ref{tabneig}).  The two-dimensional eigenspaces for T and PS contain a one-dimensional subspace that contributes to the KB spectrum.  (Note that the T and KB spectra also contain other one-dimensional eigenspaces unrelated to the other spectra.)

All these coincidences are easily explained by symmetry considerations.  In all the examples we may extend functions to be periodic of period 1 in $x$ and $y$.  In addition we have the following identities:

\begin{itemize}
\item Case N:
\begin{description}
\item[(N1)] $u(-x,y)=u(x,y)$
\item[(N2)]$u(x,-y)=u(x,y)$
\end{description}
\item Case T:
\begin{description}
\item[(T1)] $u(x+1,y)=u(x,y)$
\item[(T2)]$u(x,y+1)=u(x,y)$
\end{description}
\item Case KB:
\begin{description}
\item[(KB1)]$u(x+1,1-y)=u(x,y)$
\item[(KB2)]$u(x,y+1)=u(x,y)$
\end{description}
\item Case PS:
\begin{description}
\item[(PS1)] $u(x+1,1-y)=u(x,y)$
\item[(PS2)] $u(1-x,y+1)=u(x,y)$
\end{description}
\end{itemize}
In addition, if $u$ is symmetric with respect to horizontal and vertical reflection then 
\begin{equation}\label{4.10}
u(1-x,y)=u(x,y)\quad\text{ and }\quad u(x,1-y)=u(x,y).
\end{equation}
It is straightforward to show that under \eqref{4.10} all four sets of conditions are equivalent.  This explains why the $(1++)$ and $(1-+)$ N-eigenfunctions are also eigenfunctions on T, KB, and PS.  On the other hand, if $u$ is skew-symmetric with respect to horizontal and vertical reflections then in place of \eqref{4.10} we have
\begin{equation}\label{4.11}
-u(1-x,y)=u(x,y)\quad\text{ and }\quad -u(x,1-y)=u(x,y).
\end{equation}
Under this assumption we have
\begin{equation*}
u(x+1,1-y)=-u(-x,1-y)=u(-x,y)
\end{equation*}
so {\bf(N1)} is equivalent to {\bf (PS1)}, and similarly {\bf(N2)} is equivalent to {\bf (PS2)}.  Thus the type $(1+-)$ and $(1--)$ eigenfunctions in the N and PS examples are the same.

Next consider a two-dimensional eigenspace for example T.  In this eigenspace we can find an eigenfunction symmetric with respect to vertical reflections, $u(x,1-y)=u(x,y)$.  This makes {\bf (T1)} and {\bf (KB1)} equivalent, and of course {\bf (T2)} and {\bf (KB2)} are identical equations, so this eigenfunction is also an eigenfunction for KB.  Similarly, in a two-dimensional eigenspace for PS there is an eigenfunction that is symmetric with respect to horizontal reflection, $u(1-x,y)=u(x,y)$.  This makes {\bf(PS2)} and {\bf (KB2)} equivalent, and {\bf(PS1)} and {\bf (KB1)} are identical equations, so this eigenfunction is also an eigenfunction for KB.

Of course the identical remarks are valid in the non-fractal case of the square with Neumann boundary conditions, the torus, the Klein bottle, and the projective space, where they are presumably well-known, and can also be explained by the expressions for the eigenfunctions in terms of sines and cosines.

\begin{figure}[p]
\begin{center}
\includegraphics[scale=.22]{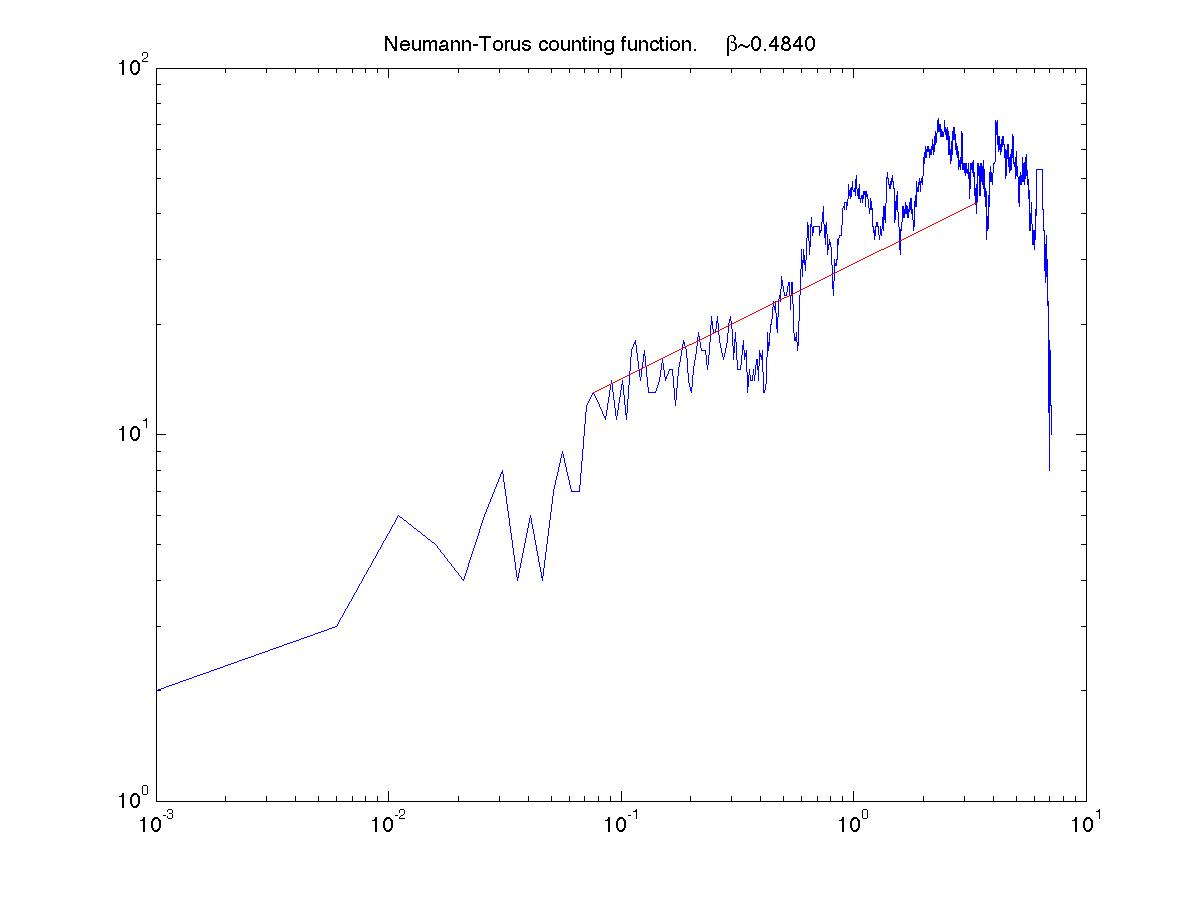}
\caption{The difference between the eigenvalue counting functions for Neumann and torus boundary conditions.  We estimate the exponential growth rate (shown with a red line) and obtain $\beta \approx 0.4769$.}  \label{fig4.13}
\end{center}
\end{figure}

\begin{figure}[p]
\begin{center}
\includegraphics[scale=.22]{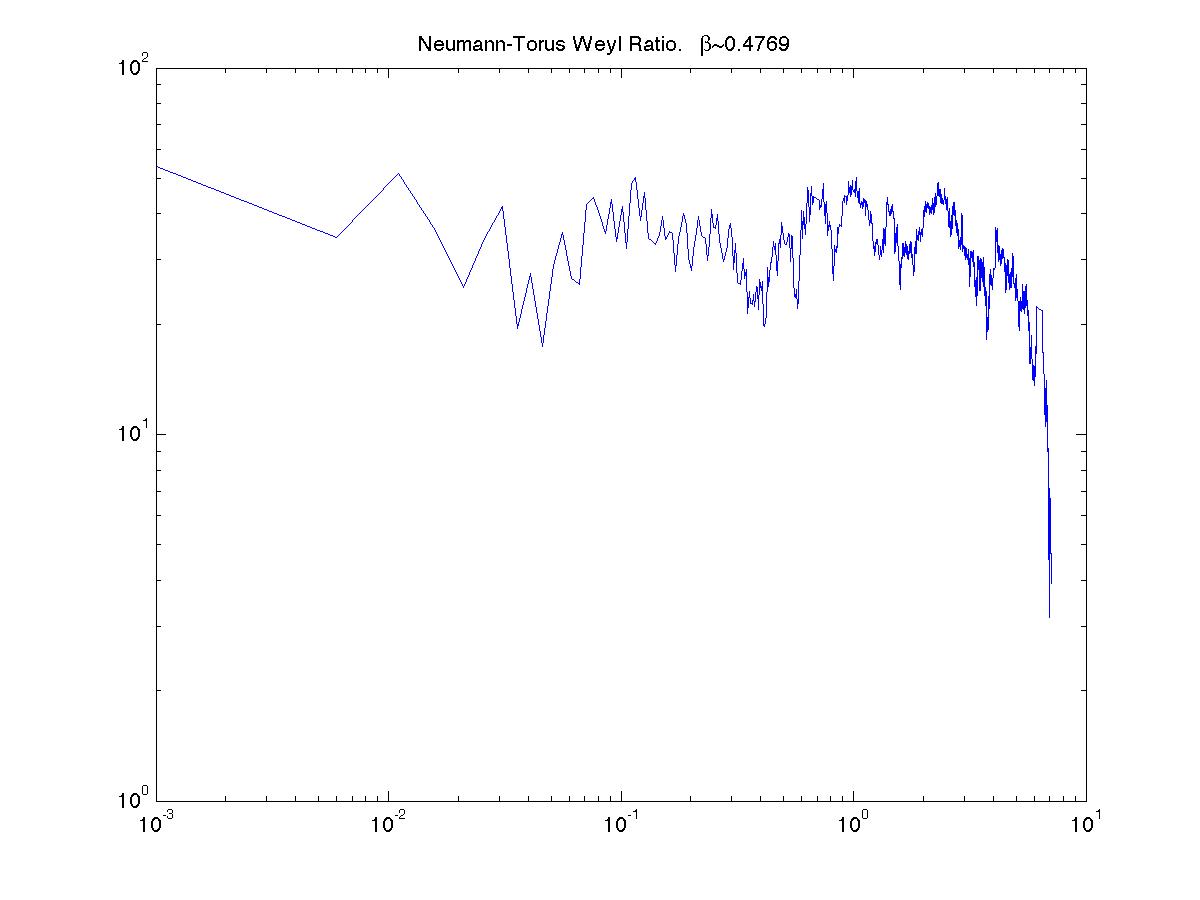}
\caption{The Weyl ratio for the difference between the eigenvalue counting functions for Neumann and torus boundary conditions with $\beta \approx 0.4769$.}  \label{fig4.14}
\end{center}
\end{figure}

\begin{figure}[p]
\begin{center}
\includegraphics[scale=.25]{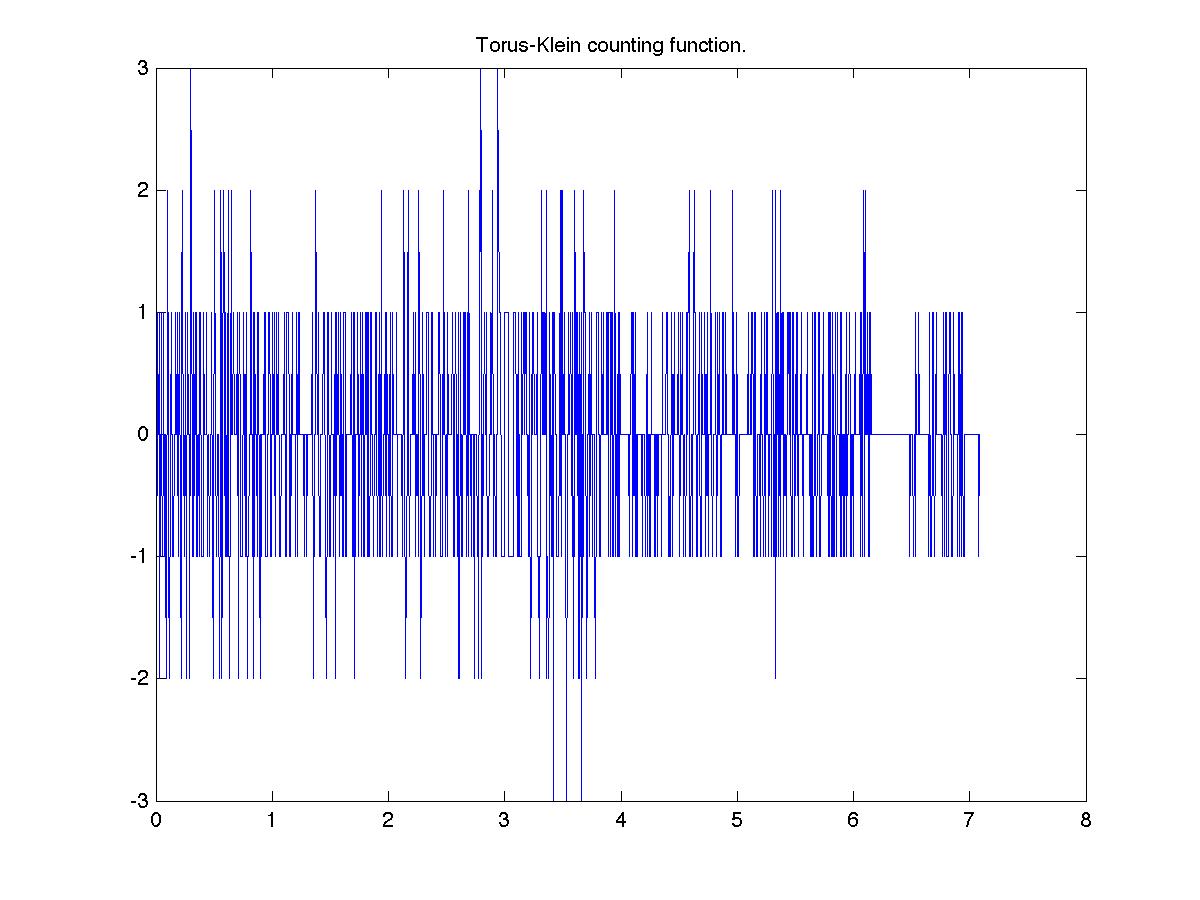}\\
\caption{The difference between the eigenvalue counting functions for torus and Klein bottle boundary conditions.}\label{fig4.15}
\end{center}
\end{figure}

\begin{figure}[p]
\begin{center}
\includegraphics[scale=.25]{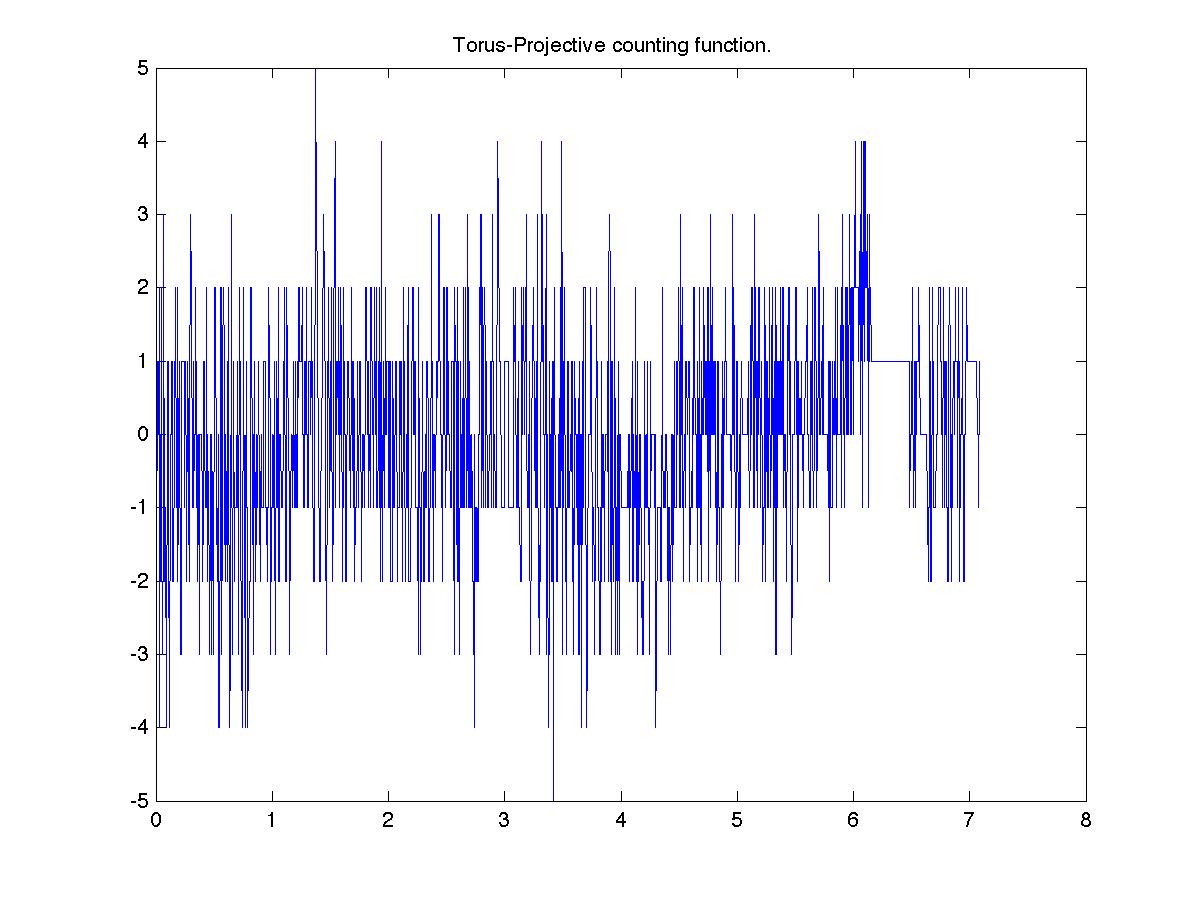}\\
\end{center}
\caption{The difference between the eigenvalue counting functions for torus and projective space boundary conditions.}\label{fig4.16}
\end{figure}

Another interesting observation is the behavior of the differences between the eigenvalue counting function for these examples.  In Figure \ref{fig4.13} we show the difference between the eigenvalue counting functions for Neumann boundary conditions and the torus, showing a growth rate of $t^\beta$ with $\beta\approx 0.4769$, and in Figure \ref{fig4.14} we show the Weyl ratio.  These results are consistent with the results for $N^{(N)}(t)-N^{(D)}(t)$ shown in Figures \ref{figNn-Nd} and \ref{figNn-NdWeyl}(similar results, not shown here, are obtained for the Klein bottle and projective space).  In Figures \ref{fig4.15} and \ref{fig4.16} we show the differences between the eigenvalue counting functions for torus and Klein bottle (respectively projective space).  Here we see a dramatically different behavior, leading us to conjecture that the difference is uniformly bounded, (Note that we do not have enough data to distinguish between boundedness and $\log t$ growth).  Moreover there is rapid oscillation between positive and negative values.

We note that for the standard Laplacian on the torus and Klein bottle, the difference of the eigenvalue counting functions is expected to be on the order of $t^{1/4}$.

\section{Heat kenrel}

\begin{figure}[p]
\begin{center}
\includegraphics[scale=.5]{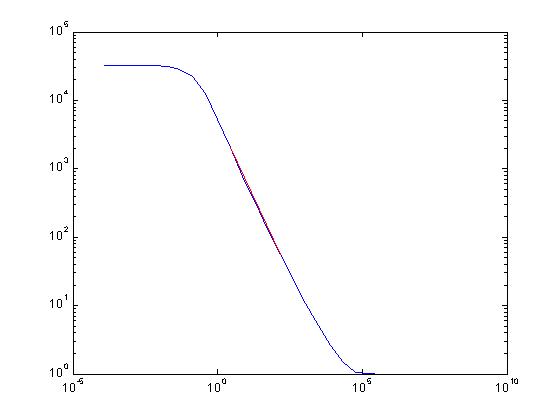}
\end{center}
\caption{Trace of the heat kernel $Z(t)$ on a log-log scale using all level-5 Neumann eigenvalues.  The slope for small $t$ is -0.895 (shown in red)}\label{figNtraceheat}
\end{figure}

\begin{figure}[p]
\begin{center}
\includegraphics[scale=.5]{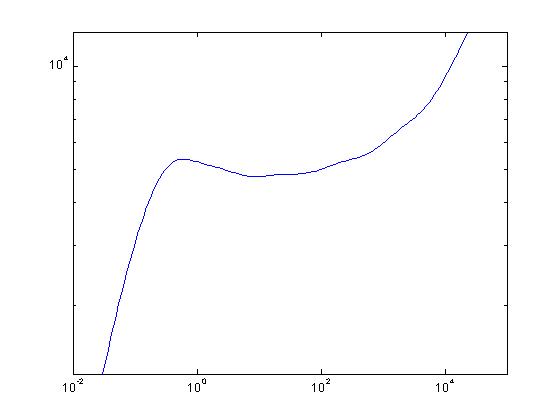}
\end{center}
\caption{$t^\alpha Z(t)$ on a log-log scale using all level-5 Neumann eigenvalues.}\label{figNtracetimesta}
\end{figure}

\begin{figure}[p]
\begin{center}
\subfloat[ ]{\includegraphics[scale=.35]{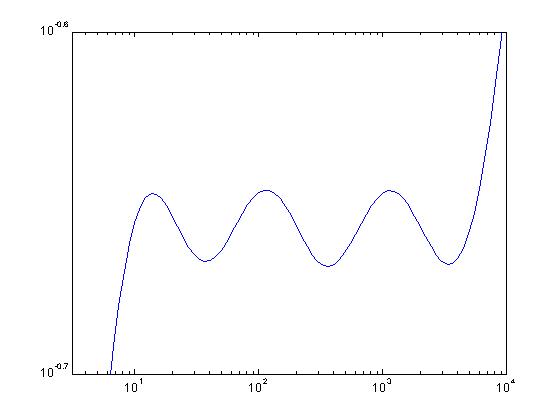}}
\subfloat[]{\includegraphics[scale=.35]{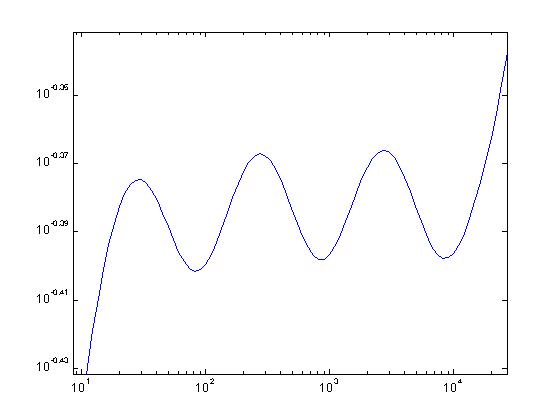}}\\
\subfloat[]{\includegraphics[scale=.35]{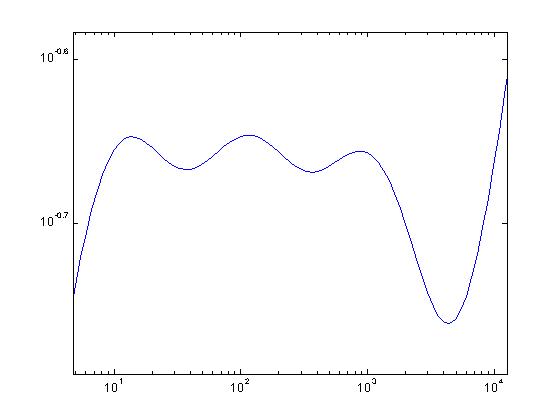}}
\subfloat[]{\includegraphics[scale=.35]{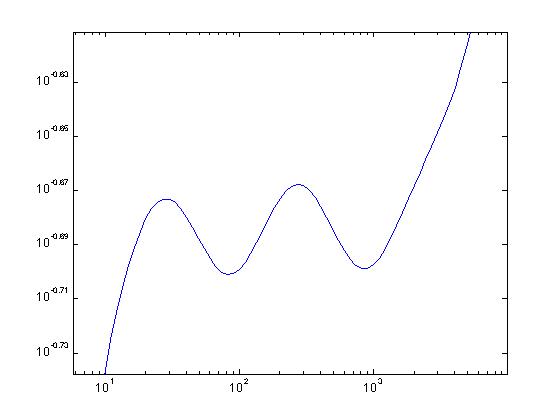}}\\
\subfloat[]{\includegraphics[scale=.35]{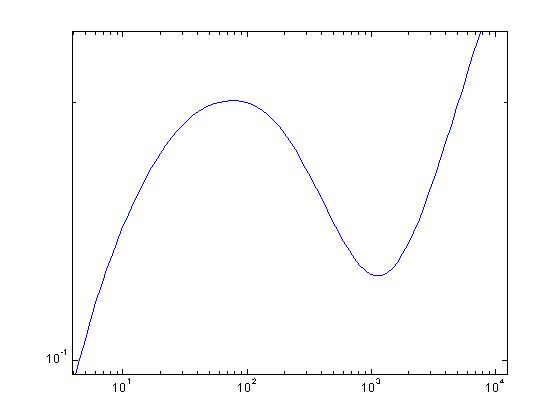}}
\end{center}
\caption{$t^\alpha h(x,x)$ on a log-log scale using the first 1500 level-5 Neumann eigenvalues and eigenfunctions.  In (a) $x=`11111'$, the top-center cell; (b) $x=`00000'$, the top-left corner cell; (c) $x=15555$, the center cell immediately above largest hole of SC; (d) $x=`10000'$, the level-1 junction point along the top edge of SC; (e) $x=`05271'$, an arbitrary interior cell.}\label{figNheattimesta}
\end{figure}

The heat kernel $h_t(x,y)$ (for either Dirichlet or Neumann boundary conditions) is defined to be 
\begin{equation}\label{5.1}
h_t(x,y)=\Sum_{j=1}^\infty e^{-\lambda_j t} \varphi_j(x)\varphi_j(y)
\end{equation}
where $\{\varphi_j\}$ is an orthonormal basis of eigenfunctions with corresponding eigenvalues $\{\lambda_j\}$.  By integrating against the heat kernel
\begin{equation}\label{5.2}
u(x,t)=\int_{\text{SC}}h_t(x,y)f(y)\,d\mu(y)
\end{equation}
you can solve the heat equation
\begin{equation}\label{5.3}
 \left\{ \begin{array}{l}
 \Delta_x u=\frac{\partial}{\partial t} u \\
  u(x,0)=f(x)
       \end{array} \right.
\end{equation}
with the appropriate boundary conditions.  Moreover, $h_t(x,y)$ gives the transition probabilities of a Brownian motion path moving from $x$ to $y$ in time $t$.  Using the probabilistic interpretation, Barlow and Bass \cite{BB} were able to prove subGaussian bounds (above and below) for the heat kernel for small times,
\begin{equation}\label{5.4}
h_t(x,y)\approx t^{-\alpha} exp\left( -c \frac{R(x,y)^\gamma }{t^\delta} \right) \quad 0<t\leq 1,
\end{equation}
where $\alpha = \frac{\log 8}{\log \rho}$ as before, and $\gamma$ and $\delta$ are other constants.  But this is not the end of the story.  We would like to use our numerical data to shed some light on the more precise behavior of the heat kernel.  In particular, the trace of the heat kernel (also called the \emph{partition function})
\begin{equation}\label{5.5}
Z(t)=\Sum_{j=1}^\infty e^{-\lambda_j t}. 
\end{equation}
Figure \ref{figNtraceheat} shows the trace of the heat kernel with Neumann eigenvalues.  Note that \eqref{5.4} implies that $Z(t)\approx t^{-\alpha}$ as $t\to 0$.  We obtain a slope of $-\alpha\approx-0.895$ which is reasonably close to our predicted $\alpha =  \frac{\log 8}{\log \rho}\approx0.9026$.  But again more is known, $t^\alpha Z(t)$ converges to a multiplicatively periodic function bounded and bounded away from zero of period $\rho$ as $t\to 0$.  In analogy with the case of SG we expect that $t^\alpha Z(t)$ as a function of $\log t$ to resemble a sine curve with very small error.  Figure \ref{figNtracetimesta} shows $t^\alpha Z(t)$.  We might also expect $t^\alpha h_t(x,x)$ for any point $x$ to have a similar behavior, but the sine curve should show a phase shift that depends on $x$.  Figure \ref{figNheattimesta} shows $t^\alpha h_t(x,x)$ for four choices of $x$.


\begin{figure}[p]
\begin{center}
\includegraphics[scale=.5]{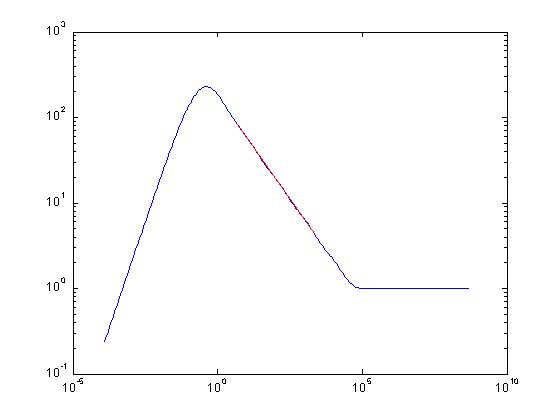}
\end{center}
\caption{Difference of Neumann and Dirichlet traces of the heat kernel $Z^{(N)}(t)-Z^{(D)}(t)$ on a log-log scale using all level-5 eigenvalues.  The slope for small $t$ is -0.4797 (shown in red).  The apparent change of behavior for very small values of $t$ is a simple consequence of the fact that we have not computed enough eigenvalues.}\label{figZN-ZD}
\end{figure}

\begin{figure}[p]
\begin{center}
\includegraphics[scale=.63]{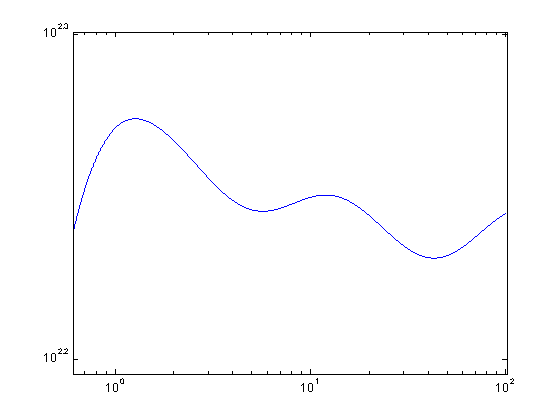}
\end{center}
\caption{${t^\beta}(Z^{(N)}(t)-Z^{(D)}(t))$ on a log-log scale using all level-5 eigenvalues.}\label{figZN-ZDtimestb}
\end{figure}

Just as we studied the counting function and Weyl ratios of the difference between Neumann and Dirichlet boundary conditions, we do the same for the heat kernel.   Again we expect $Z^{(N)}(t)-Z^{(D)}(t)\sim t^{-\beta}$ where $\beta=\log 3/\log \rho \approx 0.4769$.  Figure \ref{figZN-ZD} shows the graph of the difference of the heat kernels and has an approximate slope -0.4797 which is satisfyingly close to $-\beta$.  If we examine $t^\beta(Z^{(N)}(t)-Z^{(D)}(t))$ we see a similar periodic behavior as shown in Figure \ref{figZN-ZDtimestb}.

\begin{figure}[p]
\begin{center}
\includegraphics[scale=.4]{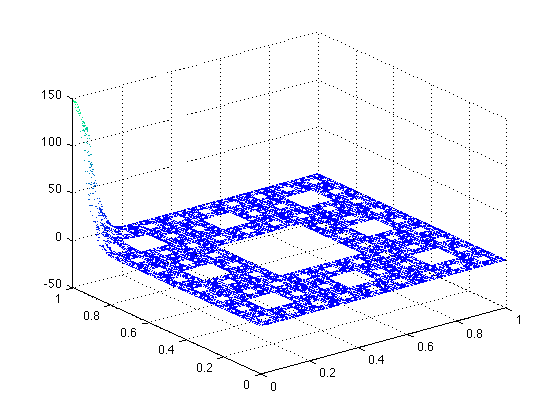}\includegraphics[scale=.4]{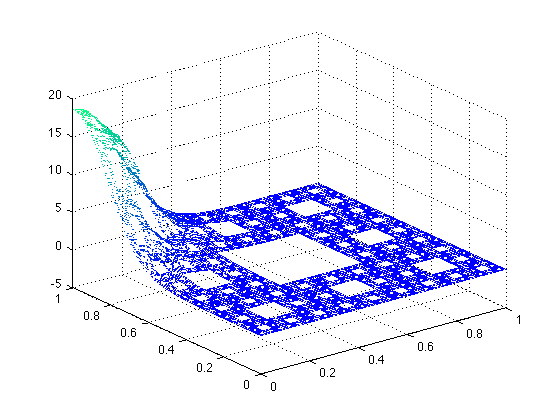}\\
\includegraphics[scale=.4]{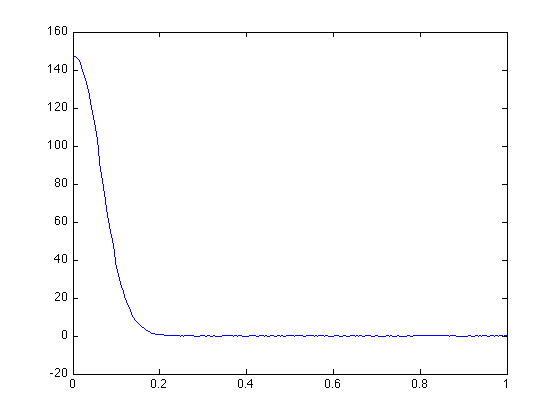}\includegraphics[scale=.4]{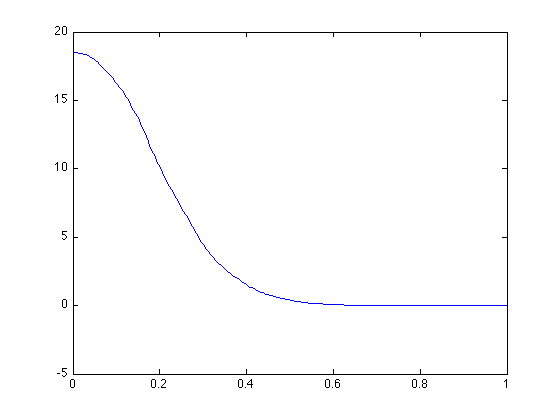}\\
\includegraphics[scale=.4]{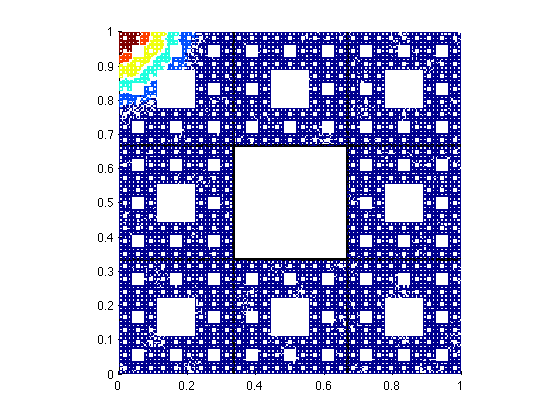}\includegraphics[scale=.4]{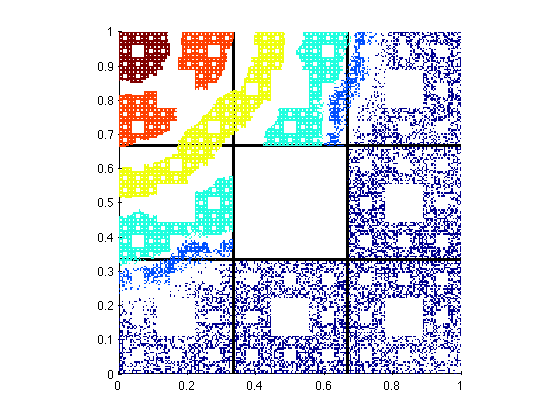}
\end{center}
\caption{Graphs of the off-diagonal heat kernel $h_t(\cdot,y)$ for $y$ fixed as the corner point `00000' with $t=1/100$ (left) and $t=1/10$ (right).  The first row shows $h_t(\cdot, y)$ in its entirety.  The second row shows the restriction of $h_t(\cdot,y)$ to the horizontal line containing $y$.  The third row shows the kernel with level sets colored differently.}\label{figheatoffdiagfirst}
\end{figure}

Figures \ref{figheatoffdiagfirst}-\ref{figheatoffdiaglast} shows graphs of the off-diagonal heat kernel $h_t(\cdot,y)$ for a few choices of fixed $y$.  Note in particular that the level sets in Figure \ref{figheatoffdiaglast} resemble the level sets of the resistance metric in Figure \ref{figresistance1111rings}, even though the functions are so different.

\begin{figure}[p]
\begin{center}
\includegraphics[scale=.4]{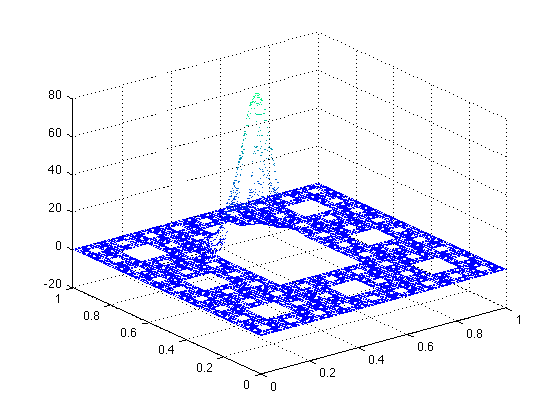}\includegraphics[scale=.4]{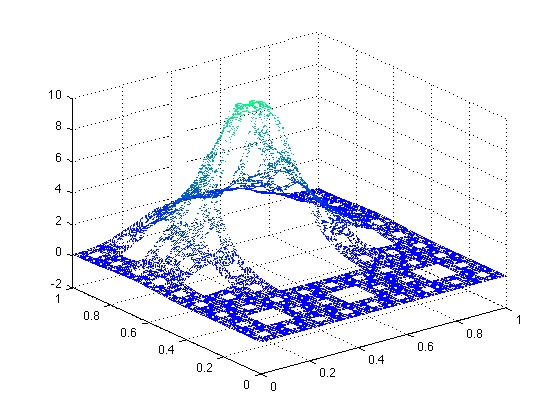}\\
\includegraphics[scale=.4]{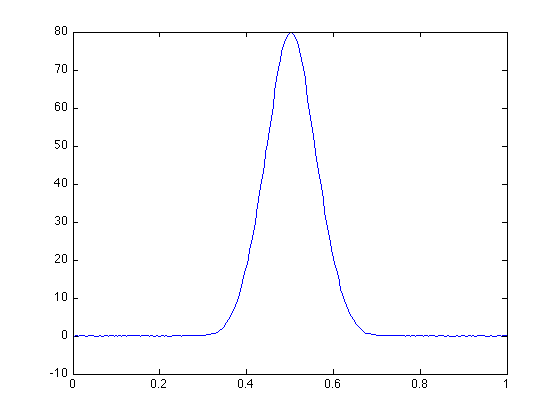}\includegraphics[scale=.4]{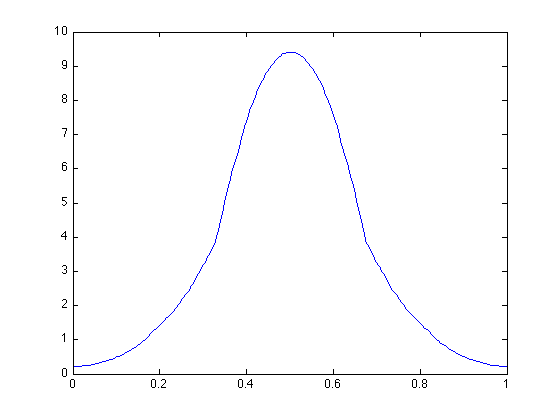}\\
\includegraphics[scale=.4]{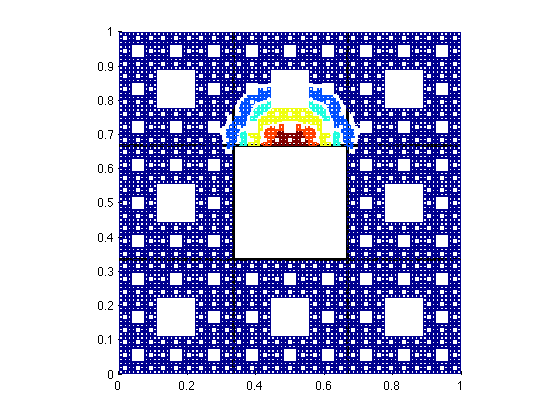}\includegraphics[scale=.4]{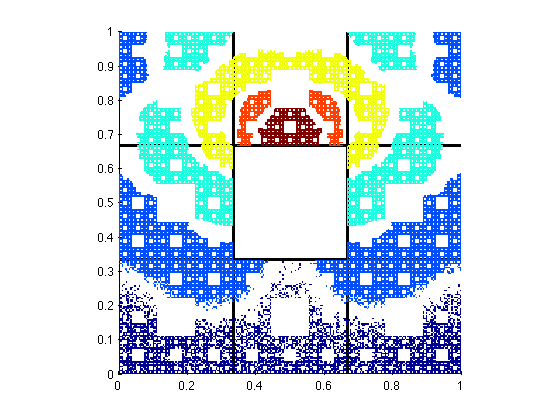}
\end{center}
\caption{Graphs of the off-diagonal heat kernel $h_t(\cdot,y)$ for $y$ fixed as the center point `15555' and $t=1/100$ (left) and $t=1/10$ (right).  The first row shows $h_t(\cdot, y)$ in its entirety.  The second row shows the restriction of $h_t(\cdot,y)$ to the horizontal line containing $y$.  The third row shows the kernel with level sets colored differently.}\label{figheatoffdiag}
\end{figure}

\begin{figure}[p]
\begin{center}
\includegraphics[scale=.4]{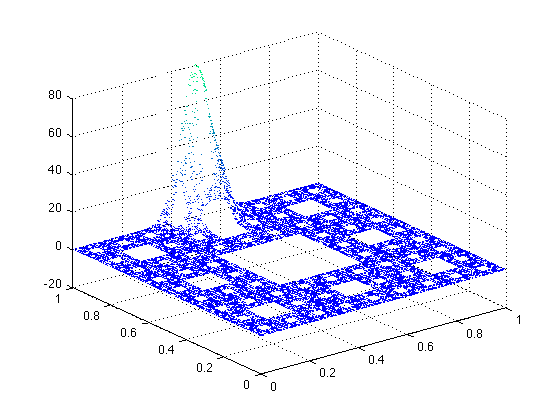}\includegraphics[scale=.4]{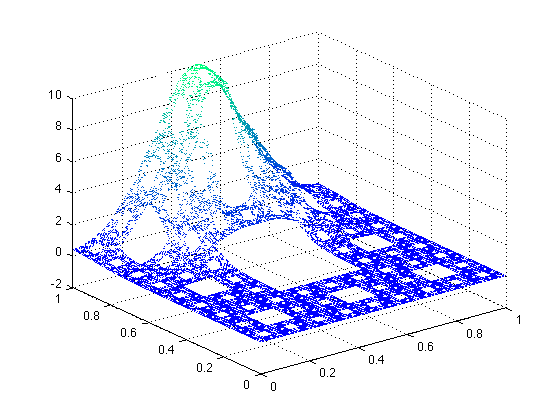}\\
\includegraphics[scale=.4]{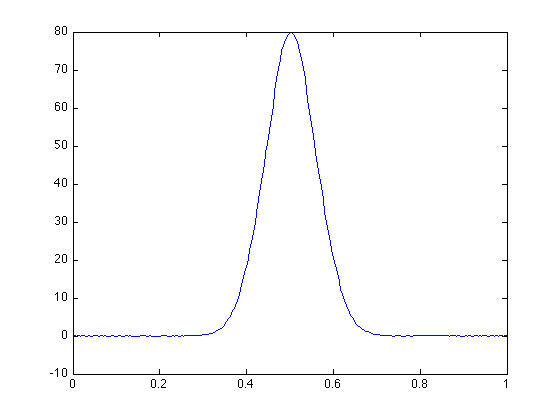}\includegraphics[scale=.4]{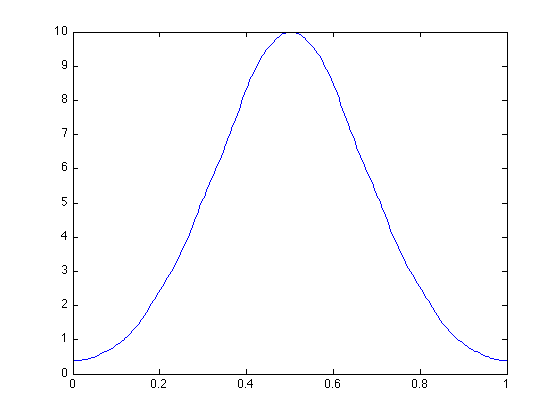}\\
\includegraphics[scale=.4]{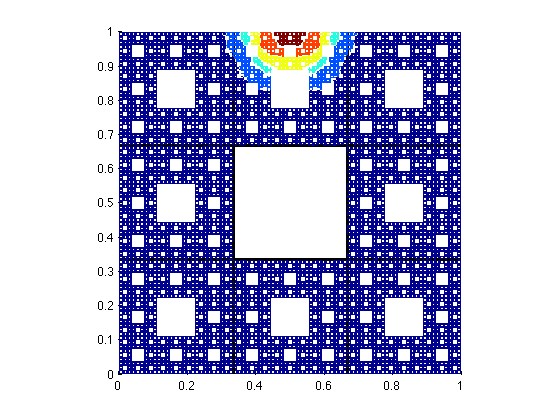}\includegraphics[scale=.4]{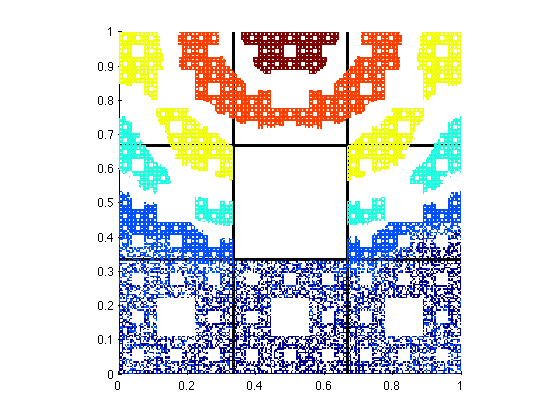}
\end{center}
\caption{Graphs of the off-diagonal heat kernel $h_t(\cdot,y)$ for $y$ fixed as the top point `11111' and $t=1/100$(left) and $t=1/10$ (right).  The first row shows $h_t(\cdot, y)$ in its entirety.  The second row shows the restriction of $h_t(\cdot,y)$ to the horizontal line containing $y$.  The third row shows the kernel with level sets colored differently.}\label{figheatoffdiag}
\end{figure}

\section{Boundary Behavior}

It is easy to obtain the analog of the Gauss-Green formula for functions $u$ satisfying Neumann boundary conditions.  We simply write
\begin{eqnarray}\label{Bob6.1}
\mathcal{E}_m(u,v)&=& \frac{1}{\rho^m}\Sum_x v(x) \Sum_{y\sim x}\left( u(x)-u(y)\right) \\
&=& - \Sum_x 8^{-m} v(x) \Sum_{y\sim x} \left(\frac{8}{\rho}\right)^m \left( u(y)-u(x)\right) \notag \\
&=& - \Sum_x 8^{-m} v(x) \Delta_m u(x). \notag
\end{eqnarray}
If $u$ satisfies Neumann boundary conditions then we obtain
\begin{equation}\label{Bob6.2}
\mathcal{E}(u,v)=-\int_{\text{SC}} v\Delta u\, d\mu
\end{equation}in the limit as $m\to \infty$.  
If $u$ does not satisfy Neumann boundary conditions then we do not have the correct expression for $\Delta_m u(x)$ in \eqref{Bob6.1} when $x$ is a boundary cell, because we need to take into account the values when $y=x^*$, the virtual neighbor (two of these at corner points). 

Thus we have 
\begin{equation}
\mathcal{E}_m(u,v)= -\Sum_x 8^{-m} v(x) \Delta_m u(x) + \Sum_{x\in \partial \text{ SC}_m} v(x) \rho^{-m} \left( u(x)-u(x^*)\right).\tag{6.1'}
\end{equation}

In the limit we hope to obtain
\begin{equation}
\mathcal{E}(u,v)=-\int_{\text{SC}} v\Delta u\, d\mu + \int_{\partial\text{ SC}} v \partial_n u \, d\nu  \tag{6.2'}
\end{equation}
for some appropriate measure on the boundary, and some appropriate definition of the normal derivative $\partial_n u$.  Since the boundary of SC is the same as the boundary of the square, one might speculate that $d\nu$ could just be Lebesgue measure, but our data does not support this supposition, and indeed there is no reason to believe that the measure should be so place independent, since the geometry of SC near boundary points varies considerably from point to point. 

We are not prepared to put forward a conjectural definition of the normal derivative.  Instead we investigate the preliminary question: what is the behavior of $u(x)$ as $x$ approaches a boundary point $z$?  When $u(z)=0$, as it does in a number of our computed examples, this becomes the question of the decay rate of $u(x)$ as $x\to z$.  

  First, we will outline the methods used to measure decay rates.  Let $z$ be a point on $\partial$SC$_n$.  Suppose there exist three $n$ cells in a line perpendicular to the edge containing $z$.  We denote them by $x_j$ for $j=1,2,3$ with $z\in x_1$.  Thus, we exclude points $z$ that lie near an $n$-level hole and for any boundary edge of SC$_n$, we can calculate decay rates for exactly $\frac{2}{3} 3^n$ of the $n$-cells intersecting that edge.  Note that the average distance to the boundary for the cell $x_j$ is $\frac{(j-1/2)}{3^n}$ for $j=1,2,3$.  To find the decay rates we will be measuring at what rate 
$u(x_j)\to 0 $ for $j=1,2,3$, when $u(z)=0$.

A naive guess would be that
\begin{equation}\label{6.2}
u(x_j)-u(z)\approx A\left(\frac{(j-1/2)}{3^n}\right)^{\alpha}
\end{equation}
for some power $\alpha$ and $A \neq 0$.  In that case, we would set $A=\partial_n u(z)$.

We present decay rates of various functions that vanish along $\partial$SC starting with the harmonic functions from Figure 3.1 
, namely, the harmonic functions, $h_k$, defined by $\sin(k\pi t)$ along the top edge and zero along the other edges along $\partial$SC.  We will look at the decay rates given by \eqref{6.2} where $z$ is along the bottom edge (furthest from $\sin(k\pi t)$).  We use the first 6 harmonic functions, $h_k$, at the highest level available (for us, level-6 harmonic functions).  For each of the $\frac{2}{3}\cdot 3^6$ acceptable cells along the bottom boundary, we can compute estimates for $A$ and $\alpha$ in \eqref{6.2}. 

\begin{figure}[p]
\begin{center}
\subfloat[$h_1$]{\includegraphics[scale=.4]{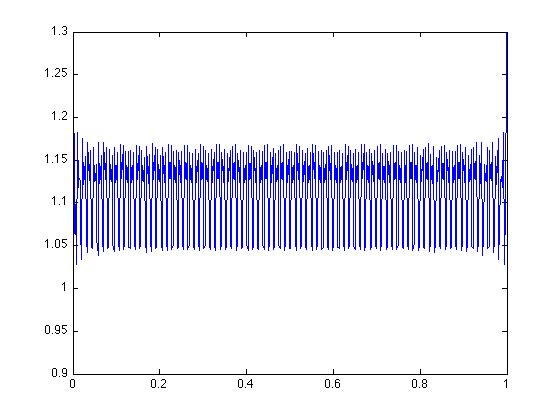}}
\subfloat[$h_2$]{\includegraphics[scale=.4]{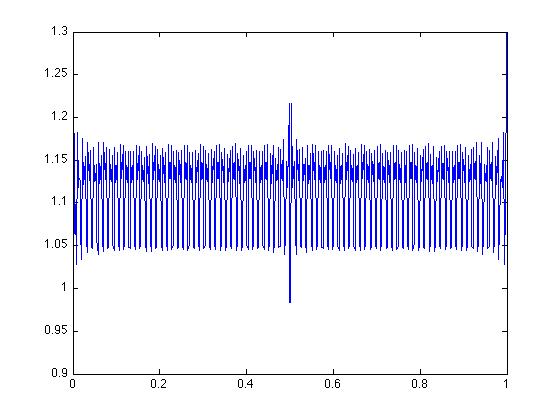}}\\
\subfloat[$h_3$]{\includegraphics[scale=.4]{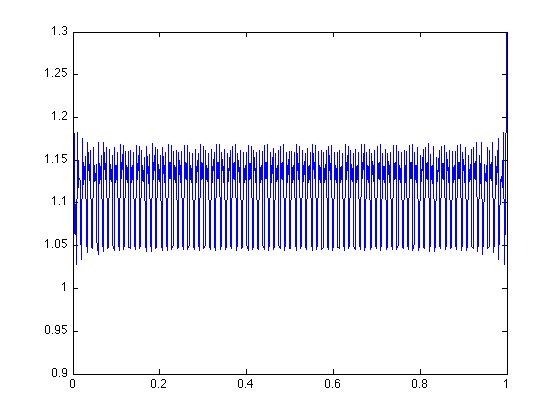}}
\subfloat[$h_4$]{\includegraphics[scale=.4]{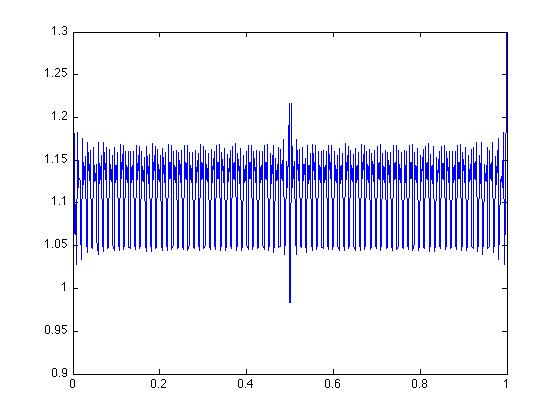}}\\
\subfloat[$h_5$]{\includegraphics[scale=.4]{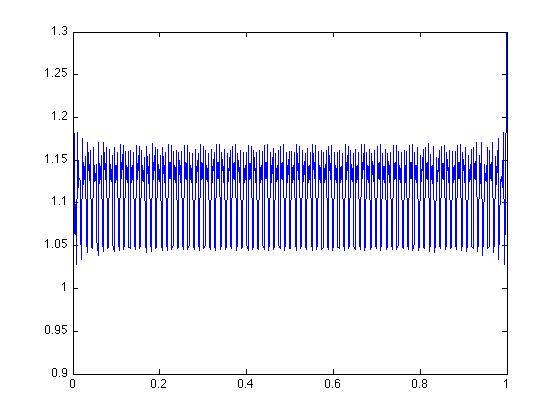}}
\subfloat[$h_6$]{\includegraphics[scale=.4]{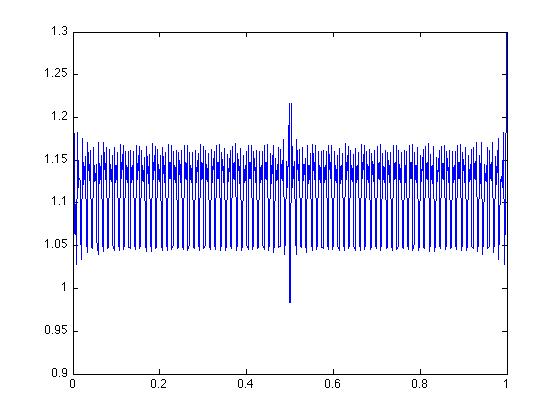}}
\end{center}
\caption{Estimates of $\alpha$ for $h_k$.}\label{fig6.1}
\end{figure}

Figure \ref{fig6.1} shows estimates for $\alpha$ for each of the $\frac{2}{3}\cdot 3^6$ cells along the bottom boundary of SC$_6$.  The graphs of $\alpha$ for $h_k$ for $k$ odd are nearly identical to each other; the same is true for $k$ even (they differ by at most $10^{-4}$).  It is clear from the data that $\alpha$ is not constant, but depends on the point $z$, with an average value of 1.1135.  The behavior for $z$ near a corner point is different as might be expected, but otherwise the variation of $\alpha$ with $z$ follows a distinctive pattern.  

\begin{figure}[p]
\begin{center}
\subfloat[$h_1$]{\includegraphics[scale=.4]{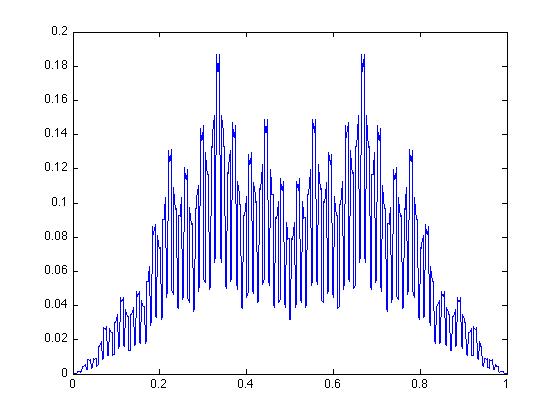}}
\subfloat[$h_2$]{\includegraphics[scale=.4]{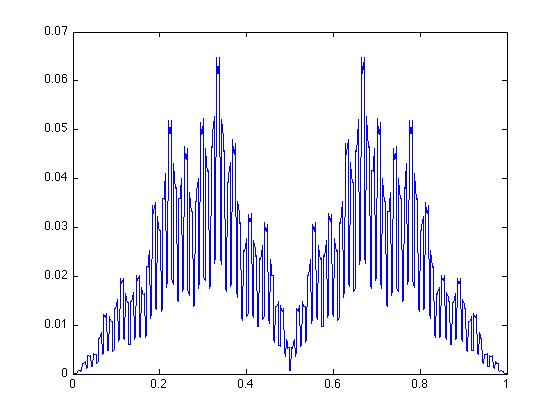}}\\
\subfloat[$h_3$]{\includegraphics[scale=.4]{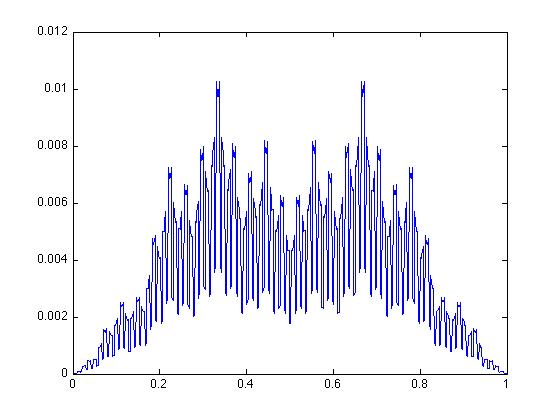}}
\subfloat[$h_4$]{\includegraphics[scale=.4]{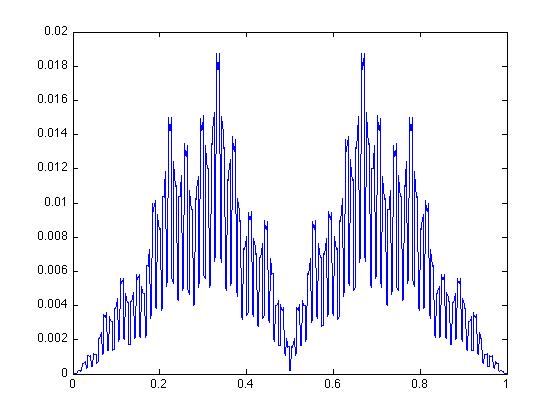}}\\
\subfloat[$h_5$]{\includegraphics[scale=.4]{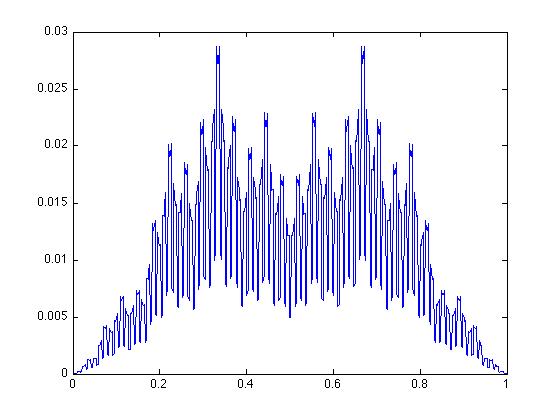}}
\subfloat[$h_6$]{\includegraphics[scale=.4]{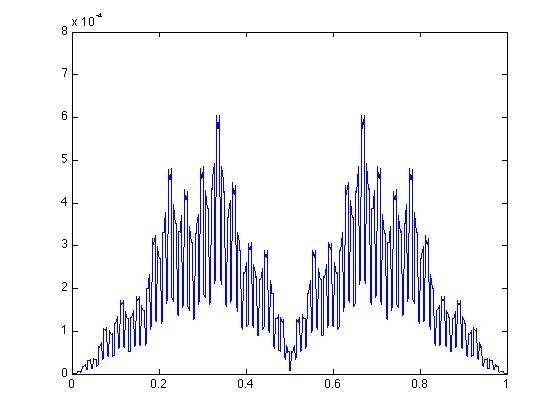}}
\end{center}
\caption{Estimates of $A$ for $h_k$.}\label{fig6.3}
\end{figure}

Figure \ref{fig6.3} shows the estimates for $A$ for each of the $h_k$.  Just as with our estimates for $\alpha$, the graph shapes for $k$ odd and even are virtually indistinguishable.  

\begin{figure}[p]
\begin{center}
\subfloat[$\varphi_1$]{\includegraphics[scale=.4]{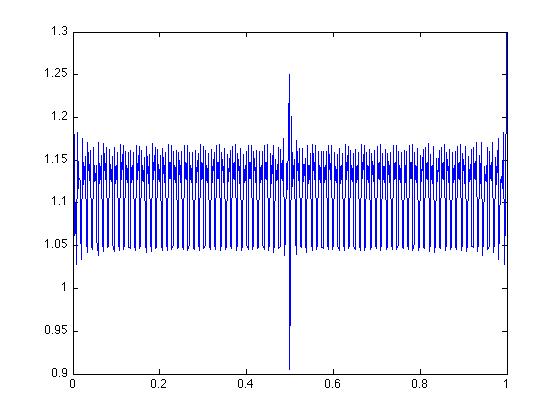}}
\subfloat[$\varphi_2$]{\includegraphics[scale=.4]{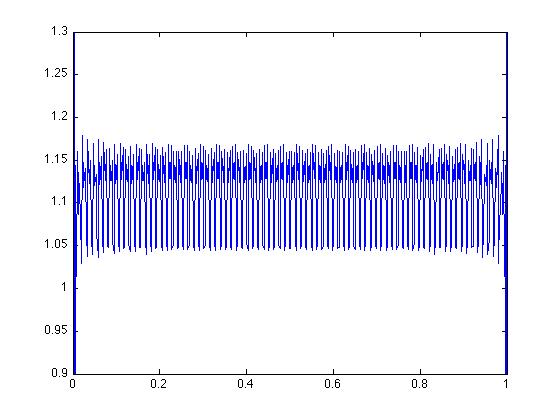}}\\
\subfloat[$\varphi_3$]{\includegraphics[scale=.4]{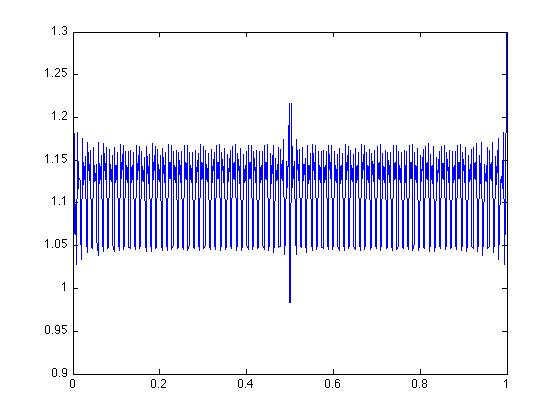}}
\subfloat[$\varphi_4$]{\includegraphics[scale=.4]{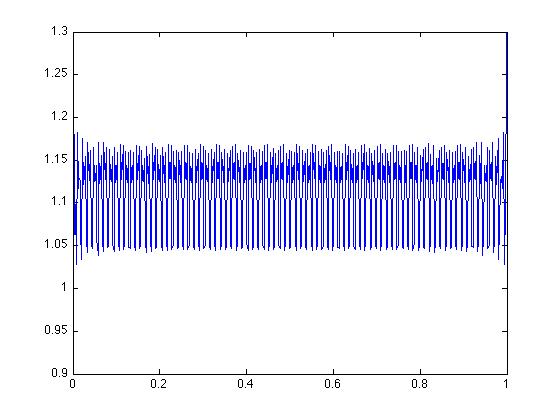}}\\
\subfloat[$\varphi_5$]{\includegraphics[scale=.4]{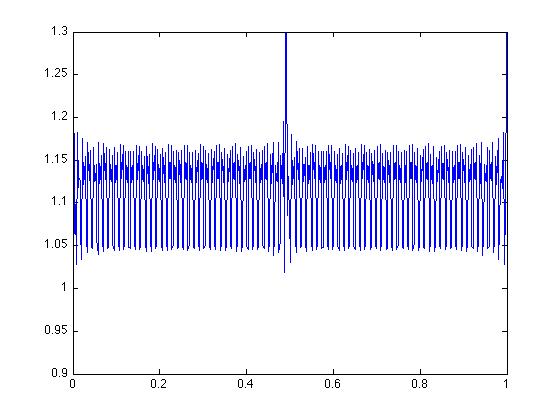}}
\subfloat[$\varphi_6$]{\includegraphics[scale=.4]{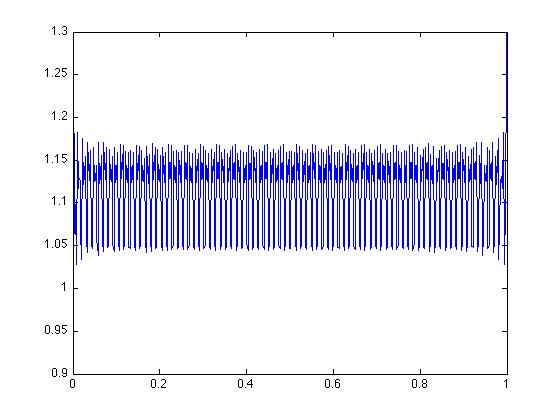}}
\end{center}
\caption{Estimates of $\alpha$ for $\varphi_k$.}\label{fig6.4}
\end{figure}

\begin{figure}[p]
\begin{center}
\subfloat[$\varphi_1$]{\includegraphics[scale=.4]{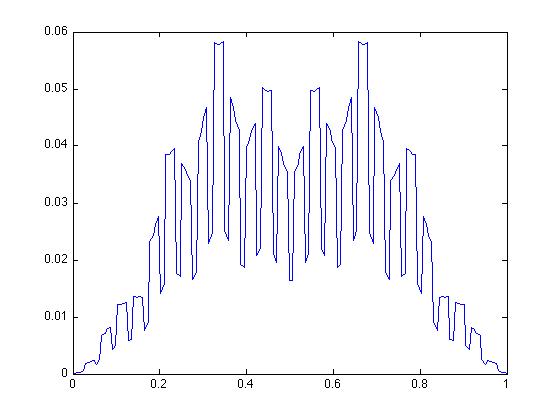}}
\subfloat[$\varphi_2$]{\includegraphics[scale=.4]{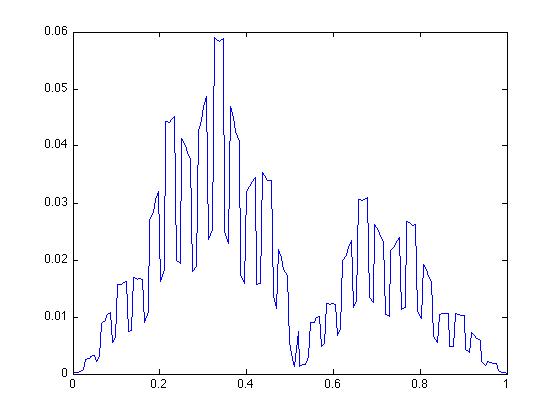}}\\
\subfloat[$\varphi_3$]{\includegraphics[scale=.4]{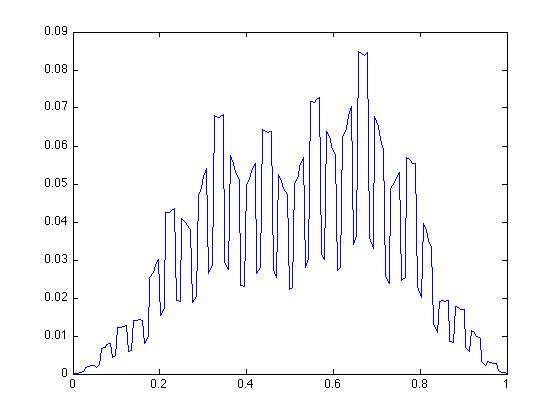}}
\subfloat[$\varphi_4$]{\includegraphics[scale=.4]{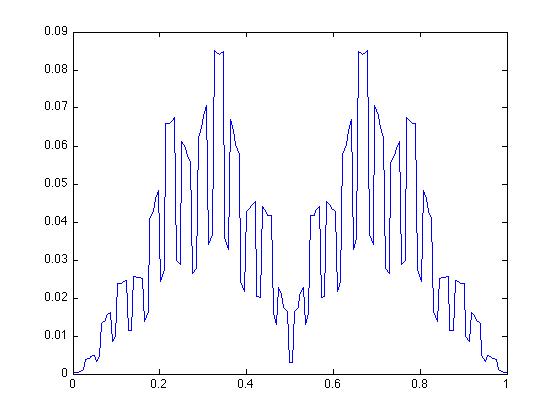}}\\
\subfloat[$\varphi_5$]{\includegraphics[scale=.4]{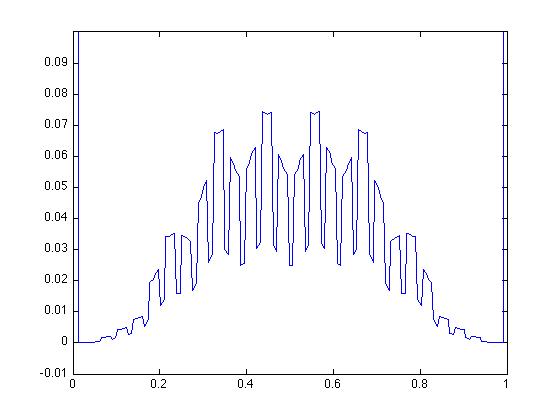}}
\subfloat[$\varphi_6$]{\includegraphics[scale=.4]{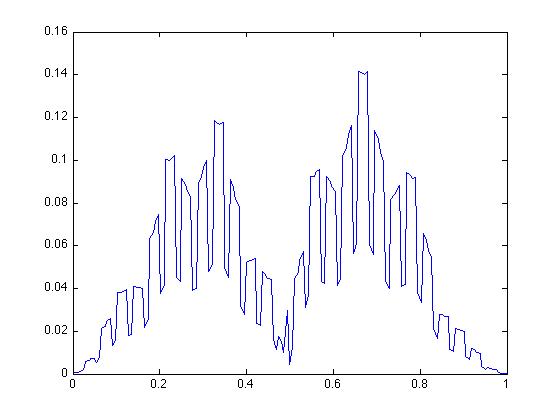}}
\end{center}
\caption{Estimates of $A$ for $\varphi_k$.}\label{fig6.5}
\end{figure}

\begin{figure}[p]
\hspace{-2.5cm}\includegraphics[scale=.4]{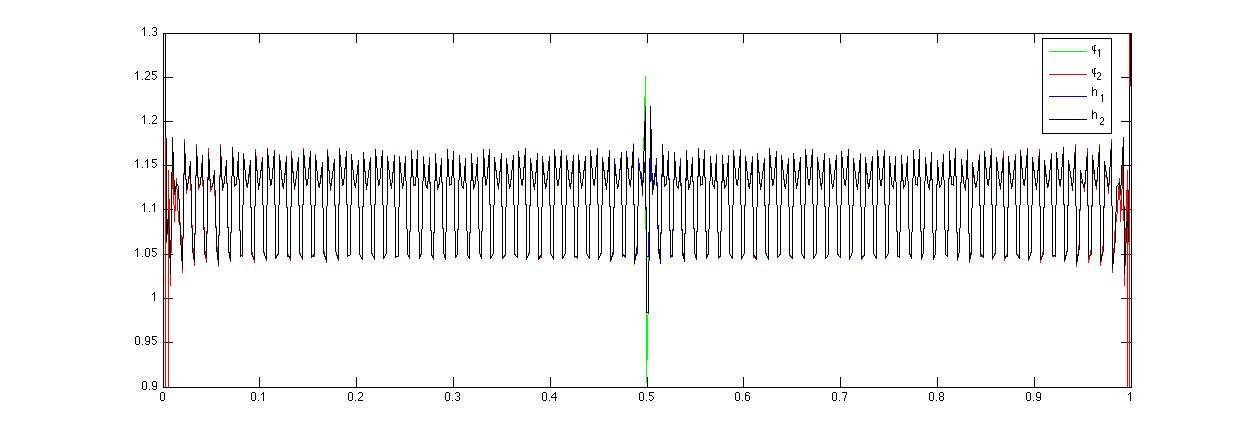}\hspace{3cm}\\ \\
\hspace{13in}\includegraphics[scale=.4]{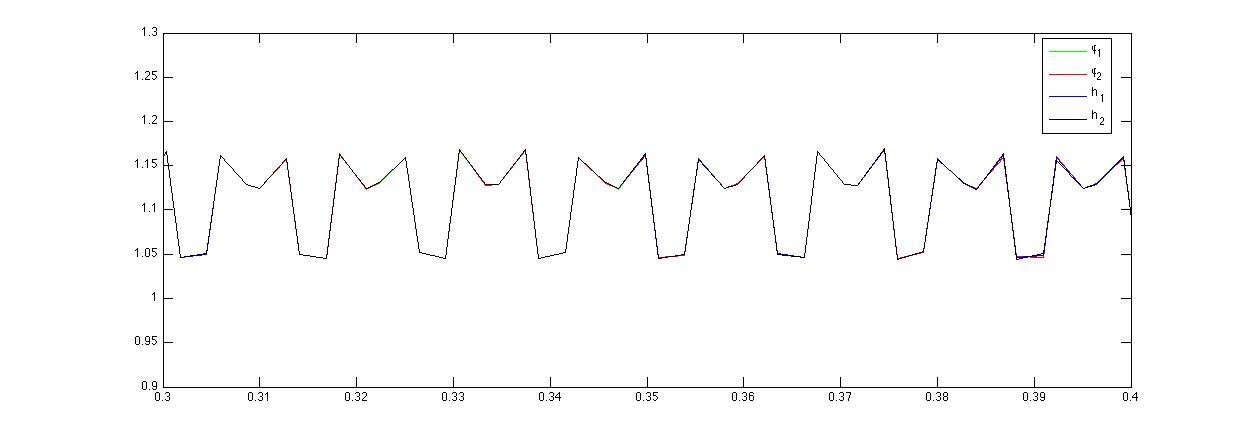}\hspace{29cm} \\
\caption{Estimates for $\alpha$ for $h_1, h_2, \varphi_1$, and $\varphi_2$ for the whole interval and zoomed in to a small interval away from the corners and center point.  Note that in the zoom the four graphs are virtually indistinguishable.}\label{fig6.6}
\end{figure}

We next present the same results for Dirichlet eigenfunctions.  We present data using the first six level-5 Dirichlet eigenfunctions $\varphi_k$.  We compute $\alpha$ and $A$ exactly as was done for the harmonic functions above.  Figure \ref{fig6.4} shows the estimates for $\alpha$ for the first six $\varphi_k$.  Just as with the harmonic functions, the behavior of $\alpha$ with $z$ is similar with the exception of the corners and the centers for the even $\varphi_k$.  These estimates for $\alpha$ vary slightly more than for the harmonic functions but we obtain an average value of $\alpha \approx 1.1152$.  This is slightly higher than the $\alpha\approx 1.1135$ but this is most likely due to the higher variation at the corners.  Figure \ref{fig6.5} shows the estimates for $A$.  Just as with the harmonic functions, the general shape for the $\varphi_{odd}$ and $\varphi_{even}$ are preserved.  Figure \ref{fig6.6} shows the simultaneous graphs of $\alpha$ for the four examples $h_1, h_2, \varphi_1, \varphi_2$.


As we saw, the boundary behavior acts quite differently at the corners of $\partial$SC.  We now focus on corner behavior and present some results.

We calculate the convergence at corner points exactly as before except for $j=2$ and $3$, $u(x_j)$ becomes the average value of the two cells that are $(j-1)$ cells away from $z\subset x_1$.  That is, we calculate the corner decay rates from the two perpendicular directions at the corner.  Again, we guess that 
\begin{equation*}
u(x_j)-u(z)\approx A\left(\frac{(j-1/2)}{3^n}\right)^{\alpha}.
\end{equation*}

First, take the harmonic functions from before, $h_k$.  There are two types of corner points for these functions.  There are `top' corners, which have one edge equal to zero and one with $\sin(\pi k x)$; and there are bottom corners, which have both edges set to zero.  For this reason, it is natural to expect the bottom corners to exhibit a higher value of $\alpha$ which we observe as shown in Tables \ref{tab6.1} and \ref{tab6.2}.  Also note that each of the harmonic functions have the same `bottom' $\alpha$.

\begin{table}[h]\begin{center}
\begin{tabular}{| r | c | c | c | c | c | c |}
\hline
& \multicolumn{6}{|c|}{$k$}\\ \hline
& 1 & 2 & 3 & 4 & 5 & 6 \\ \hline
$\alpha$ &  0.6884    &0.6882    &0.6879   & 0.6875    &0.6871 &  0.6867\\  \hline
$A$ &     0.2802   & 0.5595    &0.8374    &1.1134   &1.3872    &1.6587 \\ \hline
\end{tabular}\end{center}
\caption{Values of $\alpha$ and $A$ for the top corner of the harmonic functions $h_k$.}\label{tab6.1}
\end{table}

\begin{table}[h]\begin{center}
\begin{tabular}{| r | c | c | c | c | c | c |}
\hline
& \multicolumn{6}{|c|}{$k$}\\ \hline
& 1 & 2 & 3 & 4 & 5 & 6 \\ \hline
$\alpha$ &   2.6903   & 2.6903    &2.6903   & 2.6903   & 2.6903  &  2.6903\\  \hline
$A\times 10^{-3}$ &         0.1356   & 0.0623&    0.0077   & 0.0178   & 0.0207   & 0.0005 \\ \hline
\end{tabular}\end{center}
\caption{Values of $\alpha$ and $A$ for the bottom corner of the harmonic functions $h_k$.}\label{tab6.2}
\end{table}

We do the same computation for the Dirichlet eigenfunctions.  However, for these functions, we do not need to test top and bottom corners.  Eigenfunctions corresponding to an eigenvalue with multiplicity 1 will automatically be symmetric from any corner.   Eigenfunctions corresponding to an eigenvalue with multiplicity 2 will have both corners accounted for since the two eigenfunctions are just a 90 degree rotation of one another.  All values in Table \ref{tab6.3} are taken from the top left corner of SC.

\begin{table}[h]
\begin{center}
\begin{tabular}{| r | c | c | c | c | c | c |}
\hline
& \multicolumn{6}{|c|}{$k$}\\ \hline
& 1 & 2 & 3 & 4 & 5 & 6 \\ \hline
$\alpha$ &    1.4119    &1.4119   & 1.4119 &   1.4119  &  1.4119  &  1.4119\\  \hline
$A\times 10^{-4}$ &     0.3131   & 0.3240   & 0.2795 &   0.1298  &  0.1881  &  0.1385 \\ \hline
\end{tabular}\end{center}
\caption{Values of $\alpha$ and $A$ for the corner of the Dirichlet eigenfunctions $\varphi_k$.}\label{tab6.3}
\end{table}

\section{Periodic covering spaces}

There are many covering spaces of SC.  We are interested in spaces which have a $\mathbb{Z}$-action of isometries, for then we may use ordinary Fourier series to describe the spectrum of the Laplacian in terms of eigenfunctions on SC satisfying certain periodic boundary conditions.  Analogous examples for Julia sets are discussed in \cite{ADS}.  We discuss two examples.  The first is the \emph{staircase} in which the eight 1-cells of SC are cut along one edge and the cut edges are joined up to the next level (up and down) of a spiral staircase.  The $\mathbb{Z}$ action moves up and down the levels, and the projection onto SC just maps each 1-cell of the staircase to the corresponding 1-cell in SC.  The second example, the \emph{strip}, consists of an infinite horizontal row of copies of SC identified along their vertical boundary edges.  This is strictly speaking not a covering space because there is no covering map, but it is close enough in spirit.  It is a fractafold \cite{S1}, in that every point has a neighborhood that is isometric to an open set in SC so in particular the definition of the Laplacian lifts to the strip.  The $\mathbb{Z}$-action is just horizontal translation.  

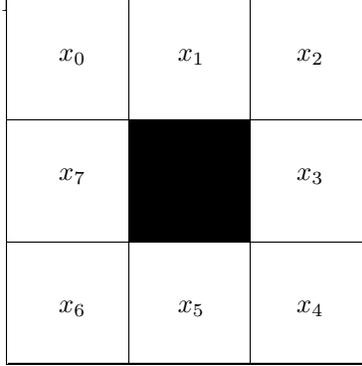
\begin{figure}[ht]
\begin{center}
\begin{picture}(20,120)(0,0)
\put(-60,0){\includegraphics[scale=.4]{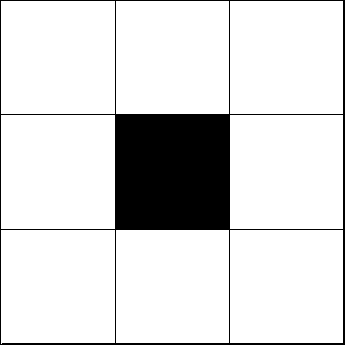}}
\put(-40,20){$x_6$}
\put(-40,70){$x_7$}
\put(-40,115){$x_0$}
\put(5,115){$x_1$}
\put(50,115){$x_2$}
\put(50,70){$x_3$}
\put(50,20){$x_4$}
\put(5,20){$x_5$}
\end{picture}
\end{center}
\caption{The eight level-1 cells of SC used in describing the strip covering space.}\label{figstrip8}
\end{figure}

We describe the situation for the strip first, since it is simpler.  The strip is a subset of $\mathbb{R}\times [0,1]$.  For any $\theta$ in $[0,1]$ we look for eigenfunctions satisfying
\begin{equation}
u(x+1,y)=e^{2\pi i \theta} u(x,y) \label{7.1}
\end{equation}
and Neumann boundary conditions on the top and bottom lines of the strip.  We call this the $\theta$-spectrum.  To be specific, on level 1 we have the eight 1-cells $x_0,...,x_7$ (see Figure \ref{figstrip8}).

At $x_1$ and $x_5$ the eigenvalue equations are the same as for SC
\begin{eqnarray*}
2u(x_1)-u(x_0)-u(x_2)=\lambda u(x_1)\\
2u(x_5)-u(x_4)-u(x_6)=\lambda u(x_5)
\end{eqnarray*}

However, the other 1-cells have neighbors in the strip that are translates by $\pm 1$ in the $x$ direction of other 1-cells, so

\begin{eqnarray*}
3u(x_0)-u(x_1)-u(x_7)-e^{-2\pi i \theta} u(x_2)=\lambda u(x_0)\\
3u(x_2)-u(x_1)-u(x_3)-e^{-2\pi i \theta} u(x_0)=\lambda u(x_2)\\
3u(x_3)-u(x_2)-u(x_4)-e^{-2\pi i \theta} u(x_7)=\lambda u(x_3)\\
\text{etc.}
\end{eqnarray*}
Note that the complex conjugate of a $\theta$-eigenfunction will be a $(1-\theta)$-eigenfunction, so the $\theta$-spectrum is real and invariant under reflection about $\theta=\frac{1}{2}$.  Vertical reflections about $y=\frac{1}{2}$ in the strip maps $\theta$-eigenfunctions to $\theta$-eigenfunctions, so the $\theta$-spectrum splits into the even and odd eigenfunctions.  Horizontal reflections map $\theta$-eigenfunctions to $(1-\theta)$-eigenfunctions.  Since the symmetry group of the strip is abelian, we do not expect multiplicities greater than one in any of the $\theta$-spectra, except for coincidences. 

\begin{figure}[p]
\begin{center}
\begin{picture}(100,400)(0,0)
\put(-210,-20){\includegraphics[scale=.9]{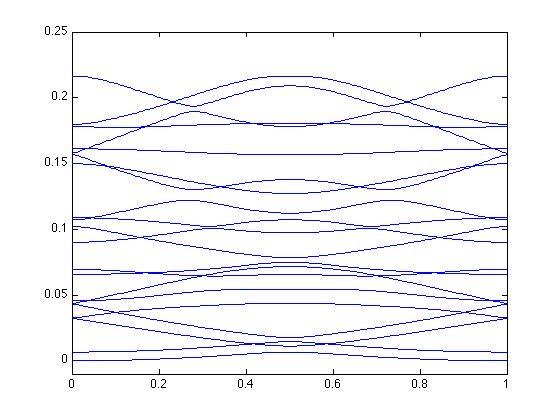}}
\end{picture}
\end{center}
\caption{Graph of the lowest twenty eigenvalues of the $\theta$-spectrum for the strip covering space with the horizontal axis being $\theta$ and the vertical axis being $\lambda$.}\label{figstripspectrum}
\end{figure}

\begin{figure}[p]
\begin{center}
\begin{picture}(100,400)(0,0)
\put(-210,-20){\includegraphics[scale=.9]{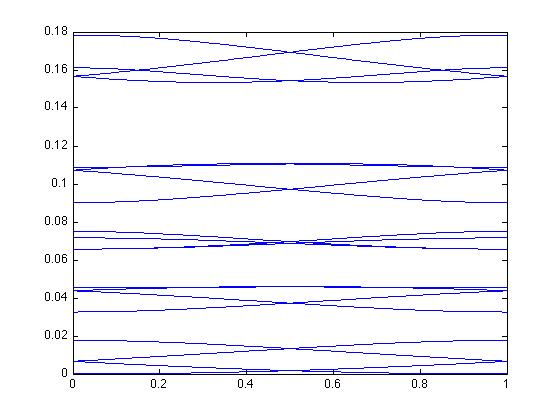}}
\end{picture}\end{center}
\caption{Graph of the lowest twenty eigenvalues of the $\theta$-spectrum for the staircase covering space with the horizontal axis being $\theta$ and the vertical axis being $\lambda$.}\label{figstaircasespectrum}
\end{figure}

\begin{figure}[p]
\begin{center}
\begin{picture}(100,400)(0,0)
\put(-240,-20){\includegraphics{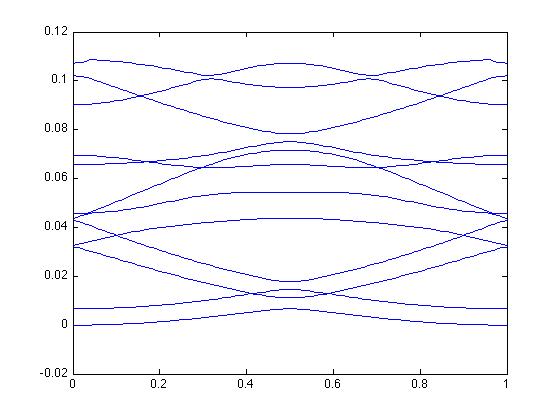}}
\put(-190,70){$+$}
\put(-190,90){$-$}
\put(-190,148){$+$}
\put(-190,154){$+$}
\put(-190,175){$-$}
\put(-190,180){$-$}
\put(-190,190){$+$}
\put(-190,235){$-$}
\put(-190,245){$+$}
\put(-190,295){$+$}
\put(-190,325){$-$}
\put(-190,340){$+$}

\put(275,70){$+$}
\put(275,90){$-$}
\put(275,148){$+$}
\put(275,154){$+$}
\put(275,175){$-$}
\put(275,180){$-$}
\put(275,190){$+$}
\put(275,235){$-$}
\put(275,245){$+$}
\put(275,295){$+$}
\put(275,325){$-$}
\put(275,340){$+$}

\put(45,92){$+$}
\put(45,98){$+$}
\put(45,112){$-$}
\put(45,120){$-$}
\put(45,185){$+$}
\put(45,210){$+$}
\put(45,230){$+$}
\put(45,251){$-$}
\put(45,255){$-$}
\put(45,270){$-$}
\put(45,315){$+$}
\put(45,340){$+$}

\end{picture}
\end{center}
\caption{First 12 eigenvalues of the spectrum of the strip with the symmetry types labeled.}\label{figstripsymmetry}
\end{figure}

The spectrum of the Laplacian on the strip is continuous, and consists of the union of all the $\theta$-spectra.  In Figures \ref{figstripspectrum}-\ref{figstaircasespectrum} we show graphs of the bottom of the $\theta$-spectra, the horizontal axis being $\theta$ and the vertical axis being $\lambda$, the $\theta$-eigenvalues.  It is easy to identify the vertical symmetry type of the eigenfunctions for $\theta=0$ and $\theta=\frac{1}{2}$.  Of course the $\theta$-spectrum varies continuously with $\theta$, so we see a set of curves going from $\theta=0$ to $\theta=1$ with some crossings and near crossings.  Generically, we do not expect to see crossings for pairs of eigenfunctions of the same symmetry type, but crossings may occur for pairs with opposite symmetry types.  It is difficult to distinguish crossings and near crossings from the data, but tracking symmetry types is helpful in deciding.  See Figure \ref{figstripsymmetry}


\begin{figure}[h]
\begin{center}
\begin{picture}(100,100)(0,0)
\put(-210,0){\includegraphics[scale=.5]{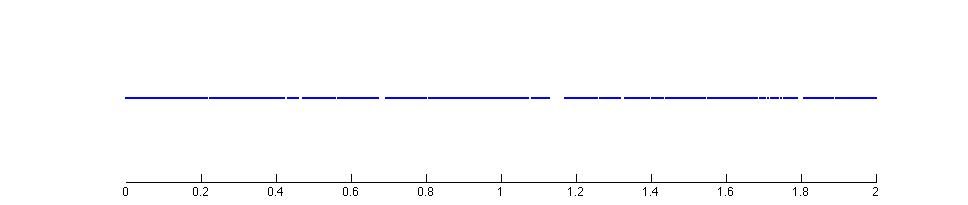}}
\end{picture}\end{center}
\caption{The $\theta$-spectra of the strip projected onto the $\lambda$ axis.  It is a union of intervals.}\label{figstripintervals}
\end{figure}

In Figure \ref{figstripintervals} we show the bottom portion of the spectrum of the strip, obtained by projection the $\theta$-spectra onto the $\lambda$-axis (shown horizontally in the figure).  We observe that it is a union of intervals.  It is not clear whether or not we should expect this pattern to persist for larger $\lambda$ values, or if the spectrum contains an infinite interval $[a,\infty)$.

For the staircase, we assume that the cut is placed between 1-cells $x_7$ and $x_0$.  The $\theta$-eigenvalue equations on level 1 are the usual away from the cut,
\begin{equation*}
2 u(x_j)-u(x_{j-1})-u(x_{j+1})=\lambda u(x_j), \quad j=1,...,6,
\end{equation*}
and
\begin{eqnarray*}
2u(x_0)-u(x_1)-e^{-2\pi i \theta} u(x_7)=\lambda u(x_0)\\
2u(x_7)-u(x_6)-e^{2\pi i \theta} u(x_0)=\lambda u(x_7)\\
\end{eqnarray*}
along the cut.  Again the complex conjugate interchanges $\theta$-eigenfunctions and $(1-\theta)$-eigenfunctions, so we have $\theta$-spectra symmetric about $\theta=\frac{1}{2}$.

In this example we have an additional symmetry of $\frac{1}{4}$ rotation, denoted $R$.  (Note that we could have used $R$ to generate a more refined $\mathbb{Z}$-action, but this would require a choice of fundamental domain that does not split into a simple union of 1-cells.)  For a $\theta$-eigenfunction we have
\begin{equation}
u(R^4x)=e^{2\pi i \theta} u(x) \label{7.2}
\end{equation}
as the analog of \eqref{7.1}, so we can further split the eigenspaces according to the symmetry types
\begin{equation}
u(Rx)=e^{2\pi i (\frac{\theta}{4}+\frac{k}{4})} u(x), \quad k=0,1,2,3 \label{7.3}.
\end{equation}

For generic $\theta$ there will be four independent families (for each choice of $k$), but for $\theta=0$ or $\theta=\frac{1}{2}$ there will be families of multiplicity two.  Of course $\theta=0$ corresponds with the Neumann spectrum of SC, and the cases

\begin{equation}\label{7.4}
u(Rx)=u(x), \quad \text{ for } k=0,
\end{equation}
and
\begin{equation}\label{7.5}
u(Rx)=-u(x), \quad \text{ for } k=2,
\end{equation}
give the multiplicity one spaces, while the $k=1,3$ cases combine to give a multiplicity two space satisfying
\begin{equation}\label{7.6}
\left\{ \begin{array}{r c l}
u(Rx)&=&-v(x) \\
v(Rx)&=&u(x) \end{array}\right. .
\end{equation}

For $\theta=\frac{1}{2}$ we can group the four cases into groups of two as follows:
\begin{equation}\label{7.7}
\left\{ \begin{array}{r c l}
u(Rx)&=&e^{\frac{\pi i}{4}}u(x)=\frac{\sqrt 2}{2} (1+i) u(x), \quad k=0 \\
u(Rx)&=&e^{\frac{7\pi i}{4}}u(x)=\frac{\sqrt 2}{2} (1-i) u(x), \quad k=3
 \end{array}\right. 
\end{equation}

\begin{equation}\label{7.8}
\left\{ \begin{array}{r c l}
u(Rx)&=&e^{\frac{3\pi i}{4}}u(x)=\frac{\sqrt 2}{2} (-1+i) u(x), \quad k=1 \\
u(Rx)&=&e^{\frac{5\pi i}{4}}u(x)=\frac{\sqrt 2}{2} (-1-i) u(x), \quad k=2
 \end{array}\right. .
\end{equation}

Replacing the complex valued $u$ by $u+iv$ for real-valued $u$ and $v$ yields
\begin{equation}\tag{7.7'}
\left\{ \begin{array}{r c l}
u(Rx)&=&\frac{\sqrt 2}{2} (u(x)-v(x)) \\
v(Rx)&=&\frac{\sqrt 2}{2} (u(x)+v(x)) 
 \end{array}\right.,
\end{equation}
and
\begin{equation}\tag{7.8'}
\left\{ \begin{array}{r c l}
u(Rx)&=&\frac{\sqrt 2}{2} (-u(x)-v(x)) \\
v(Rx)&=&\frac{\sqrt 2}{2} (u(x)-v(x)) 
 \end{array}\right..
\end{equation}

\begin{figure}[h]
\begin{center}
\begin{picture}(100,100)(0,0)
\put(-210,0){\includegraphics[scale=.6]{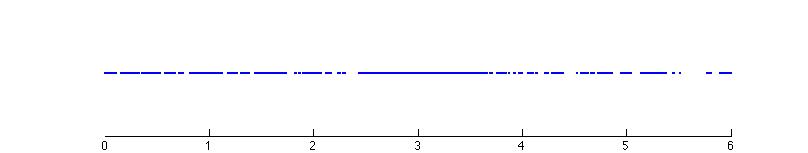}}
\end{picture}
\end{center}
\caption{The $\theta$-spectra of the staircase projected onto the $\lambda$ axis.  It is a union of intervals.}\label{figstaircaseintervals}
\end{figure}

Thus at $\theta=\frac{1}{2}$ all eigenspaces have multiplicity 2 (or a multiple of 2), and they come in two symmetry types that we denote $(0,3)$ and $(1,2)$.  Moreover, as $\theta$ varies between 0 and $\frac{1}{2}$, the multiplicity two eigenspace of type \eqref{7.6} will split and one curve will join up at $\theta=\frac{1}{2}$ with one curve starting from \eqref{7.4} to form a $(0,3)$ eigenspace, while the other curve will join up with one from \eqref{7.5} to form a $(1,2)$ eigenspace.  Thus we expect to see groups of 4 curves (or $4n$ if there are crossings) in the $\theta$-spectra.  This is apparent in Figure \ref{figstaircasespectrum}, where we see groups of 4,4,4,4,8,... starting from the bottom, and sizable gaps between groups.  Figure \ref{figstaircaseintervals} shows the spectrum of the staircase, and in this example it appears likely that the pattern will persist and yield an infinite disjoint union of finite intervals.

\section{Conclusions}

In this paper we have presented a substantial body of numerical computations of basic functions on SC.  While there is no theoretical estimate for the rate of convergence of the algorithms we used, the results obtained give strong experimental evidence that we are in fact obtaining reliable accuracy.  For some of the computations it would have been desirable to continue to greater depth in the approximation to SC, for the most part the error is acceptably small.  So we achieved a ``picture book'' glimpse of basic calculus on SC.

One general conclusion is that calculus on SC is somewhat more regular than on SG and other PCF fractals.  Functions such as harmonic functions, eigenfunctions of the Laplacian, Poisson kernels, and heat kernels appear more regular when graphed against the Euclidean embedding of SC in the plane.  The spectrum of the Laplacian, as observed in the Weyl ratio of the eigenvalue counting function, is somewhat smoother.  $L^2$-normalized eigenfunctions of the Laplacian appear to remain uniformly bounded throughout the spectrum, or at worst have logarithmic growth with the eigenvalue.  On the other hand, since there are apparently no large gaps in the spectrum, there are no nice Dirichlet kernels with approximate identity behavior, so partial sums of Fourier series are not well-behaved.  

The geometry of SC is very far from being homogeneous.  In particular there are ``corner points'', lying at the corners of each of the deleted squares in the construction of SC.  These points have the same local geometry when the scaled is reduced, as compared with generic points, where the local geometry varies with the scale.  It is not surprising that the behavior of functions at corner points is somewhat different from the behavior at generic points.  Further work is needed to elucidate these differences.  But perhaps the difference is not as pronounced as in the junction point/nonjunction point dichotomy for PCF fractals. 

What is the ``natural'' metric on SC?  There are several competing metrics to consider: the metric inherited from the embedding of SC in the plane, the geodesic metric inherited from that embedding, and the effective resistance metric.  There are also more qualitative geometric objects, like the level sets of the heat kernel, that are roughly comparable to spheres in the effective resistance metric.  We would need to be able to compute effective resistances to greater resolution in order to draw more precise conclusions.

We have also been successful in understanding the spectrum of several fractafolds based on SC.  In the compact case we have looked at boundary identifications corresponding to those that produce the torus, Klein bottle, and projective plane when applied to the square that contains SC.  Our most striking conjecture based on experimental evidence is that the differences of the eigenvalue counting functions for these three fractafolds remains uniformly bounded, a statement that is evidently false in the nonfractal case.  In the noncompact case we have looked at two periodic covering spaces, the strip (not technically a true covering space) and the staircase.  We have experimental evidence that the spectrum is continuous and consists of a countable union of intervals. 

We have not been successful in defining a normal derivative on the boundary of SC with a  corresponding Gauss-Green formula.  However, we have presented a wealth of data on the decay rates of functions near the boundary, and this should be useful in further work on this interesting and important problem.

\pagebreak

\end{document}